\documentclass[a4paper,11pt]{scrartcl}

\usepackage[utf8]{inputenc}
\usepackage[T1]{fontenc}

\usepackage[british]{babel}

\usepackage[noblocks]{authblk}

\usepackage{etoolbox}
\usepackage{xstring}
\usepackage[shortlabels]{enumitem}

\usepackage{graphicx}
\usepackage{xcolor}
\definecolor{darkgreen}{rgb}{0.2,0.8,0.55}

\usepackage{mathtools,amssymb,mathrsfs}
\allowdisplaybreaks

\usepackage{bm} 					

\usepackage{braket}					

\usepackage{nicefrac}				

\usepackage{adjustbox}

\usepackage{longtable}

\usepackage{amsthm}
\usepackage{thmtools,thm-restate}

\usepackage[linesnumbered,ruled,nokwfunc,longend]{algorithm2e}
\DontPrintSemicolon
\SetKwInOut{Input}{Input}\SetKwInOut{Output}{Output}

\usepackage[style=alphabetic,maxbibnames=99,maxalphanames=4]{biblatex}

\newrobustcmd*{\citefullauthor}{\AtNextCite{\DeclareNameAlias{labelname}{given-family}}\citeauthor}

\usepackage{tikz}
\usetikzlibrary{calc} 
\usetikzlibrary{decorations.pathreplacing} 
\usetikzlibrary{positioning,calc,arrows}
\usetikzlibrary{nfold}

\usepackage{pgfplots} 

\tikzstyle{vertex} = [draw,circle]
\tikzstyle{edge} = [draw,thick,->]

\usepackage{zref-clever}
\zcsetup{nameinlink=false,cap=true}

\usepackage[bookmarks=false,colorlinks=true, linkcolor=blue, urlcolor=blue, citecolor=blue, breaklinks=true,hypertexnames=false]{hyperref}

\theoremstyle{definition}

\newtheorem{definition}{Definition}[section]
\AtBeginEnvironment{definition}{} 
\newtheorem{example}[definition]{Example}
\AddToHook{env/example/begin}%
{\zcsetup{countertype={definition=example}}}

\theoremstyle{plain}

\AddToHook{env/proposition/begin}%
{\zcsetup{countertype={definition=proposition}}}
\newtheorem{lemma}[definition]{Lemma}
\AddToHook{env/lemma/begin}%
{\zcsetup{countertype={definition=lemma}}}
\zcRefTypeSetup{lemma}{
	Name-sg-ab = Lem. 
}	
\newtheorem{theorem}[definition]{Theorem}
\AddToHook{env/theorem/begin}%
{\zcsetup{countertype={definition=theorem}}}
\zcRefTypeSetup{theorem}{
	Name-sg-ab = Thm. 
}

\AddToHook{env/conjecture/begin}%
{\zcsetup{countertype={definition=conjecture}}} 
\zcRefTypeSetup{conjecture}{
	Name-sg = Conjecture ,
	name-sg = conjecture ,
	Name-pl = Conjectures ,
	name-pl = conjectures ,
}	
 
\AddToHook{env/question/begin}%
{\zcsetup{countertype={definition=question}}} 
\zcRefTypeSetup{question}{
	Name-sg = Question ,
	name-sg = question ,
	Name-pl = Questions ,
	name-pl = questions ,
}	
\newtheorem{corollary}[definition]{Corollary}
\AddToHook{env/corollary/begin}%
{\zcsetup{countertype={definition=corollary}}}
\zcRefTypeSetup{corollary}{
	Name-sg-ab = Cor. 
}	
\newtheorem{claim}{Claim}
\AddToHook{env/claim/begin}%
{\zcsetup{countertype={definition=claim}}} 
\zcRefTypeSetup{claim}{
	Name-sg = Claim ,
	name-sg = claim ,
	Name-pl = Claims ,
	name-pl = claims ,
	Name-sg-ab = Cl. 
}	

\newenvironment{proofClaim}[1][]{\ifthenelse{\equal{#1}{}}{\begin{proof}}{\begin{proof}[#1]}}{\end{proof}}
	
\theoremstyle{remark}

\AddToHook{env/remark/begin}%
{\zcsetup{countertype={definition=remark}}}

\AddToHook{env/observation/begin}%
{\zcsetup{countertype={definition=observation}}}
\zcRefTypeSetup{observation}{
	Name-sg = Observation ,
	name-sg = observation ,
	Name-pl = Observations ,
	name-pl = observations ,
}	

\AddToHook{env/notation/begin}%
{\zcsetup{countertype={definition=notation}}} 
\zcRefTypeSetup{notation}{
	Name-sg = Notation ,
	name-sg = notation ,
	Name-pl = Notations ,
	name-pl = notations ,
}	

\zcRefTypeSetup{algocf}{
	name-sg=algorithm,
	Name-sg=Algorithm
}

\newcommand{\thmpartlabel}[1]{%
	{\zcsetup{reftype=thmpart}%
	\label{#1}}%
}
\zcRefTypeSetup{thmpart}{ 
	Name-sg = Item(?) ,
	name-sg = Item(?) ,
	Name-pl = Item(?) ,
	name-pl = Item(?) ,
}	
\makeatletter
\NewDocumentCommand{\zcreftp}{s o m}{%
	\IfSubStr{#3}{,}{
		\StrBefore[1]{#3}{,}[\FirstThm]%
	}{
		\StrLeft{#3}{1000}[\FirstThm]%
	}%
	\ifthenelse{\equal{\zref@extract{\FirstThm}{zc@type}}{thmpart}}{%
		\StrCount{\FirstThm}{:}[\NumberColons]%
		\StrBefore[\NumberColons]{\FirstThm}{:}[\ThmLabel]%
		\IfBooleanTF{#1}{%
			\zcref*[#2]{\ThmLabel}~\zcref*[noname]{#3}%
		}{%
			\zcref[#2]{\ThmLabel}~\zcref[noname]{#3}%
		}%
	}{%
		\IfBooleanTF{#1}{%
			\zcref*[#2]{#3}%
		}{%
			\zcref[#2]{#3}%
		}%
	}%
}
\makeatother

\newcommand{\proofitem}[1]{\def\pidescr{#1}%
	\item}
\newlist{structuredproof}{enumerate}{3}
\setlist[structuredproof]{
	leftmargin=1em,
    labelwidth=.3em,
    parsep=0pt,
    listparindent=\parindent,
	label=\boldmath\bfseries\sffamily\protect\pidescr:,
	align=left
}

\newcommand{\caseitem}[1]{\def\casedescr{#1}%
	\item}
\newlist{proofbycases}{enumerate}{1}
\setlist[proofbycases]{
	leftmargin=3em,
	labelwidth=1.5em,
	label=\boldmath\bfseries\sffamily\arabic*. Case \protect\casedescr:,
	ref=\arabic*,
	align=left
}

\newcommand*{\myproofname}{Proof}

\newcommand{\toverset}[2]{\overset{\text{#1}}{#2}}
\newcommand{\zoverset}[2]{\toverset{\zcref[abbrev]{#1}}{#2}}
\newcommand{\coverset}[3][]{\toverset{\ifthenelse{\equal{#1}{}}{\cite{#2}}{\cite[#1]{#2}}}{#3}}

\newcommand{\refsym}[1]{%
	\ensuremath{(%
		\ifthenelse{\equal{#1}{1}}{\ast}%
		{\ifthenelse{\equal{#1}{2}}{\#}%
		{\ifthenelse{\equal{#1}{3}}{\triangle}%
		{\ifthenelse{\equal{#1}{4}}{\bigcirc}%
		{\ifthenelse{\equal{#1}{5}}{\diamond}
		{\color{red}NaN}}}}}%
	)}%
}
\newcommand{\symoverset}[2]{\toverset{\refsym{#1}}{#2}}

\NewDocumentCommand{\stpath}{O{s}O{t}}{\ensuremath{#1,#2\text{-path}}}
\NewDocumentCommand{\stsubpath}{O{s}O{t}}{\ensuremath{#1,#2\text{-subpath}}}
\NewDocumentCommand{\stwalk}{O{s}O{t}}{\ensuremath{#1,#2\text{-walk}}}
\NewDocumentCommand{\stnetwork}{O{s}O{t}}{\ensuremath{#1,#2\text{-network}}}

\newcommand{\IR}{\mathbb{R}}
\newcommand{\IRnn}{\IR_{\geq 0}}
\newcommand{\IRp}{\IR_{>0}}

\newcommand{\INo}{\mathbb{N}_0}

\newcommand{\eps}{\varepsilon}
\newcommand{\e}{\mathrm{e}}

\DeclareMathOperator{\dist}{dist}

\newcommand{\deriv}[2][]{%
	\ifthenelse{\equal{#1}{}}%
		{\partial #2}%
		{\partial_{#1} #2}}

\newcommand{\diff}{\mathrm{d}}

\newcommand{\abs}[1]{\left|#1\right|}

\NewDocumentCommand{\pLocInt}{O{p}O{\IR}}{L^{#1}_{\mathrm{loc}}(#2)}
\NewDocumentCommand{\pInt}{O{p}O{\IR}}{L^{#1}(#2)}

\NewDocumentCommand{\CharF}{O{}}{%
	\ifthenelse{\equal{#1}{}}
	{\mathbf{1}}
	{\mathbf{1}_{#1}}
}

\usepackage{twoopt}
\newcommand{\eTTime}[1][e]{\tau_{#1}} 
\newcommand{\eWTime}[1][e]{q_{#1}} 
\newcommand{\eCap}[1][e]{\nu_{#1}} 
    \newcommand{\eCapmin}{\eCap[\min]} 
    \newcommand{\eCapmax}{\eCap[\max]} 
\newcommand{\eToll}[1][e]{p_{#1}} 
\newcommand{\sumTolls}{\hat p} 
\newcommand{\uptosumTolls}{[\hat p]_0} 
\newcommand{\sourceTime}{\theta}
\newcommand{\netTime}{\xi}
\newcommand{\sinkTime}{\varphi}
\newcommand{\qt}{\tilde q}

\newcommand{\lpref}[1]{(\ref{#1})}
\newcommand{\mincut}{\nu_C}

\newcommand{\setActiveEdgesR}{E'_r}
\newcommand{\setActiveTimesR}[1][e]{A_{#1}}

\newcommandtwoopt{\crem}[2][v][\netTime]{\tilde c_{#1}(#2)}

\newcommand{\qmNT}[1][e]{z_{#1}} 
\newcommand{\wtST}[1][e]{q_{#1}} 

\NewDocumentCommand{\wtNT}{O{e}D(){\sinkTime}}{\frac{\qmNT[#1](#2)}{\eCap[#1]}} 

\newcommand{\UpD}{U}

\newcommand{\edgesTo}[2][]{\delta\ifthenelse{\equal{#1}{}}{}{_{#1}}^-(#2)}
\newcommand{\edgesFrom}[2][]{\delta\ifthenelse{\equal{#1}{}}{}{_{#1}}^+(#2)}

\newcommand{\head}[1][]{\mathop{\mathrm{head}}\ifthenelse{\equal{#1}{}}{}{_{#1}}}
\newcommand{\tail}[1][]{\mathop{\mathrm{tail}}\ifthenelse{\equal{#1}{}}{}{_{#1}}}
\newcommand{\PathSet}[1][]{\ifthenelse{\equal{#1}{}}{\mathcal{P}}{\mathcal{P}_{#1}}}

\newcommand{\potCCO}{\Phi^{\mathrm{CCO}}}
\newcommand{\potBasic}{\Phi}    
\newcommand{\potVar}{\Phi_{\mathrm{D}}}  
\newcommand{\potCorr}{B}        
\newcommand{\fullpot}{\Psi}     
\newcommand{\fullpotVar}{\bar\Psi}

\newcommand{\dualObj}{\Phi_{\mathrm{D}}}

\usepackage{booktabs}
\usepackage{array}
\newlength{\hsp}\newlength{\colw}
\newcommand{\dfill}{\leaders\hbox{\rule[0.55ex]{0.1pt}{0.4pt}}\hfill}

\newcommand{\MainbodyOrAppendix}[2]{#1} 

\tikzset{
  vtxp/.style={circle,draw,thick,minimum size=5mm,inner sep=0pt,font=\small},
  R/.style={-{Stealth[length=2mm]},line width=1.8pt,red!75!black},          
  A/.style={-{Stealth[length=2mm]},line width=0.9pt,black},                 
  I/.style={-{Stealth[length=2mm]},line width=0.5pt,gray!55,dashed},        
  N/.style={-{Stealth[length=2mm]},line width=0.8pt,black},                 
  albl/.style={font=\scriptsize,inner sep=1pt}}
\newcommand{\phasepic}[7]{%
\begin{tikzpicture}[x=0.95cm,y=0.95cm,vtxs/.style={circle,draw,thick,minimum size=4mm,inner sep=0pt,font=\scriptsize}]
\node[vtxs](s)at(0,0){$s$};\node[vtxs](v)at(1.2,0){$v$};\node[vtxs](w)at(2.4,0){$w$};\node[vtxs](t)at(3.4,0){$t$};
\draw[#2](s)to[bend left=35](v);
\draw[#3](s)to[bend right=35](v);
\draw[#5](v)to[bend left=35](w);
\draw[#6](v)to[bend right=35](w);
\draw[#7](w)--(t);
\draw[#4](s)to[out=70,in=110,looseness=.55](t);
\node[font=\scriptsize,anchor=north]at(1.7,-0.55){#1};
\end{tikzpicture}}
\newcommand{\instancepic}{%
\begin{tikzpicture}[x=1cm,y=1cm]
\tikzset{dl/.style={font=\scriptsize,inner sep=1pt}}
\node[vtxp](s)at(0,0){$s$};\node[vtxp](v)at(3.4,0){$v$};\node[vtxp](w)at(6.8,0){$w$};\node[vtxp](t)at(9.6,0){$t$};
\draw[N](s)to[bend left=30]node[dl,above]{$e_1\!:(1,2,0)$}(v);
\draw[N](s)to[bend right=30]node[dl,below]{$e_2\!:(1,1,1)$}(v);
\draw[N, color=blue!75!black, line width=1.8pt](v)to[bend left=30]node[dl,above]{$a\!:(3,2,0)$}(w);
\draw[N, color=blue!75!black, line width=1.8pt](v)to[bend right=30]node[dl,below]{$b\!:(1,1,1)$}(w);
\draw[N, color=blue!75!black, line width=1.8pt](w)--node[dl,above]{$c\!:(1,2,0)$}(t);
\draw[N](s)to[out=60,in=120,looseness=.55]node[dl,above]{$e_3\!:(\tfrac{25}{4},6,0)$}(t);
\draw[{Stealth[length=2mm]}-,thick](s.west)--++(-.7,0)node[left,font=\small]{$u_0=6$};
\node[font=\small,anchor=north]at(4.8,-0.95){Instance with arc data $(\tau_e,\nu_e,p_e)$, echo gadget in blue.};
\end{tikzpicture}}

\title{Long-term Behavior of Nash Flows over Time with Tolls}
\author{Lukas Graf}
\affil{University of Passau}
\author{Laura Ritter}
\author{Laura Vargas Koch\footnote{L.V. is supported by a JPI Fellowship of RWTH Aachen University, funded under the Excellence Strategy of the Federal Government and the Länder.}}
\affil{RWTH Aachen}
\date{\today}

\begin{document}

\maketitle             
\begin{abstract}

We consider Nash flows over time with tolls where the arcs in the network are equipped with three constant parameters: a capacity, a free-flow transit time and a toll. Infinitesimally small users travel from a common source to a common sink, each minimizing the sum of their travel time and paid tolls.  
We show that a change of perspective to sink-based reference times allows the derivatives of equilibria in this setting to be described by a system of sink-based thin flow equations with time shifts. 
Restricting to the setting where the capacity of a minimum cut in the underlying network strictly exceeds the inflow rate, we obtain several results on the long-term behavior of Nash flows over time with tolls. First, we show that throughout the evolution of the Nash flow over time, queues remain bounded.
Moreover, we consider a primal and dual linear program (a special case of those considered in \cite{TolledDEStructure}) and show that the optimal dual solutions characterize steady states. We finally show that the distance between queues in a tolled dynamic equilibrium and the set of optimal dual solutions tends to 0.
\end{abstract}

\section{Introduction}

Nash flows over time are a well-studied model which is motivated by modeling road traffic as a continuous flow consisting of infinitesimally small particles. These particles aim to travel as quickly as possible through a network from their source to their destination.
Flow propagation along arcs is based on the Vickrey bottleneck model as proposed in~\cite{VickreyQueuingModel}. There, each arc is equipped with a free-flow transit time and a capacity. Whenever flow wants to enter an arc at a higher rate than the capacity allows for, a queue builds up and subsequent particles experience an additional waiting time on that arc.

We consider Nash flows over time with tolls, an extension of the base model where every arc is additionally equipped with a price. Instead of only minimizing travel time, particles now aim to minimize total cost, i.e., the sum of their travel time and the tolls paid on their path. This model has been studied in \cite{TolledDEStructure}, while more general tolled models were investigated in \cite{ImplementabilityCharacterization,ImplementabilityOfSystemOptima}.
A related extension is bicriteria Nash flows over time, where players minimize their cost while aiming to arrive before a certain deadline \cite{BicriteriaNashFlows}. Another variation of tolled equilibria, studied in~\cite{FOFlowOverTimewithSchedulingCosts}, allows particles to also choose their departure time at the source.

In the single-source-single-destination setting, the structure of equilibria in the base Nash flow over time model is well understood \cite{CCLDynEquil,KochSkutellaDEandPoA}.
A crucial property in these works is the network-wide FIFO property of equilibrium flows: Particles and their path choices can only be influenced by the path choices of particles entering the network ahead of them. This is no longer true in the tolled setting where particles can overtake others. As a consequence, the classical thin flow characterization of equilibria, which measures time at the source, does not carry over. As our first result, we show that tolled equilibria can still be characterized by thin flows if time is measured at the sink instead.

In game-theoretical settings, the quality of equilibrium flows is studied, as equilibria can be very different from (system) optimal solutions. One line of research here is to compare the quality of equilibria to that of optimal solutions, e.g., in terms of throughput or makespan. 
With respect to the former, \cite{KochSkutellaDEandPoA} showed that the Price of Anarchy can be unbounded while for the latter the Price of Anarchy is known to be bounded for certain simple classes of graphs \cite{TopologiesExcludingBraessParadox,DEPoA} while bounds for general network topologies are only known if one is allowed to adjust some arc capacities \cite{BFAStacklebergStrategyRoutingFlowOverTime,DEPoA}. A monotonicity conjecture which would have implied the bound to hold without changes to the underlying network has recently been refuted~\cite{josemonotonicity}.
An alternative approach for studying the quality of equilibrium flows is to analyze their long term behavior under a constant network inflow rate and, in particular, whether queues are bounded and stabilize. A state in which queues do not change anymore is then known as a \emph{steady state}. Here, \cite{LongTermBehaviorOfDE} showed in the base model that when the capacity $\mincut$ of a minimum cut in the underlying static network is larger than or equal to the network inflow rate $u_0$, queues remain bounded. Moreover, they present a primal and dual LP and show that steady state solutions are characterized by the dual LP and are always reached in finite time. In \cite{DEContUniqLT}, it was shown that the constraint on the inflow can be dropped when we generalize the notion of steady state to allow for linear growth rate of the queues. 
Recently, \cite{TolledDEStructure} generalized the LP introduced in \cite{DEContUniqLT} to the tolled setting.

\paragraph{Our contributions.} 
First, we show that tolled single-source-single-sink dynamic equilibria are characterized by a system of sink-based thin flow equations with time shifts (\zcref{lem:ThinFlowCharacterization}).
Then, we investigate the long-term behavior of Nash flows over time with tolls in the setting where $u_0<\mincut$. In this setting, the LP pair used in \cite{TolledDEStructure} simplifies and we can show the following three results:
\begin{itemize}
    \item In a tolled dynamic equilibrium queues remain bounded (\zcref{thm:queues_bounded}).
    \item Optimal dual solutions characterize steady states (\zcref{thm:Q*steadystates}).
    \item The distance between queues in a tolled dynamic equilibrium and the set of optimal dual solutions converges to 0 (\zcref{thm:convergence_queues}).
\end{itemize}

In contrast to the base model, where \cite{LongTermBehaviorOfDE} show that a steady state is always reached in finite time, we present an example (\zcref{fig:nfc,thm:non_finite_convergence}) where even though $u_0 < \mincut$ the steady state is not reached in finite time.
This complements the example presented in \cite{TolledDEStructure} for the setting where  $u_0 \geq \mincut$.

\paragraph{Main challenges and techniques.} One difficulty in characterizing equilibrium flows via thin flows is that one has to define a common reference time at which flow and label derivatives are evaluated.
In previous works, the thin flow formulation for the base model was extended to more complex settings such as spillback and time-varying free-flow transit times \cite{SpillbackSeringKoch,DETimeVarying}. 
However, all these existing thin flow formulations are based on earliest arrival labels. The earliest arrival label of a node~$v$ for time $\theta$ denotes the earliest time that a particle entering the network at time $\theta$ can reach~$v$. Here, the common reference time is just the network entry time of particles, that is, time is measured at the source. This notion is not helpful in the tolled setting since two particles entering the network at completely different points in time can arrive at~$v$ at the same instant by paying differing amounts of tolls. 
The second difficulty lies in incorporating tolls into node labels. We solve both these problems by switching the perspective from the source to the sink in two ways: by considering paths from~$v$ to the sink $t$ and anchoring time at $t$. Only doing the former and considering the total remaining cost of the best \stpath[v] when starting at a certain time in node $v$ as in \cite{TolledDEStructure} does not suffice for a thin flow structure. In this perspective, every node is treated as a source which does not result in a unifying common reference time. In contrast, additionally measuring the time required to reach~$t$ by a certain sink-based time is what allows us to define a cumulative flow in the network. In this sink-based view, we ask for latest departure times instead of earliest arrival times at nodes. We define $r_v(\sinkTime)$ as the latest time a particle can leave node~$v$ and reach sink~$t$ with the sum of the sink arrival time and the remaining tolls to be paid being at most $\sinkTime$. 

In the untolled setting, it was shown in \cite{LongTermBehaviorOfDE} that a Nash flow over time reaches a steady state in finite time. They do so by considering a primal-dual pair of linear programs and defining a potential function based on the objective function of the dual. This potential is not only bounded, its derivative is positive and strictly bounded away from~$0$ outside the steady state. 

We consider the same primal and dual LPs as \cite{LongTermBehaviorOfDE} and extend them by tolls. However, the natural potential function derived from the dual objective is no longer monotone in our tolled setting.
A crucial difficulty is that as our reference time is a deadline including tolls still to pay, different network times interact on tolled arcs both in the primal and dual LP, coupling label values for different sink times.
This complicates the translation of results from \cite{LongTermBehaviorOfDE} to our setting. 
To recover monotonicity of the potential function while retaining boundedness, we introduce a correction term taking care of the time shift. This correction is inspired by Lyapunov-Krasovskii potentials from the theory of delay systems (\cite{krasovskii1963stability}, see \cite{FRIDMAN2014271} for an introduction).

However, we cannot hope to give a lower bound in terms of network constants on the derivative of the potential outside of steady state as we provide an example in which the steady state is not reached in finite time.
Instead, we first show that queues remain bounded as monotonicity of the potential implies a lower bound on the dual objective function. While primal feasibility already yields an upper bound, we need a lower bound here as feasible dual solutions with unbounded queues would drive the dual objective function to negative infinity. 
In this part, we use the strictness of our assumption $u_0 < \mincut$ and give an argument based on static flows.

Next, we exploit that due to monotonicity and boundedness of the potential the integral of its derivative is bounded. Any change of queues contributes to this integral. We use this to show that queue changes decrease over time. 

With this at hand, we follow the basic structure of the proof of \cite[Theorem 1]{LongTermBehaviorOfDE}, where a steady state flow is used as certificate of optimality of an optimal dual solution. However, we cannot show complementary slackness and, thus, optimality directly. This is again due to time shifts in our flow which average out as we integrate over an interval extending ever further into the future. Only in the limit, as the interval length goes to infinity, is the resulting average flow a primal certificate for dual optimality. Boundedness of queues yields a sequence of times going to infinity along which the queues in the tolled equilibrium converge to a queue vector of some optimal dual solution. 
This implies that the distance between the tolled dynamic equilibrium and the set of optimal dual solutions converges to 0. We only obtain convergence to a fixed vector of queues if the dual optimal solution is unique.
Finally, while \cite{TolledDEStructure} showed that every optimal dual solution corresponds to a steady state we now complement this with a result saying that a steady state always yields a point in the set of optimal dual solutions.
In this sense we characterize the set of steady states in the setting where $u_0 < \mincut$.

\paragraph{Paper organization.} The remainder of this work is structured as follows: We first introduce the model in \zcref{sec:model} and provide key technical lemmata, regularity properties and a thin flow characterization of tolled dynamic equilibria in \zcref{sec:thin_flow_characterization}. 
We then introduce our potential function and sketch proofs for boundedness and monotonicity in \zcref{sec:potential}. Finally, we explain the implications for long term behavior in \zcref{sec:queuesconverge}.
All missing proofs and additional technical lemmata can be found in the appendix. \zcref[S]{fig:Overview} in \zcref{app:PaperOverview} provides a schematic overview over the main proof structure.
\newpage\section{Model}
\label{sec:model}

We consider a directed graph $G=(V,E)$ with a designated source node $s \in V$ and sink node $t \in V$ as well as for every arc $e \in E$ a free-flow transit time $\eTTime \in \IRp$, a capacity $\eCap \in \IRp$ and a toll $\eToll \in \INo$. 
Flow particles enter the network at the source~$s$ at a constant rate of~$u_0 \in \IRp$ starting from time~$0$. 
A \emph{flow over time} in this network is then a vector $f = (f^+_e,f^-_e) \in L_{\text{loc}}^1(\IR)^{E \times \set{\pm}}$ where $f^+_{e}(\netTime)$ and $f^-_{e}(\netTime)$ denote the in- and outflow rate, respectively, of arc~$e$ at time~$\netTime \in \IR$. Note that, while we use the whole of~$\IR$ as our time domain, we will only consider flows with support in~$\IRnn$ (consistent with the network inflow only starting at time~$0$).

For any flow~$f$ the cumulative inflow into~$e$ is denoted by $F_e^+(\netTime)\coloneqq \int_0^\netTime f_e^+(z)dz$; the cumulative outflow~$F_e^-$ is defined analogously. 
We call a flow~$f$ \emph{feasible} if it satisfies flow conservation at all nodes $v \in V$, i.e., 
\[\sum_{e\in \delta^+(v)} f_e^+(\netTime)-\sum_{e\in \delta^-(v)}f_e^-(\netTime)\begin{cases}
    = 0 & \text{if }v\in V\setminus\{s,t\}\\
    =u_0 & \text{if }v=s\\
    \leq 0 & \text{if }v=t
\end{cases} \quad\text{for almost all } \netTime\in \IRnn \]
and if $f$ is a Vickrey flow on every arc $e \in E$, i.e.,
	\[f^-_e(\netTime) = \begin{cases}
		\eCap[e], & \text{if } \qmNT(\netTime-\tau_e) > 0 \\
		\min\set{f^+_e(\netTime-\eTTime[e]),\eCap[e]}, &\text{else}
	\end{cases} \quad\text{for almost all } \netTime\in \IRnn,\]
where $\qmNT(\netTime) \coloneqq F^+_e(\netTime) - F^-_e(\netTime+\eTTime)$ is the volume of particles waiting in the queue at the tail of arc~$e$ at time~$\netTime$.

Under a given flow, a particle entering an arc~$e$ at time~$\netTime$ exits this arc by time $T_e(\netTime)\coloneqq \netTime +\tau_e +\wtNT(\netTime)$. Correspondingly, a particle entering a path $P=e_1,e_2,\dots,e_m$ at time $\netTime$, reaches its end by time $T_P(\netTime) \coloneqq (T_{e_m} \circ \dots \circ T_{e_1})(\netTime)$. For any node $v \in V$
    \[\crem \coloneqq  \min_{P \in \PathSet[v,t]}\Bigl(T_P(\netTime)+\sum\nolimits_{e \in P}\eToll[e]\Bigr)\]
denotes the minimum sum of arrival time and cost for reaching the sink from~$v$ when starting at time~$\netTime$. Here, $\PathSet[v,t]$ denotes the set of all simple paths from~$v$ to~$t$. It is easy to see that $T_e$ is Lipschitz-continuous and non-decreasing (see, e.g., \cite[Proposition~3.56]{GrafThesis}) and, hence, so are $T_P$ and $\tilde c_v$.

With this we can define tolled dynamic equilibria as in \cite[Definition~1]{TolledDEStructure}, except that they use labels~$c_v$ denoting the minimum sum of \emph{travel} time and cost for reaching the sink, whereas we consider the minimum sum of \emph{arrival} time and cost. However, this change in reference point is for notational convenience and does not affect the definition.
\begin{definition}\label{def:TolledDE}
    A feasible flow $f$ is a \emph{tolled dynamic equilibrium} if for all $e=vw \in E$ and almost all $\netTime \in \IR$ we have
		\begin{align}\label{eq:DefTolledDE}
			f^+_e(\netTime) > 0 \implies \tilde c_v(\netTime) = \eToll[e] + \tilde c_w(T_e(\netTime)).
		\end{align}
\end{definition}

\section{Thin Flow Characterization}
\label{sec:thin_flow_characterization}

In previous models, Nash flows over time have been characterized in terms of their derivatives, yielding different variants of so-called thin flows with resetting. 
We give a thin flow characterization for tolled equilibria. 
A crucial ingredient of this is a common reference time assigning a network time to each node $v$ at which all particles at $v$ have the same set of optimal outgoing arcs. 
For untolled dynamic equilibria, this is typically done using so-called $\ell$-labels where $\ell_v(\sourceTime)$ denotes the earliest time that a particle starting at~$s$ at time~$\sourceTime$ can reach~$v$. For the tolled setting, we introduce sink-based $r$-labels. The label $r_v(\sinkTime)$ denotes the latest time a particle can depart at node $v$ and still reach sink $t$ for a combined sum of arrival time and paid tolls of at most~$\sinkTime$. Although $\sinkTime$ therefore denotes a sum of arrival time and cost, we will usually refer to it simply as a (sink) time.
 In order to more easily distinguish where time is measured, we will typically use $\sinkTime$ for times measured at the sink~$t$ and $\netTime$ for snapshot times elsewhere in the network.

\begin{definition}\label{def:rLabel}
    For $v \in V$ and $\sinkTime \in \IR$ we denote by
        \[r_v(\sinkTime) \coloneq \max_{P \in \PathSet[v,t]}\sup\Set{\netTime | T_P(\netTime) + \sum\nolimits_{e \in P}\eToll \leq \sinkTime}\]
    the latest time to leave $v$ and arrive at the sink $t$ at sink time~$\sinkTime$.
\end{definition}
A direct computation (see \zcref{app:rLabelsBasicProperties}) then shows that our latest-departure time labels also satisfy the following natural Bellman-type equations:
\begin{restatable}{lemma}{restaterLabelsAlternativeEquation}\label{lem:rLabelsAlternativeEquation}
    The labels $r_v$ for any feasible flow satisfy
        \[r_v(\sinkTime) = \begin{cases}
            \sinkTime, &\text{if } v = t \\
            \max_{e=vw \in \edgesFrom{v}}\bigl(r_w(\sinkTime-\eToll)-\eTTime-\wtST(\sinkTime)\bigr), &\text{else}
        \end{cases}\MainbodyOrAppendix{,}{.}\]
    \MainbodyOrAppendix{where $\wtST(\sinkTime) \coloneqq \wtNT(r_v(\sinkTime))$ denotes the waiting time on arc~$e=vw$ at sink-time~$\sinkTime$ (i.e., at network time~$\netTime=r_v(\sinkTime)$)}{}.
\end{restatable}

A first observation is that the labels~$\tilde c_v$ are left-inverses of our node labels $r_v$ (\zcref{lem:cvLeftinverseOfrv}). This, in turn, lets us prove that, for tolled dynamic equilibria, the $r$-labels are Lipschitz-continuous and have positive derivatives that are strictly bounded away from~$0$ and exist almost everywhere:
\begin{restatable}{lemma}{restaterBound}\label{cor:r Lipschitz}\label{cor:r LowerBoundGrowth}\label{cor:r AbsCont}\label{lem:r Lipschitz}
    For any given network with constant network inflow rate, there are constants $c,C > 0$ such that for every tolled dynamic equilibrium, all $v \in V$ and $\sinkTime_1,\sinkTime_2 \in \IR$,
        \begin{align*}
            c\abs{\sinkTime_2-\sinkTime_1} \leq \abs{r_v(\sinkTime_2)-r_v(\sinkTime_1)} \leq C\abs{\sinkTime_2 - \sinkTime_1}.
        \end{align*}
    In particular, $r_v$ is Lipschitz-continuous and, therefore, locally absolutely continuous with $c \leq r'_v \leq C$ almost everywhere.    
\end{restatable}
We prove this \zcref[noref,nocap]{cor:r Lipschitz} in \zcref{app:rLabelsContinuity} by first showing corresponding bounds for the node labels~$\tilde c_v$. The upper bound on the growth rate of~$\tilde c_v$ (and the corresponding lower bound for $r_v$) is a direct consequence of Lipschitz-continuity of the path-travel time functions~$T_P$. To establish a lower bound for $\tilde c_v$ (and the corresponding upper bound for~$r_v$) we mainly need to show that in an equilibrium with constant network inflow rate, queues cannot deplete too fast. 
Crucially, the latter properties (and even the continuity of~$r_v$) do not hold for general feasible flows.

The second ingredient for our definition of thin flows for tolled dynamic equilibria is the function $x_e: \IR \to \IRnn, \sinkTime \mapsto F^+_e(r_v(\sinkTime))$. That is, $x_e(\sinkTime)$ denotes the amount of flow that entered arc~$e$ up to the latest time such that it was still possible to arrive at the sink at cost plus arrival time~$\sinkTime$. Clearly, $x_e$ is a non-decreasing function (as concatenation of two such functions). Moreover, for tolled dynamic equilibria, this function is even Lipschitz-continuous. This is a direct consequence of the Lipschitz-continuity of $r_v$ and the boundedness of all flow rates.

\begin{restatable}{corollary}{restatexLipschitz}\label{lem:xLipschitz}\label{cor:xLipschitz}
    For any tolled dynamic equilibrium and any arc~$e$, the function $x_e$ is Lipschitz continuous.
\end{restatable}

Transferring the usual approach for Nash flows over time \cite[Theorem~1 (a)$\iff$(b)]{CCLDynEquil} to our setting with time shifts, we show that for any tolled dynamic equilibrium~$f$, every arc $e=vw \in E$ and every time $\sinkTime \in \IR$ we have $F^+_e(r_v(\sinkTime)) = F^-_e(r_w(\sinkTime-\eToll))$. This immediately implies the following corollary for the functions~$x_e$:

\begin{restatable}{corollary}{restatexeFlowConservation}\label{cor:Fe+andFe-}\label{cor:xeFlowConservation}
    For any tolled equilibrium, any node $v \neq t$ and any time $\sinkTime \in \IR$ we have 
        \[ \sum_{e \in \edgesFrom{v}}x_e(\sinkTime)=u_0 r_s(\sinkTime) \CharF[v=s, \sinkTime \geq \crem[s][0]]  + \sum_{e \in \edgesTo{v}}x_e(\sinkTime+\eToll).\]
    The same holds for the derivatives of $x_e$ and $r_s$ for almost all~$\sinkTime \in \IR$.
\end{restatable}

This now allows us to present our thin flow equations for tolled dynamic equilibria based on latest departure labels. 

\begin{definition}
    A pair of vector-valued functions $x': \IR \to \IR^E$ and $r': \IR \to \IR^V$ satisfies the \emph{sink-based thin-flow equations} at time~$\sinkTime \in \IR$ with respect to some sets $E^* \subseteq E' \subseteq E$ if
    \begin{align}
        \sum_{e \in \edgesFrom{v}}x'_e(\sinkTime) &= u_0r'_s(\sinkTime) \CharF[v=s, \sinkTime \geq \crem[s][0]] +\sum_{e \in \edgesTo{v}}x'_e(\sinkTime+\eToll)  & v \in V\setminus\set{t}  \tag{TF-1}\label{eq:TF_FlowCons}\\
        x'_e(\sinkTime) & = 0 & e \notin E' \tag{TF-2}\label{eq:TF_InactiveEdges}\\
        r_t'(\sinkTime) & = 1 & \tag{TF-3}\label{eq:TF_Sink} \\
        r_v'(\sinkTime) & = \max_{e=vw\in E' \cap \edgesFrom{v}} \rho_e(r'_v(\sinkTime),r'_w(\sinkTime-\eToll),x_e'(\sinkTime)) & v\in V\setminus \{t\} \tag{TF-4}\label{eq:TF_NodeLabel}\\
        r_v'(\sinkTime) & = \rho_e(r'_v(\sinkTime),r'_w(\sinkTime-\eToll),x_e'(\sinkTime)) & \mathllap{e=vw \in E': x_e'(\sinkTime) > 0} \tag{TF-5}\label{eq:TF_NodeLabelUsedEdge}
    \end{align}
    where
    \begin{align*}
        \rho_e(r'_v(\sinkTime),r'_w(\sinkTime-\eToll),x_e'(\sinkTime))&\coloneqq \begin{cases}
            r'_w(\sinkTime-\eToll) - \frac{x_e'(\sinkTime)}{\nu_e} +r_v'(\sinkTime) & e\in E^*\\
            r'_w(\sinkTime-\eToll) - \max\Set{\frac{x'_e(\sinkTime)}{\nu_e} - r_v'(\sinkTime),0} & e\in E'\setminus E^*.
        \end{cases}
    \end{align*}
\end{definition}

\begin{restatable}{lemma}{restateThinFlowCharacterization}\label{lem:ThinFlowCharacterization}
    For any tolled dynamic equilibrium $f$ the corresponding functions $x'$ and $r'$ satisfy the sink-based thin flow equations for almost all $\sinkTime \in \IR$ with respect to
    \begin{align*}
        E' \coloneqq \setActiveEdgesR(\sinkTime) &\coloneqq \Set{e = vw \in E | T_e(r_v(\sinkTime)) = r_w(\sinkTime-\eToll)} \\
        E^* \coloneqq E^*_r(\sinkTime) &\coloneqq \Set{e = vw \in E | \wtST(\sinkTime) > 0} 
    \end{align*}
   Conversely, given a feasible flow over time $f$ with $x$ and $r$ absolutely continuous such that $x'$ and $r'$ satisfy the sink-based thin flow equations w.r.t $E_r'(\sinkTime)$ and $E_r^*(\sinkTime)$ for almost all $\sinkTime$, then $f$ is a tolled dynamic equilibrium.  
\end{restatable}

From the thin flow properties, we can deduce the following bounds on $x'_e$;

\begin{restatable}{corollary}{restateBoundsOnX}\label{cor:bounds on x}
    For any tolled dynamic equilibrium~$f$ we have for every arc $e=vw \in E$ and almost every time~$\sinkTime \in \IR$ that
        \MainbodyOrAppendix{$0\leq x_e'(\sinkTime)\leq \eCap r_w'(\sinkTime-\eToll).$}%
        {\[0\leq x_e'(\sinkTime)\leq \eCap r_w'(\sinkTime-\eToll).\]}        
    Moreover, for $e \in E \setminus \setActiveEdgesR(\sinkTime)$ we have $x'_e(\sinkTime) = 0$ while for $e \in E^*_r(\sinkTime)$ we have $x'_e(\sinkTime) = \eCap r_w'(\sinkTime-\eToll)$.
\end{restatable}

In the following \zcref[noref,nocap]{lem:queuepropertiesmain}, we summarize two properties that describe the relation of queues and $r$ labels that will be relevant in the remaining paper. The first property is restated and proven in \zcreftp{lem:CharaActiveAndResettingEdges:QueueViaR}, the second one in \zcref{cor:WaitingTimeDerivative}.

\begin{lemma}\label{lem:queuepropertiesmain}
	For any tolled dynamic equilibrium and any arc $e=vw \in E$ we have
	\begin{enumerate}[label=(\alph*)]
		\item $\wtST(\sinkTime) = [r_w(\sinkTime-\eToll) - r_v(\sinkTime) - \eTTime]^+$ for any time $\sinkTime \in \IR$. Here, $[a]^+ \coloneqq \max\set{a,0}$.\label{lem:CharaActiveAndResettingEdges:QueueViaR}
        \item $\wtST$ is absolutely continuous and its derivative satisfies, for almost all times $\sinkTime \in \IR$\[\wtST'(\sinkTime) = \begin{cases}
        r_w'(\sinkTime-\eToll) - r_v'(\sinkTime), &\text{if } \wtST(\sinkTime) > 0 \\
        0, &\text{else}
    \end{cases}.\]
	\end{enumerate}
\end{lemma}

Finally, we now formally define a steady state. 

\begin{definition}
   We say a tolled dynamic equilibrium is in \emph{steady state} at time $\sinkTime \geq \crem[s][0]$, if
   $q_e$ is constant on $[\sinkTime,\infty)$ for every $e\in E$.
 \end{definition}

Note that, with \zcref{lem:queuepropertiesmain}, one can deduce that this is equivalent to $r_v'(\sinkTime')=1$ for all $v \in V$ and almost all sufficiently large $\sinkTime'$. 
\section{The Potential Function} 
\label{sec:potential}
For the remainder, we restrict to instances where the inflow rate $u_0$ is less than the capacity $\mincut$ of a minimum cut in the (static) network $(G,\eCap[])$. Analogously to Cominetti, Correa and Olver \cite{LongTermBehaviorOfDE}, we characterize the steady states by a (static) minimum cost flow LP. Tolls are added to the arc costs and, to fit our latest departure time labels, we normalize the node potentials in the dual by setting the node potential at the sink instead of the source to 0.
Here, $\mathcal{F}_{u_0}$ refers to the set of vectors corresponding to an $s$-$t$-flow of value $u_0$.

\noindent\begin{minipage}[t]{0.35\textwidth}
\begin{align*}
  \min_{y}\ & \sum_{e\in E}(\eTTime+\eToll)\, y_e \tag{P}\label{P}\\
  \text{s.t.}\       & y\in\mathcal F_{u_0}\\
                     & 0\le y_e\le \eCap \quad \forall e\in E
\end{align*}
\end{minipage}%
\hfill
\begin{minipage}[t]{0.6\textwidth}
\begin{align*}
  \max_{d\!,\ \qt}\ & u_0 d_s-\sum_{e\in E}\nu_e\,\qt_e \tag{D}\label{D}\\
  \text{s.t.}\ & d_t=0\\
               & d_v-d_w\le \eTTime+\eToll+\qt_e \quad &\forall e=vw\in E\\
               & \qt_e\ge 0 \quad &\forall e\in E
\end{align*}
\end{minipage}

\vspace{\belowdisplayskip}

 A generalization of these LPs was introduced in~\cite{DEContUniqLT} and applied to the tolled setting in~\cite{TolledDEStructure}. Both allow for network inflow rates that exceed the network's minimum cut capacity requiring a more general definition of steady states. In our setting, where $u_0<\mincut$, the primal-dual pair studied in~\cite{TolledDEStructure} reduces to the pair we consider here, the only difference being that we normalize $d_t=0$ explicitly whereas in~\cite{TolledDEStructure} this is only done when necessary. 
 
 Let $Q^*$ be the set of vectors $q^*$ such that $(d^*,q^*)$ is an optimal dual solution to~\eqref{D} for some $d^*$. This set $Q^*$ is closely linked to steady states: In \cite[Section~4.2]{TolledDEStructure} it is shown that placing a vector of queues $q^*\in Q^*$ on the arcs and considering an equilibrium starting in this pre-loaded network yields an equilibrium flow which is directly in steady state. That is, the queues stay constant. 
Their result translates directly to our setting. 
We show that the reverse direction also holds and, thus, steady states are characterized by the set of dual optimal solutions. We will state and prove this in \zcref{sec:queuesconverge} after having shown boundedness of queues. 
The reason is that our proof uses a complementary slackness argument (as \cite[Theorem~1]{LongTermBehaviorOfDE}), but we cannot simply use the thin flow as primal solution. Instead we construct an approximately feasible primal solution by averaging over a suitable time horizon and then establish the existence of feasible primal solution certifying optimality as its limit. The averaging exploits boundedness of queues.

Unfortunately, the potential function used to show convergence to steady state in \cite{LongTermBehaviorOfDE}, namely the objective function of the dual \lpref{D},  cannot be applied to the tolled setting without modification, as it is not monotone (see \zcref{fig:CCO_not_monotone} in \zcref{app:CCONonMonotone}).

Instead, we use the following, adjusted potential function:\[\fullpot(\sinkTime) \coloneqq \underset{\potBasic(\sinkTime)\text{, analogue of the potential in \cite{LongTermBehaviorOfDE}}}{\underbrace{u_0\cdot(\sinkTime-r_s(\sinkTime)) - \sum_{e=vw \in E}\nu_e\cdot \wtST(\sinkTime) }}- \underset{B(\sinkTime)}{\underbrace{\sum_{e=vw\in E}\int_\sinkTime^{\sinkTime+\eToll}x'_e(z)\log r_w'(z-\eToll)\diff z}}.\]

In \zcref{app:CCONonMonotone}, we show that this potential is indeed a generalization of the potential in \cite{LongTermBehaviorOfDE} in the sense that both potential functions coincide on networks without tolls (when considering the same reference time).

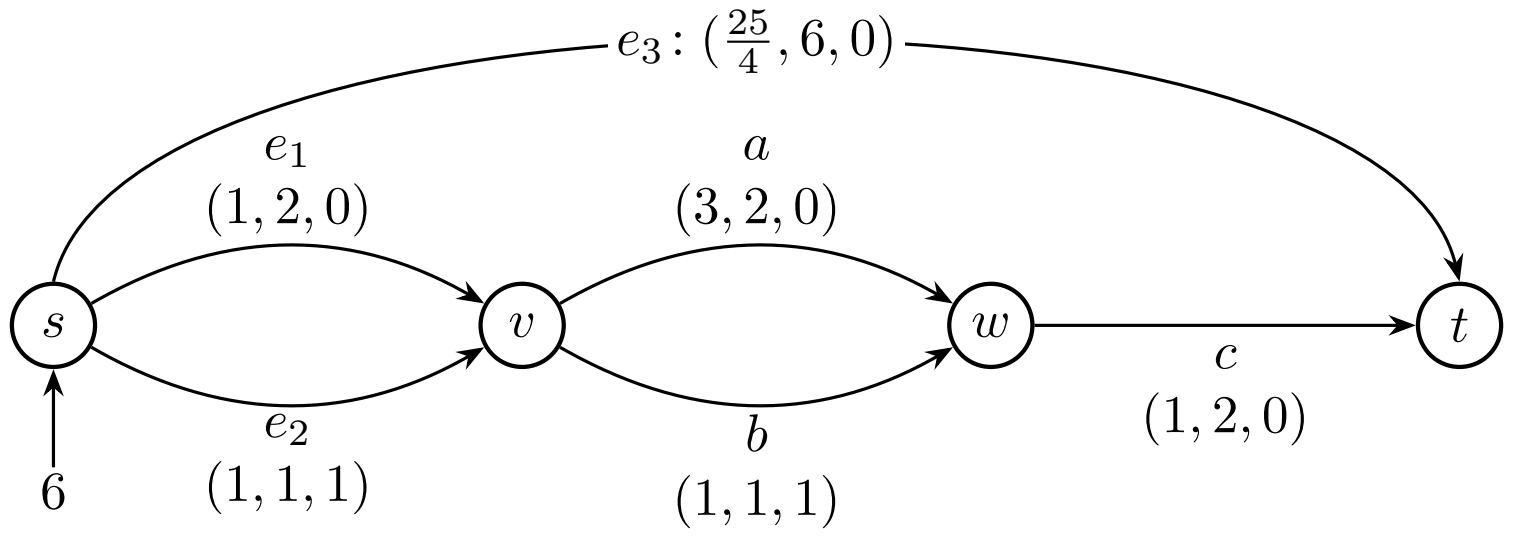
\begin{figure}[t]
\centering
\begin{tikzpicture}[x=2cm,y=1.3cm,font=\small,
  vtx/.style={circle,draw,thick,minimum size=5.5mm,inner sep=0pt},
  N/.style={-{Stealth[length=1.8mm]},line width=0.6pt},
  dl/.style={align=center,inner sep=1.5pt,font=\small}]
\node[vtx](s)at(0,0){$s$};\node[vtx](v)at(1.55,0){$v$};\node[vtx](w)at(3.1,0){$w$};\node[vtx](t)at(4.65,0){$t$};
\draw[N](s)to[bend left=30]node[dl,above]{$e_1$\\$(1,2,0)$}(v);
\draw[N](s)to[bend right=30]node[dl,below]{$e_2$\\$(1,1,1)$}(v);
\draw[N](v)to[bend left=30]node[dl,above]{$a$\\$(3,2,0)$}(w);
\draw[N](v)to[bend right=30]node[dl,below]{$b$\\$(1,1,1)$}(w);
\draw[N](w)--node[dl,below=2pt]{$c$\\$(1,2,0)$}(t);
\draw[N](s.north) .. controls (0.25,1.85) and (4.65-0.25,1.85) .. node[dl,fill=white]{$e_3\!:(\tfrac{25}4,6,0)$}(t.north);
\draw[{Stealth[length=1.8mm]}-,line width=0.6pt](s.south)--++(0,-.5)node[below,inner sep=1pt]{$6$};
\end{tikzpicture}

\caption{The network shown in \zcref{fig:nfc} consists of a loading gadget (arcs $e_1,e_2,e_3$) to
place queues on $b$ and $c$, as well as to provide capacity exceeding $u_0=6$ (via arc~$e_3$). Infinite time convergence is then due
to an ever decreasing but never vanishing amount of queue mass shifting from~$b$ to~$c$. The edge properties are given as $(\tau_e, \nu_e, p_e)$. We describe the equilibrium flow in more detail in \zcref{sec:nonfinite_conv}.}\label{fig:nfc}
\end{figure}

We first note that $\fullpot$ is absolutely continuous, due to Lipschitzness of its constituent parts and the resulting integrability of the integrand in $B$. Moreover, $\fullpot$ satisfies both monotonicity on $[\crem[s][0],\infty)$ and boundedness, the proof of which we will sketch below. 
However, this alone does not suffice for convergence to $Q^*$. While the potential derivative in the untolled setting is bounded away from~$0$ until the equilibrium trajectory has reached the steady state after some finite time \cite{LongTermBehaviorOfDE}, convergence to a steady state may take infinite time for tolled dynamic equilibria  (see \zcref{fig:nfc,sec:nonfinite_conv}). We only obtain that the total potential increase $\int_{\crem[s][0]}^\infty \fullpot'$ is finite. Nevertheless, this suffices to show that queues are both bounded and approach the set $Q^*$ (see \zcref{sec:queuesconverge}). 

The motivation for the correction term $B$ comes from the theory of delay systems where a differential equation has terms depending on both present and past \cite{FRIDMAN2014271}. While our setting is far from classical differential equation theory, our potential function needs to take into account the time shift between $r_v(\sinkTime)$ and $r_w(\sinkTime-\eToll)$ which breaks monotonicity. 
Considering the monotonicity proof in \cite{LongTermBehaviorOfDE}, the crucial step which fails in our setting is the use of flow conservation for~$x_e$ (which only holds with an additional time shift for tolled equilibria, see \zcref{cor:Fe+andFe-}). This problem occurs in several places in our proofs, requiring additional averaging arguments.   

We now give an intuition for the choice of $B$ and why it restores monotonicity of $\fullpot$. In this discussion, let $\sinkTime \geq \crem[s][0]$.
    
The function $\potBasic$ is the dual objective function evaluated at $d_s=\sinkTime-r_s(\sinkTime)$ and $q$ and, as such, the analogue of the potential function in \cite{LongTermBehaviorOfDE}. It has the following derivative, where the second equality uses \zcref{lem:queuepropertiesmain,cor:bounds on x}:
\[\potBasic'(\sinkTime) = u_0(1-r_s'(\sinkTime))- \sum_e\nu_eq_e'(\sinkTime)= u_0(1-r_s'(\sinkTime))+ \sum_{e=vw} x_e'(\sinkTime)\left(\frac{r_v'(\sinkTime)}{r_w'(\sinkTime-\eToll)}-1\right)\]
Morally, we can only consider the terms in the summation, as we can interpret network inflow at~$s$ as entering via an arc~$ts$ with $\nu_{ts}=u_0$, so that the first term can be subsumed in the sum.
While $x_e'$ is always non-negative, the expression $\frac{r_v'(\sinkTime)}{r_w'(\sinkTime-\eToll)}-1$ 
can be both positive, if a queue depletes on~$vw$, or negative, if it grows. Now, consider the function $g:\mathbb R\to \mathbb R, z \mapsto  z- 1 -\log(z)$ which is nonnegative on $\mathbb R_{> 0}$ as it is convex and attains its minimum 0 at $z=1$. With it, we can rewrite $\frac{r_v'(\sinkTime)}{r_w'(\sinkTime-\eToll)}-1$ in the potential function as $g\bigl(\frac{r_v'(\sinkTime)}{r_w'(\sinkTime-\eToll)}\bigr) + \log\bigl(\frac{r_v'(\sinkTime)}{r_w'(\sinkTime-\eToll)}\bigr)$. This restricts the critical part to the artificially introduced logarithmic term which we can rewrite as $R(\sinkTime)\coloneqq \sum_e x_e'(\sinkTime)[\log(r_v'(\sinkTime))-\log(r_w'(\sinkTime-\eToll)]$ which again evaluates the label derivatives at different sink times. This is exactly what subtracting $B'$ remedies, as 
\[-B'(\sinkTime)=\sum_e[x_e'(\sinkTime)\log r_w'(\sinkTime-\eToll)-x_e'(\sinkTime+\eToll)\log r_w'(\sinkTime)],\]
the first part of which cancels the subtrahend $x_e'(\sinkTime)\log(r_w'(\sinkTime-\eToll)$. The remaining label derivatives in $R-B'$ are now evaluated at $\sinkTime$ and the time shift in the evaluation times of $x'$ yields terms canceling by flow conservation (\zcref{eq:TF_FlowCons}). Thus, $\fullpot'=\potBasic'-B' \geq 0$ and the resulting potential is non-decreasing.

\begin{restatable}{theorem}{thmPotentialNondecreasing} 
\label{thm:PotNondecreasing}
    The potential $\fullpot$ is non-decreasing on $[\crem[s][0], \infty)$.
\end{restatable}

We record part of the previous discussion as a corollary for later use.
\begin{corollary}\label{not:potential_deriv_e}
    Let $g(z) = z- 1 -\log(z)$. The contribution $\fullpot_e'(\sinkTime) \coloneqq x'_e(\sinkTime) g\bigl(\frac{r_v'(\sinkTime)}{r_w'(\sinkTime-\eToll)}\bigr)$ of each arc $e$ to the potential increase evaluated is non-negative for almost all $\sinkTime$. The same is true for the contribution $u_0\cdot r'_s(\sinkTime)g(1/r'_s(\sinkTime))$ at the source.
\end{corollary}

Our next goal is to establish an upper bound for our potential function~$\fullpot$:

\begin{restatable}{theorem}{thmpotbounded}\label{thm:pot bounded}
    The potential $\fullpot$ is upper bounded.
\end{restatable}

For the proof, we first show by a case distinction on the value of $r'_w$ and a logarithm inequality that $-B$ is upper bounded.
To show that $\potBasic= \fullpot + B$ is upper bounded as well, we consider $d$ and $\qt$ as in the following lemma.

\begin{restatable}{lemma}{lemdualfeasible}
    \label{lem:dual_feasible}
        For every flow over time and any time $\sinkTime \in \IR$, the vector $(d(\sinkTime), \tilde q(\sinkTime))$ with $ d_v(\sinkTime) \coloneqq \sinkTime - r_v(\sinkTime)$ and $\tilde q_e(\sinkTime) \coloneqq [r_w(\sinkTime)-r_v(\sinkTime)-\eTTime-\eToll]^+$ is a feasible solution to~\eqref{D}.
     Moreover, \eqref{D} is upper bounded.
\end{restatable}

Feasibility can be shown by a simple check of the constraints. The upper bound on \eqref{D} follows as our assumption $u_0\leq \mincut$ implies that \lpref{P} is feasible. Thus, the function
    \[\potVar: \IR \to \IR, \sinkTime \mapsto u_0\cdot(\sinkTime-r_s(\sinkTime)) - \sum_{e=vw \in E}\nu_e[r_w(\sinkTime)-r_v(\sinkTime)-\eTTime-\eToll]^+\]
is upper bounded.
This function almost coincides with~$\potBasic$ except the terms in the sum are $\qt_e(\sinkTime) = [r_w(\sinkTime)- r_v(\sinkTime)-\eTTime-\eToll]^+$ in $\potVar$, whereas in $\potBasic$ they are $q_e(\sinkTime) = [ r_w(\sinkTime-\eToll)- r_v(\sinkTime)-\eTTime]^+$. 
The difference between the two is that the contribution of the toll $\eToll$ shifts from the arc itself (in $\potVar$) to the sink time, i.e., into the argument of $r_w$ in $\potBasic$. We can, therefore, split the potential $\fullpot= \potBasic-\potCorr$ into three terms, the first two of which we already bounded above:
    \begin{equation}\label{eq:3_part_potential}
        \fullpot (\sinkTime) = \underset{\text{$\potVar$}}{\underbrace{u_0d_s(\sinkTime)- \sum_{e}{\nu_e}\cdot \qt_e(\sinkTime) }}-B(\sinkTime)+ \underset{\text{$\potBasic-\potVar$}}{\underbrace{\sum_e \nu_e (\qt_e(\sinkTime) -q_e(\sinkTime))}}.
    \end{equation}

Hence, the final step of the proof is to bound the difference~$\potBasic-\potVar$ which is done via Lipschitz-continuity of the $r$-labels (\zcref{lem:r Lipschitz}).

In the next section, we show that equilibria approach their steady state set. The proof extensively uses the following simple corollary of monotonicity and boundedness of $\fullpot$.
\begin{corollary}\label{cor:dissipation_finite}
    The total potential increase $\int_{\crem[s][0]}^\infty \fullpot'(\sinkTime)\diff\sinkTime $ is finite.
\end{corollary}

\section{Bounded Queues and Convergence to \texorpdfstring{$Q^*$}{Q*}}
\label{sec:queuesconverge}

As our first step, we use strictness of $u_0<\mincut $ combined with boundedness of the potential to show that queues remain bounded. Here, the bound on the potential helps us bound the dual objective function from below.
In the second step, we show that the variation of queues tends to~$0$. Lastly, we obtain convergence of the queues to $Q^*$.

\begin{theorem}\label{lm:queues_bounded}\label{thm:queues_bounded}
    Assuming $u_0<\mincut$, for any tolled dynamic equilibrium, $q_e(\sinkTime)$ is bounded for every $e\in E$. Moreover, there exists a constant $\UpD$ such that $d_v(\sinkTime)\leq \UpD$ for all $v \in V$. 
\end{theorem}

\begin{proof}
    We first consider $\sinkTime \geq \crem[s][0]$, so that $\fullpot(\sinkTime) \geq \fullpot (\crem[s][0])$ by monotonicity (\zcref{thm:PotNondecreasing}).
    Let $\varepsilon\coloneqq 1- \frac{u_0}{\mincut}$. By the Max-Flow-Min-Cut-Theorem, there exists a flow of value $\mincut$ in the network. Scaling this flow down by $\frac{u_0}{\mincut}$ yields a flow $y$ of value $u_0$ such that $y_e\in [0,(1-\varepsilon)\nu_e]$ for all $e \in E$. Now observe that 
        \[u_0d_s = \sum_{e=vw} y_e(d_v-d_w) \leq \sum_e y_e (\eTTime+\eToll+\qt_e)\leq \sum_e y_e (\eTTime+\eToll) + (1-\varepsilon)\sum_e \nu_e\qt_e.\]
    The first equality holds as all terms in the telescoping sum except those for $s$ and $t$ cancel (due to flow conservation) and $d_t=0$. The first inequality holds by dual feasibility and the second by the bounds on the scaled down flow $y$.

    Therefore, $\dualObj(\sinkTime) = u_0d_s(\sinkTime) - \sum_e \nu_e \qt_e(\sinkTime) \leq \sum_e y_e (\eTTime+\eToll) -\varepsilon\sum_e \nu_e\qt_e(\sinkTime).$
    Note that $\dualObj = \Psi + B - (\Phi-\Phi_{\mathrm{D}})$ is bounded from below since all three terms on the right are lower bounded. $\Psi$ is lower bounded by case distinctions, bounds on $B$ and  $-(\Phi-\Phi_{\mathrm{D}})$ are discussed in \zcref{sec:potential} and formally proven in  \zcref{lm:B_bound,lem:DifferenceBetweenPots}). Thus, $\varepsilon\sum_e \nu_e\qt_e$ is bounded above.
    Thus, also all $\qt_e$ are bounded above individually as they are non-negative. Now, $r_w$ being non-decreasing lets us derive an upper bound for $\wtST$ from there:
        \begin{align*}
            \qt_e(\sinkTime)
                &=[r_w(\sinkTime)-r_v(\sinkTime)-\eTTime-\eToll+r_w(\sinkTime-\eToll)-r_w(\sinkTime-\eToll)]^+ \\
                &\geq [r_w(\sinkTime-\eToll)-r_v(\sinkTime)-\eTTime]^+-\eToll
                    = q_e(\sinkTime)-\eToll.
        \end{align*}

    With all queues being bounded, so are all travel times and, therefore, all $d_u$.
    Finally, the earlier times $\sinkTime< \crem[s][0]$ are covered by continuity and the fact that queues vanish for negative sink times.
\end{proof}

Next, we show that the queues change less and less over fixed-length time intervals:

\begin{restatable}{lemma}{lemfreezing}
\label{lm:freezing}
    For any tolled dynamic equilibrium, arc $e = vw\in E$ and $h > 0$ we have
        \[\int_{\sinkTime}^{\sinkTime+h} \abs{\wtST'(z) } \diff z \xrightarrow{\sinkTime \to \infty} 0.\]
\end{restatable}

This \zcref[noref,nocap]{lm:freezing}
is essentially a consequence of $\Psi_e'$ going to zero on average (by \zcref{not:potential_deriv_e} together with \zcref{cor:dissipation_finite}). From there, we get that $g\bigl(\frac{r_v'(\sinkTime)}{r_w'(\sinkTime-\eToll)}\bigr)$ goes to zero and, therefore, $\frac{r_v'(\sinkTime)}{r_w'(\sinkTime-\eToll)}$ to one (both on average). Since the node label derivatives at the tail and head of an arc are bounded by \zcref{cor:r Lipschitz}, their difference must decrease as time goes to infinity. As the waiting time can be expressed via those node labels, this yields the desired result. The formal proof always uses the average instead of pointwise rate of change, since we only have convergence of the former for $\Psi_e$.

With this, we show that in a tolled dynamic equilibrium the distance between the queues and $Q^*$ decreases to 0. Before we do this, we want to state that $Q^*$ \emph{describes} the steady states in a certain sense. As shown in \cite{TolledDEStructure} and discussed in \zcref{sec:potential}, an equilibrium starting in a point in $Q^*$ is in steady state. We show the reverse direction.

\begin{restatable}{theorem}{thmQsteadystates}
\label{thm:Q*steadystates}
    Consider an instance with $u_0 < \nu_C$ and a tolled dynamic equilibrium that is in steady state at $\sinkTime - \max_{e \in E} \eToll-\sum_e \eToll$. Then, $\bar q \coloneqq q_e(\sinkTime')$ and $d_v \coloneqq \sinkTime' - r_v(\sinkTime')$ form a dual optimal solution for all $\sinkTime' \geq \sinkTime$. In particular, $\bar q \in Q^*$.
\end{restatable}

Structurally, the proof follows that of \cite[Theorem 1]{LongTermBehaviorOfDE} which considers a feasible primal and dual solution and shows complementary slackness. In the tolled setting, significant additional effort is needed as a thin flow does not yield a feasible primal solution due to the time shift. Crucially, however, we can exploit boundedness of queues to obtain that the time shift vanishes when averaging flow over a sufficiently long time horizon. We show that every sequence of time points diverging to infinity has a subsequence along which feasibility and optimality hold in the limit. 

Next, we show that the distance of every equilibrium to the set of steady states $Q^*$ converges to 0. For this, we start by observing that due to the boundedness of the queues any sequence of times diverging to infinity must contain a subsequence of times such that the corresponding queues converge. We then show that such a limit point is an optimal dual solution in a similar fashion to the proof of \zcref{thm:Q*steadystates}, but using \zcref{lm:freezing} as replacement for the queues already being constant.

\begin{restatable}{theorem}{thmqueuesconverge}
\label{lm:queues_converge}\label{thm:queues_converge_subsequences}
    Consider an instance with $u_0 < \nu_C$ and an equilibrium flow in this instance.
    For any sequence of deadlines $\sinkTime_n\to \infty$,  there exist a subsequence $(\sinkTime_{n_k})_k$ and $\bar q \in Q^*$ such that $q(\sinkTime_{n_k}) \to \bar q$.    
\end{restatable}

Our main theorem, convergence of dynamic equilibria to the set of steady states, is now an immediate consequence of the above \zcref[noref,nocap]{thm:queues_converge_subsequences}:

\begin{theorem}\label{thm:convergence_queues}
    Consider an instance with $u_0 < \mincut$ and a tolled dynamic equilibrium.
    The queue lengths approach those in the set of optimal solutions~$Q^*$ of~\lpref{D}, that is, $\dist( q(\sinkTime),Q^*)\overset{\sinkTime\to \infty}{\longrightarrow}  0$.     
    In particular, if $Q^*=\set{q^*}$, then $ q(\sinkTime)\overset{\sinkTime\to \infty}{\longrightarrow}q^*$.
\end{theorem}
If $Q^*$ contains more than one solution, the queue lengths could slowly drift towards and along the convex solution set, slowing down, but never converging to a single solution.

\section{Conclusion}

We developed a sink-based thin flow characterization for tolled dynamic equilibria. A challenge here was to find a common time as FIFO does not hold in tolled dynamic equilibria. Using this sink based perspective we constructed a monotone and bounded potential function.  This allowed us to show that throughout the evolution of the tolled dynamic equilibrium, queues remain bounded if the network inflow rate is strictly below the minimum-cut capacity. Moreover, under the same condition, the distance of the queue vector to the set of optimal dual queue vectors converges to zero. In contrast to the base setting without tolls, we cannot hope to show that a steady state is reached in finite time as we presented an example showing that this is not true.
Two natural questions that remain open are: 
First, can we show convergence to a vector in $Q^*$ or find an example where drifting along $Q^*$ actually happens? Second, can we show that the same results still hold when the minimum cut capacity equals the inflow rate?  Furthermore, it would be interesting to extend the approach to the more complex case without the inflow constraint. 

\subsection*{Acknowledgement}
We thank Neil Olver, Shaul Rosner, Marc Schröder and Julian Schwarz for fruitful discussions on the general topic.

\subsection*{AI declaration}

We used Astra 6 to find the potential function and for a first version of the proofs in \zcref{sec:potential,sec:queuesconverge}. Moreover, we used Opus 5.5 for finding the examples in \zcref{sec:nonfinite_conv,app:CCONonMonotone}. We improved and simplified the proofs drastically and restructured the presentation. 
Most significantly, the AI-generated proofs worked with labels of bounded variation and therefore required a tedious analysis of jumps and Cantor parts of the potential. By showing that the labels~$r_v$ are Lipschitz-continuous and the derivatives can be bounded from above and below (\zcref{lem:r Lipschitz}), we ruled out these cases, which considerably simplified all subsequent proofs.
Similarly, the AI-generated proof that queues freeze on bounded time windows relied on a chain of analytic inequalities. We replaced it by a direct argument based on \zcref{lem:r Lipschitz} (\zcref{lm:freezing}).
In particular, all proofs in \zcref{sec:thin_flow_characterization} and the corresponding parts in \zcref{sec:rlabels_appendix} are our own.
In the writing process we used Opus 5.5 and Sol 6.1 for proof reading, but wrote the text without the help of AI tools.

\printbibliography

@misc{DynamicFlowDecomposition,
	title={A Decomposition Theorem for Dynamic Flows}, 
	author={Lukas Graf and Tobias Harks and Julian Schwarz},
	year={2024},
	eprint={2407.04761},
	archivePrefix={arXiv},
	primaryClass={cs.DS},
	url={https://arxiv.org/abs/2407.04761},
    note={Preprint}
}

@article{VickreyQueuingModel,
	ISSN = {00028282},
	URL = {http://www.jstor.org/stable/1823678},
	author = {Vickrey, William S.},
	journal = {The American Economic Review},
	number = {2},
	pages = {251--260},
	publisher = {American Economic Association},
	title = {Congestion Theory and Transport Investment},
	volume = {59},
	year = {1969}
}

@article{KochSkutellaDEandPoA,
	author = {Koch, Roland and Skutella, Martin},
	title = {Nash Equilibria and the Price of Anarchy for Flows over Time},
	journal = {Theory of Computing Systems},
	date = {2011-07-01},
    year={2011},
	volume = {49},
	pages = {71--97},
	doi = {10.1007/s00224-010-9299-y},
	OPTurl = {https://doi.org/10.1007/s00224-010-9299-y},
}

@article{CCLDynEquil,
	author = {Cominetti, Roberto and Correa, José and Larré, Omar},
	title = {Dynamic Equilibria in Fluid Queueing Networks},
	journal = {Operations Research},
	volume = {63},
	number = {1},
	pages = {21--34},
	year = {2015},
	doi = {10.1287/opre.2015.1348},
	OPTURL = {https://doi.org/10.1287/opre.2015.1348},
	OPTeprint = {https://doi.org/10.1287/opre.2015.1348}
}

@article{LongTermBehaviorOfDE,
	author = {Cominetti, Roberto and Correa, Jos\'{e} and Olver, Neil},
	title = {Long-Term Behavior of Dynamic Equilibria in Fluid Queuing Networks},
	journal = {Operations Research},
	volume = {70},
	number = {1},
	pages = {516--526},
	year = {2022},
	doi = {10.1287/opre.2020.2081},	
	OPTURL = {https://doi.org/10.1287/opre.2020.2081},
	OPTeprint = {https://doi.org/10.1287/opre.2020.2081},
	OPTnote = {Also available on arXiv: \url{https://arxiv.org/abs/2112.10412v1}}
}

@article{DEContUniqLT,
	author={Olver, Neil and Sering, Leon and Vargas Koch, Laura},
	title = {Continuity, Uniqueness, and Long-Term Behavior of Nash Flows over Time},
    journal = {Operations Research},
    volume = {74},
    number = {4},
    pages = {1748-1769},
    year = {2026},
    doi = {10.1287/opre.2024.1007}
}

@article{BFAStacklebergStrategyRoutingFlowOverTime,
	author    = {Bhaskar, Umang and Fleischer, Lisa and Anshelevich, Elliot},
	title     = {A {S}tackelberg strategy for routing flow over time},
	journal   = {Games and Economic Behavior},
	volume    = {92},
	pages     = {232--247},
	year      = {2015},
	OPTurl       = {https://www.sciencedirect.com/science/article/pii/S0899825613001334},
	issn	  = {0899-8256},
    doi = {10.1016/j.geb.2013.09.004}
}

@misc{josemonotonicity,
	author    = {Jos{\'e} Correa},
	title     = {A counterexample to the monotonicity conjecture for dynamic equilibria in flows over time},
	howpublished = {Manuscript, \url{https://www.dii.uchile.cl/~jcorrea/papers/Manuscripts/2022CCOcounterexample.pdf}},
	year      = {2026}}

@article{DEPoA,
	author = {Correa, Jos\'{e} and Cristi, Andr\'{e}s and Oosterwijk, Tim},
	title = {On the Price of Anarchy for Flows over Time},
	journal = {Mathematics of Operations Research},
	volume = {47},
	number = {2},
	pages = {1394-1411},
	year = {2022},
	doi = {10.1287/moor.2021.1173},
	OPTURL = {https://doi.org/10.1287/moor.2021.1173}
}

@InProceedings{DETimeVarying,
	author={Pham, Hoang Minh and Sering, Leon},
	OPTeditor={Harks, Tobias and Klimm, Max},
	title={Dynamic Equilibria in Time-Varying Networks},
	booktitle={Algorithmic Game Theory},
	year={2020},
	publisher={Springer International Publishing},
	address={Cham},
	pages={130--145},
	isbn={978-3-030-57980-7},
	doi = {10.1007/978-3-030-57980-7_9},
	OPTnote = {full version available on arXiv: \url{https://arxiv.org/abs/2007.01525}}
}

@inbook{SpillbackSeringKoch,
	author    = {Sering, Leon and Vargas Koch, Laura},
	title     = {Nash Flows over Time with Spillback},
	year = {2019},
	booktitle = {Proceedings of the 2019 Annual ACM-SIAM Symposium on Discrete Algorithms (SODA)},
	pages = {935-945},
	doi = {10.1137/1.9781611975482.57},
	OPTURL = {https://epubs.siam.org/doi/abs/10.1137/1.9781611975482.57},
	OPTnote       = {Also available at \url{http://arxiv.org/abs/1807.05862}}
}

@article{BicriteriaNashFlows,
	author       = {Oosterwijk, Tim and Schmand, Daniel and Schr{\"{o}}der, Marc},
	title = {Bicriteria Nash flows over time},
    journal = {Games and Economic Behavior},
    volume = {147},
    pages = {19-37},
    year = {2024},
    issn = {0899-8256},
    doi = {10.1016/j.geb.2024.06.002},
    OPTurl = {https://www.sciencedirect.com/science/article/pii/S0899825624000836}
}

@misc{TopologiesExcludingBraessParadox,
      title={Network Topology That Excludes Braess's Paradox and Maintains Monotonicity in Flows Over Time}, 
      author={Chen, Xujin and Deng, Xiyuan and Wang, Changjun},
      year={2026},
      eprint={2609.32738},
      archivePrefix={arXiv},
      primaryClass={math.CO},
      url={https://arxiv.org/abs/2609.32738}, 
      note = {to be presented at WINE 2026}
}

@inproceedings{ImplementabilityCharacterization,
	author = {Graf, Lukas and Harks, Tobias and Schwarz, Julian },
	title = {Tolls for Dynamic Equilibrium Flows},
	booktitle = {Proceedings of the 2025 Annual ACM-SIAM Symposium on Discrete Algorithms (SODA)},
	year = {2025},
	pages = {2560-2606},
	doi = {10.1137/1.9781611978322.84},
	OPTURL = {https://epubs.siam.org/doi/abs/10.1137/1.9781611978322.84},
	OPTeprint = {https://epubs.siam.org/doi/pdf/10.1137/1.9781611978322.84}
}

@inproceedings{ImplementabilityOfSystemOptima,
	author = {Schwarz, Julian and Harks, Tobias and Graf, Lukas},
	title = {Are System Optimal Dynamic Flows Implementable by Tolls?},
	year = {2025},
	isbn = {9798400719431},
	publisher = {Association for Computing Machinery},
	address = {New York, NY, USA},
	OPTurl = {https://doi.org/10.1145/3736252.3742635},
	doi = {10.1145/3736252.3742635},
	booktitle = {Proceedings of the 26th ACM Conference on Economics and Computation},
	pages = {839},
	numpages = {1},
	location = {Stanford University, Stanford, CA, USA},
	series = {EC '25},
	OPTnote = {full version available at \url{https://arxiv.org/abs/2503.07387}}
}

@phdthesis {SeringThesis,
	author = {Sering, Leon},
	title = {Nash flows over time},
	school = {Technische Universität Berlin},
	year = {2020},
	type = {Doctoral Thesis},
	address = {Berlin},
	doi = {10.14279/depositonce-10640},
	OPTurl = {http://dx.doi.org/10.14279/depositonce-10640}
}

@book {GrafThesis,
	author = {Graf, Lukas},
	title = {Dynamic Network Flows with Adaptive Route Choice based on Current Information},
	year = {2024},
	publisher = {Springer Spektrum Wiesbaden},
	series = {Mathematische Optimierung und Wirtschaftsmathematik | Mathematical Optimization and Economathematics},
	doi = {10.1007/978-3-658-44948-3},
	OPTnote = {also available at \url{https://lukasmgraf.de/files/phd-thesis.pdf}}
}

@InProceedings{TolledDEStructure,
	title={Nash Flows Over Time with Tolls}, 
	author={Rosner, Shaul and Schröder, Marc and Vargas Koch, Laura},
    OPTeditor= {Bil{\`o}, Vittorio and Cai, Yang and Huang, Zhiy},
    booktitle={Web and Internet Economics},
	year={2026},
    publisher={Springer Nature Switzerland},
    address={Cham},
    pages={236--253},
	isbn={978-3-032-18660-7},
    doi={10.1007/978-3-032-18660-7_12}
}

@article{FOFlowOverTimewithSchedulingCosts,
    author = {Frascaria, Dario and Olver, Neil},
    title = {Algorithms for flows over time with scheduling costs},
    journal = {Mathematical Programming},
    year = {2022},
    volume = {192},
    pages = {177--206},
    doi = {10.1007/s10107-021-01725-z}
}

@article{krasovskii1963stability,
  title={Stability of motion},
  author={Krasovskii, Nicolai Nikolaevich},
  journal={Stanford University Press},
  year={1963}
}

@article{FRIDMAN2014271,
    title = {Tutorial on {Lyapunov-based} methods for time-delay systems},
    journal = {European Journal of Control},
    volume = {20},
    number = {6},
    pages = {271-283},
    year = {2014},
    issn = {0947-3580},
    doi = {10.1016/j.ejcon.2014.10.001},
    OPTurl = {https://www.sciencedirect.com/science/article/pii/S0947358014000764},
    author = {Fridman, Emilia}
}

\appendix

\renewcommand{\MainbodyOrAppendix}[2]{#2}

\section{Properties of \texorpdfstring{$r$}{r}-Labels and Label Derivatives}
\label{sec:rlabels_appendix}

\subsection{Basic Properties of \texorpdfstring{$r$}{r}-Labels}\label{app:rLabelsBasicProperties}

We start by showing that the sink-based labels~$r_v$ defined in \zcref{def:rLabel} satisfy the natural Bellman equations:

\begin{lemma}\label{lem:rLabelsBellmanEq}
    The labels $r_v$ for any feasible flow satisfy
        \[r_v(\sinkTime) = \begin{cases}
            \sinkTime, &\text{if } v = t \\
            \max_{e=vw \in \edgesFrom{v}}\sup\set{\netTime \in \IR | T_e(\netTime) \leq r_w(\sinkTime-\eToll)}, &\text{else}
        \end{cases}.\]
\end{lemma}

\begin{proof}
    For $v=t$ this is clear. Otherwise, we have
        \begin{align*}
            &r_v(\sinkTime)
                = \!\!\!\max_{P \in \PathSet[v,t]}\sup\Set{\netTime \in \IR | T_P(\netTime) + \sum_{e' \in P}\eToll[e'] \leq \sinkTime} \\
                &= \!\!\!\max_{e=vw \in \edgesFrom{v}, P \in \PathSet[w,t]}\sup\Set{\netTime \in \IR| T_P(T_e(\netTime)) + \eToll + \sum_{e' \in P}\eToll[e'] \leq \sinkTime} \\
                &= \!\!\!\max_{e=vw \in \edgesFrom{v}}\max_{P \in \PathSet[w,t]}\sup\Set{\netTime  \in \IR| T_P(T_e(\netTime)) + \sum_{e' \in P}\eToll[e'] \leq \sinkTime-\eToll} \\
                &= \!\!\!\max_{e=vw \in \edgesFrom{v}}\max_{P \in \PathSet[w,t]}\sup\Set{\netTime \in \IR | T_e(\netTime) \leq \sup\Set{\netTime' \in \IR | T_P(\netTime') + \sum_{e' \in P}\eToll[e'] \leq \sinkTime-\eToll}} \\
                &= \!\!\!\max_{e=vw \in \edgesFrom{v}}\sup\Set{\netTime \in \IR | T_e(\netTime) \leq \max_{P \in \PathSet[w,t]}\sup\Set{\netTime' \in \IR| T_P(\netTime') + \sum_{e' \in P}\eToll[e'] \leq \sinkTime-\eToll}} \\
                &= \!\!\!\max_{e=vw \in \edgesFrom{v}}\sup\set{\netTime \in \IR | T_e(\netTime) \leq r_w(\sinkTime-\eToll)} \qedhere
        \end{align*}
\end{proof}

\restaterLabelsAlternativeEquation*

\begin{proof}
    For any $v \neq t$ and $e=vw \in \edgesFrom{v}$ define
        \[\netTime_w \coloneqq \sup \set{\netTime \in \IR| T_e(\netTime) \leq r_w(\sinkTime-\eToll)}.\]
    Then, continuity of $T_e$ gives us $r_w(\sinkTime-\eToll) = T_e(\netTime_w)$ while \zcref{lem:rLabelsBellmanEq} gives us $r_v(\sinkTime) = \max_{vw \in \edgesFrom{v}}\netTime_w$. Since $T_e$ is non-decreasing, this implies 
        \[r_w(\sinkTime-\eToll) \leq T_e(r_v(\sinkTime)) = r_v(\sinkTime) + \eTTime + \wtST(\sinkTime)\]
    and, therefore,
        \[r_w(\sinkTime-\eToll) - \eTTime - \wtST(\sinkTime)\leq r_v(\sinkTime)\]
    for all $vw \in \edgesFrom{v}$ with equality for at least one. This shows the \zcref[noref,nocap]{lem:rLabelsAlternativeEquation}.
\end{proof}

Next, we show that $r_v$ is a right-inverse of $\tilde c_v$. Moreover, if $r_v$ is continuous (which we will later show to be true for equilibrium flows under constant network inflow -- see \zcref{lem:QueueingEdgesHaveMinimumInflow}), those two labels are even inverses of each other:

\begin{lemma}\label{lem:cvLeftinverseOfrv}
	For any feasible flow and any node $v \in V$ the function $\tilde c_v$ is a left-inverse of $r_v$, i.e.,
		\[\tilde c_v(r_v(\sinkTime)) = \sinkTime \quad\text{for all } \sinkTime \in \IR.\]
\end{lemma}

\begin{proof}
	By the definition of $r_v$ we can reach the sink when starting from node~$v$ at time~$r_v(\sinkTime)$ for a sum of arrival time and cost of~$\sinkTime$. Hence, 
		\[\tilde c_v(r_v(\sinkTime)) = \min_{P \in \PathSet[v,t]}\Bigl(T_P(r_v(\sinkTime))+\sum_{e \in P}\eToll[e]\Bigr) \leq \sinkTime.\]
	At the same time, we cannot have $\tilde c_v(r_v(\sinkTime)) < \sinkTime$ as, then, there would be a \stpath[v]~$P$ over which we can reach the sink for a generalized arrival time of strictly less than~$\sinkTime$ when starting at time~$r_v(\sinkTime)$. But then, the continuity of $T_P$ implies that there is a time $\netTime > r_v(\sinkTime)$ with $T_P(\netTime)+\sum_{e' \in P}\eToll[e'] = \sinkTime$, which is a contradiction to the definition of $r_v$. Hence, we have $\tilde c_v(r_v(\sinkTime)) = \sinkTime$.
\end{proof}

\begin{corollary}\label{cor:cvInverseOfrv}
    For any feasible flow and any node $v \in V$ where the function $r_v$ is continuous, the functions $r_v$ and $\tilde c_v$ are inverses of each other, i.e.,
        \[r_v(\tilde c_v(\netTime)) = \netTime \text{ and } \tilde c_v(r_v(\netTime)) = \netTime \quad\text{for all } \netTime \in \IR.\]
\end{corollary}

\begin{proof}
    If $r_v$ is continuous then it is surjective. As $r_v$ is always injective, this shows that $r_v$ is bijective and, thus, any left-inverse is also a right-inverse. With \zcref{lem:cvLeftinverseOfrv}, this shows the claimed result.
\end{proof}

\subsection{Continuity of and Derivative-Bounds for \texorpdfstring{$r_v$}{r\_v}}\label{app:rLabelsContinuity}

In the following we aim to show that the node labels $r_v$ are Lipschitz continuous and their derivatives (which, therefore, exist almost everywhere) are strictly bounded away from $0$. We will obtain such bounds by first showing corresponding bounds for the node labels~$\tilde c_v$ which (by \zcref{lem:cvLeftinverseOfrv}) are the left-inverses of~$r_v$. The upper bound on the growth rate of~$\tilde c_v$ (and, correspondingly, the lower bound for $r_v$) is a direct consequence of the well-known Lipschitz-continuity of the path-travel time functions~$T_P$. To establish a lower bound for $\tilde c_v$ (and, correspondingly, an upper bound for~$r_v$) we mainly need to show that, in an equilibrium, queues cannot deplete too fast. For this, we will establish a lower bound on the inflow into arcs with queues on them.

\begin{lemma}\label{lem:cv Lipschitz}
    For any given network with fixed (constant) network inflow rate, there is a constant $M > 0$ such that for all feasible flows, all $v \in V$ and all $\netTime_1,\netTime_2 \in \IR$ we have
        \begin{align*}
            \abs{\tilde c_v(\netTime_2)-\tilde c_v(\netTime_1)} \leq M\abs{\netTime_2 - \netTime_1}
        \end{align*}
    (i.e., all $\tilde c_v$ are Lipschitz-continuous).
\end{lemma}

\begin{proof}
    As all $T_e$ are Lipschitz-continuous (see, e.g., \cite[Proposition~3.56]{GrafThesis}), so is $T_P$ for any path~$P$ as the concatenation of arc exit time functions~$T_e$.
    Since $\tilde c_v$ is defined as the minimum over the travel time functions $T_P$ for $P \in \PathSet[v,t]$ (plus some constant offset), Lipschitz-continuity of the latter implies Lipschitz continuity of the former (with $M \coloneqq \max\set{L_P \text{ Lipschitz-constant of } P | P \in \PathSet[v,t], v \in V}$). 
\end{proof}

As our first step towards a lower bound on the growth rate of $\tilde c_v$, we establish such a lower bound for the label at the source node~$s$. Here, we can use that the constant network inflow rate ensures that there is always some active \stpath{} with some minimum inflow. Hence, the queues along this path cannot deplete too fast and, hence, the travel time along this path has to grow with a certain minimum speed.

\begin{lemma}\label{lem:cs LowerBoundGrowth}
    For any given network with fixed (constant) network inflow rate, there is a constant $m_s > 0$ such that for all dynamic equilibrium flows and all $\sourceTime_1,\sourceTime_2 \in \IR$ we have
        \begin{align*}
            \abs{\tilde c_s(\sourceTime_2)-\tilde c_s(\sourceTime_1)} \geq m_s\abs{\sourceTime_2-\sourceTime_1}.
        \end{align*}
\end{lemma}

\begin{proof}
    Note that it suffices here to only consider times $\sourceTime_1,\sourceTime_2 \geq 0$, as before that there is no flow in the network and, hence, $\tilde c_s$ is just a linear function. We now start by fixing some path-decomposition~$(f_P) \in (L^1(\IR))^{\PathSet[s,t]}$ of the given equilibrium flow (which exists by~\cite[Corollary~5.5]{DynamicFlowDecomposition}\footnote{Note that, while the results from \cite{DynamicFlowDecomposition} only hold for flows with bounded support whereas we study flows with unbounded support in this paper, for the purposes of this proof step it suffices to only consider a part with large enough, but finite time horizon of our given flows.}). This flow decomposition is then a path-based dynamic equilibrium flow meaning that it satisfies
        \[f_P(\sourceTime) > 0 \implies \tilde c_s(\sourceTime) = T_P(\sourceTime) + \sum\nolimits_{e \in P}\eToll \text{ for almost all } \sourceTime \in \IR \text{ and all } P \in \PathSet[s,t].\]
    In particular we have, using the monotonicity of~$\tilde c_s$, for almost all times $\sourceTime \in [\sourceTime_1,\sourceTime_2]$ that
        \[f_P(\sourceTime) > 0 \implies T_P(\sourceTime) \in \bigcup_{p \in \uptosumTolls}[\tilde c_s(\sourceTime_1)-p,\tilde c_s(\sourceTime_2)-p] \text{ for all } P \in \PathSet[s,t],\]
    where $\sumTolls \coloneqq \sum_{e \in E}\eToll$ denotes the sum of all arc tolls in the network and the set $\uptosumTolls \coloneqq \set{0,1,\dots,\sumTolls}$ includes all possible tolls along any simple path.
    With this, we get the following estimate
        \begin{align*}
            u_0\cdot(\sourceTime_2-\sourceTime_1) 
                &= \sum_{P \in \PathSet[s,t]}\int_{\sourceTime_1}^{\sourceTime_2} f_P(\sourceTime)\diff\sourceTime 
                    \leq \sum_{p \in \uptosumTolls}\sum_{e \in \edgesTo{t}}\int_{\tilde c_s(\sourceTime_1)-p}^{\tilde c_s(\sourceTime_2)-p}f^-_e(\sinkTime)\diff\sinkTime \\
                &\symoverset{1}{\leq} \!\!\sum_{p \in \uptosumTolls}\sum_{e \in \edgesTo{t}}(\tilde c_s(\sourceTime_2)-\tilde c_s(\sourceTime_1))\eCap
                    = (\tilde c_s(\sourceTime_2)-\tilde c_s(\sourceTime_1))(\sumTolls+1)\!\!\sum_{e \in \edgesTo{t}}\!\!\eCap,
        \end{align*}
    where \refsym{1} holds since Vickrey flows respect arc capacities, i.e., $f^-_e \leq \eCap$.
    Thus, setting $m_s \coloneqq \frac{u_0}{(\sumTolls+1)\sum_{e \in \edgesTo{t}}\eCap}$ gives us the desired bound
        \[\tilde c_s(\sourceTime_2)-\tilde c_s(\sourceTime_1) \geq \frac{u_0}{(\sumTolls+1)\sum_{e \in \edgesTo{t}}\eCap}(\sourceTime_2-\sourceTime_1) = m_s(\sourceTime_2-\sourceTime_1). \qedhere\]
\end{proof}

In order to transfer the above bound to all nodes in the network, we first show that arcs with queue on them always have some minimum inflow. This result implies, in particular, that in an equilibrium, arcs with queues on them, always have positive inflow. For dynamic equilibria without tolls this was shown in~\cite[Lemma~2.4]{DEContUniqLT}.
\begin{lemma}\label{lem:QueueingEdgesHaveMinimumInflow}
    For any given network with fixed (constant) network inflow rate, there is a constant $\eps > 0$ such that for any tolled dynamic equilibrium flow, every arc~$e \in E$ and any interval $(\netTime_1,\netTime_2)$ during which there is a (positive) queue on that arc (i.e., $\qmNT > 0$ on $(\netTime_1,\netTime_2)$), we have $\int_{\netTime_1}^{\netTime_2}f^+_e(\netTime)\diff\netTime \geq \eps\cdot(\netTime_2-\netTime_1)$.
\end{lemma}

\begin{proof}
    As in the proof of \zcref{lem:cs LowerBoundGrowth}, we start by some path-decomposition~$(f_P) \in (L^1(\IR))^{\PathSet[s,t]}$ of the given equilibrium flow (which exists by~\cite[Corollary~5.5]{DynamicFlowDecomposition}). We then observe that we may assume without loss of generality that there is an \stpath~$P$ and a time $\sourceTime_1$ such that $P$ is active at source time~$\sourceTime_1$ (i.e., a path consisting only of arcs that are active in the sense of \zcref{def:TolledDE} -- see \cite[Claim~5.4]{DynamicFlowDecomposition}), $e$ is contained in this path and we have $T_{P_{< e}}(\sourceTime_1)=\netTime_1$, where $P_{<e}$ denotes the subpath of~$P$ containing all arcs up to (but not including) arc~$e$. We will later also use the analogue notation $P_{>e}$ for the subpath containing all arcs (strictly) after arc~$e$.

    This is, because $e$ has a queue during $[\netTime_1,\netTime_2]$ and, hence, must have had inflow at some point and, therefore, lay on an active \stpath. 
    Let $\netTime_1'= \netTime_1$ if $\netTime_1$ belongs to the closure of used entrance times into $e$, otherwise let $\netTime_1'$ be the last time in the closure of used entrance times before~$\netTime_1$ where $e$ was active. 
    Note that $\netTime_1'$ exists because the queue was supplied by earlier inflow. By continuity, there is an active \stpath~$P$ reaching $e$ at time $\netTime_1'$: take used entrance times converging to $\netTime_1'$ and select a subsequence with the same path and convergent source times.
    As there was no inflow between $\netTime_1'$ and $\netTime_1$ and the queue remains positive on the added interval, lower bounding the arc inflow during $[\netTime_1',\netTime_2]$ also gives the same lower bound for the smaller interval $[\netTime_1,\netTime_2]$. 

    Finally, let $\sourceTime_2 \in \IR$ be such that $T_{P_{< e}}(\sourceTime_2) = \netTime_2$.

    Since $P$ is active at (source time) $\sourceTime_1$ we have $\tilde c_s(\sourceTime_1) = T_P(\sourceTime_1)+\sum_{e' \in P}\eToll[e']$. Moreover, we certainly have $\tilde c_s(\sourceTime_2) \leq T_P(\sourceTime_2) + \sum_{e' \in P}\eToll[e']$. Now, let $L \geq 1$ be some common Lipschitz-constant for all $T_{P'}$ (over all simple paths $P'$ in the network), which exists by Lipschitz-continuity of all~$T_e$ (see \cite[Corollary~3.56]{GrafThesis}), and $m_s > 0$ the constant from \zcref{lem:cs LowerBoundGrowth}. Then, we have
        \begin{align*}
            m_s\cdot(\sourceTime_2-\sourceTime_1) 
                &\leq \tilde c_s(\sourceTime_2) - \tilde c_s(\sourceTime_1) \\
                &\leq T_P(\sourceTime_2) - T_P(\sourceTime_1) 
                    = T_{P_{>e}}(T_e(T_{P_{<e}}(\sourceTime_2))) - T_{P_{>e}}(T_e(T_{P_{<e}}(\sourceTime_1))) \\
                &= T_{P_{>e}}(T_e(\netTime_2)) - T_{P_{>e}}(T_e(\netTime_1))
                    \leq L\cdot\bigl(T_e(\netTime_2)-T_e(\netTime_1)\bigr) \\
                &= L\cdot\Bigl(\netTime_2+\frac{\qmNT(\netTime_2)}{\eCap}-\netTime_1-\frac{\qmNT(\netTime_1)}{\eCap}\Bigr) \\
                &= L\cdot\biggl(\netTime_2-\netTime_1+\frac{1}{\eCap}\Bigl(\int_{\netTime_1}^{\netTime_2} f^+_e(\netTime)\diff\netTime - \int_{\netTime_1+\eTTime}^{\netTime_2+\eTTime}f^-_e(\netTime)\diff\netTime\Bigr)\biggr) \\
                &\symoverset{1}{=} L\cdot\biggl(\netTime_2-\netTime_1+\frac{1}{\eCap}\int_{\netTime_1}^{\netTime_2} f^+_e(\netTime)\diff\netTime - \frac{1}{\eCap}\cdot(\netTime_2-\netTime_1)\cdot\eCap\biggr) \\
                &= \frac{L}{\eCap}\int_{\netTime_1}^{\netTime_2} f^+_e(\netTime)\diff\netTime.
        \end{align*}
    Here, \refsym{1} uses that Vickrey queues work at capacity, that is $f^-_e = \eCap$ during $(\netTime_1+\eTTime,\netTime_2+\eTTime)$ since $e$ has a queue during $(\netTime_1,\netTime_2)$. With another application of Lipschitz-continuity of $T_{P'}$ we get from there
        \begin{align*}
            \int_{\netTime_1}^{\netTime_2} f^+_e(\netTime)\diff\netTime \geq \frac{m_s\eCap}{L}\cdot(\sourceTime_2-\sourceTime_1) \geq \frac{m_s\eCap}{L^2}\cdot\bigl(T_{P_{<e}}(\sourceTime_2)-T_{P_{<e}}(\sourceTime_1)\bigr) = \frac{m_s\eCap}{L^2}\cdot(\netTime_2-\netTime_1).
        \end{align*}
    Hence, setting $\eps \coloneqq \frac{m_s\eCapmin}{L^2}$ gives the desired bound.
\end{proof}

The above \zcref[noref,nocap]{lem:QueueingEdgesHaveMinimumInflow} ensures that the arc travel time function $T_e$ cannot grow too slowly. This, then, directly implies the same for all path travel time functions~$T_P$ and, hence, for the node labels~$\tilde c_v$:

\begin{lemma}\label{lem:cv LowerBoundGrowth}
    For any given network with fixed (constant) network inflow rate, there is a constant $m > 0$ such that for all tolled dynamic equilibrium flows, all $v \in V$ and all $\netTime_1,\netTime_2 \in \IR$ we have
        \begin{align*}
            \abs{\tilde c_v(\netTime_2)-\tilde c_v(\netTime_1)} \geq m\abs{\netTime_2-\netTime_1}.
        \end{align*}
\end{lemma}

\begin{proof}
    We start by establishing a corresponding lower bound on the average rate of change for $T_e$:
    \begin{claim}
        For all $\netTime_1,\netTime_2 \in \IR$ with $\netTime_1 < \netTime_2$ we have
            \[T_e(\netTime_2)-T_e(\netTime_1) \geq \frac{\eps}{\eCap} (\netTime_2 - \netTime_1)\]
        where $\eps > 0$ is the minimum of~$\eCapmin$ and the constant from \zcref{lem:QueueingEdgesHaveMinimumInflow}.
    \end{claim}
    \begin{proofClaim}
        Since $T_e$ is absolutely continuous, it suffices to show that $T'_e(\netTime) \geq \frac{\eps}{\eCap}$ for times $\netTime \in \IR$ where $T'_e(\netTime)$ exists. For those, we now distinguish two cases:
        \begin{proofbycases}
            \caseitem{$\qmNT(\netTime) > 0$} By continuity of~$\qmNT$ arc~$e$ then also has non-zero queue on any small enough interval $(\netTime_1,\netTime_2)$ containing~$\netTime$. \zcref{lem:QueueingEdgesHaveMinimumInflow} then implies for any such interval
                \begin{align*}
                    T_e(\netTime_2) - T_e(\netTime_1) 
                        &= \netTime_2+\frac{\qmNT(\netTime_2)}{\eCap} -\netTime_1-\frac{\qmNT(\netTime_1)}{\eCap}   \\
                        &= \netTime_2-\netTime_1+\frac{1}{\eCap}\Bigl(\int_{\netTime_1}^{\netTime_2}f^+_e(\netTime)\diff\netTime - \int_{\netTime_1+\tau_e}^{\netTime_2+\tau_e}f^-_e(\netTime)\diff\netTime\Bigr) \\
                        &\symoverset{1}{\geq} \netTime_2-\netTime_1 + \frac{\eps}{\eCap}(\netTime_2-\netTime_1) - (\netTime_2-\netTime_1) = \frac{\eps}{\eCap}(\netTime_2-\netTime_1),
                \end{align*}
            where \refsym{1} holds since Vickrey flows respect arc capacities, i.e., $f^-_e \leq \eCap$. As this holds for any small enough interval containing~$\netTime$, this implies $T'_e(\netTime) \geq \frac{\eps}{\eCap}$.
            \caseitem{$\qmNT(\netTime) = 0$} Here, we can compute $T'_e(\netTime)$ directly as follows
                \begin{align*}
                    T'_e(\netTime) 
                        &= \Bigl(\netTime + \frac{\qmNT(\netTime)}{\eCap}  + \eTTime\Bigr)'
                            = 1 + \frac{\qmNT'(\netTime)}{\eCap} 
                              \symoverset{1}{=}1 + \max\Set{\frac{f^+_e(\netTime)}{\eCap}-1,0}
                            \geq 1 \geq \frac \eCapmin \eCap,
                \end{align*}
            where, at \refsym{1}, we use \cite[Lemma~3.1~(vii)]{SeringThesis}. \qedhere
        \end{proofbycases}
    \end{proofClaim}
    
    Now, for any \stpath[v]~$P$ we get $T_P(\netTime_2)-T_P(\netTime_1) \geq (\netTime_2 - \netTime_1)\prod_{e \in P}\frac{\eps}{\eCap}$ and, therefore, $m \coloneqq \min \{ 1, \bigl(\frac{\eps}{\eCapmax}\bigr)^{\abs{V}} \}$ yields the desired lower bound.
\end{proof}

Having established both upper and lower bound on the growth rate of the labels~$\tilde c_v$, the corresponding bounds for the labels~$r_v$ are now an immediate corollary (as these two label functions are inverses of each other):

\restaterBound*

\begin{proof}
    Let $m,M > 0$ be the constants from \zcref{lem:cv Lipschitz,lem:cv LowerBoundGrowth} and define $c \coloneqq \frac{1}{M} > 0$ and $C \coloneqq \frac{1}{m} > 0$. Since we have $\tilde c_v(r_v(\sinkTime))=\sinkTime$ for any $\sinkTime \in \IR$ by \zcref{lem:cvLeftinverseOfrv}, we get
        \begin{align*}
            \abs{\sinkTime_2-\sinkTime_1} = \abs{\tilde c_v(r_v(\sinkTime_2)) - \tilde c_v(r_v(\sinkTime_1))} \begin{cases}
                \leq M\abs{r_v(\sinkTime_2)-r_v(\sinkTime_1)} \\
                \geq m\abs{r_v(\sinkTime_2)-r_v(\sinkTime_1)}
            \end{cases}
        \end{align*}
    and, therefore, 
        \[c\abs{\sinkTime_2-\sinkTime_1} = \frac{1}{M}\abs{\sinkTime_2-\sinkTime_1} \leq \abs{r_v(\sinkTime_2)-r_v(\sinkTime_1)} \leq \frac{1}{m}\abs{\sinkTime_2-\sinkTime_1} = C\abs{\sinkTime_2-\sinkTime_1}. \qedhere\]
\end{proof}

\subsection{A Sink-Based Characterization of Tolled Dynamic Equilibria}

With $r_v$ and $\tilde c_v$ being inverses of each other we can characterize tolled dynamic equilibria via sink-based labels in a way similar to the one for untolled dynamic equilibria via source-based labels in~\cite[Lemma~1]{CCLDynEquil}. For this, we introduce the following notation:

\begin{definition}
    We say that an arc $e=vw \in E$ is \emph{active at sink-time~$\sinkTime$} if $T_e(r_v(\sinkTime)) = r_w(\sinkTime-\eToll)$. Hence,
        \[\setActiveEdgesR(\sinkTime) \coloneqq \set{e = vw \in E | T_e(r_v(\sinkTime)) = r_w(\sinkTime-\eToll)} \]
    denotes the set of active arcs at sink-time~$\sinkTime$. Moreover, we denote by
        \[\setActiveTimesR \coloneqq \set{\sinkTime \in \IR | e \in \setActiveEdgesR(\sinkTime)}\]
    the set of sink times for which a given arc~$e$ is active.
\end{definition}

\begin{lemma}\label{lem:TolledEqChara}
     A feasible flow $f$ is a tolled dynamic equilibrium if and only if for every arc $e=vw \in E$ we have
        \begin{align}\label{eq:TolledEqCharaActiveTimes}
            f^+_e(\netTime) > 0 \implies \netTime \in r_v(\setActiveTimesR)
        \end{align}
    for almost all  $\netTime \in \IR$.
\end{lemma}

\begin{proof}
    \begin{structuredproof}
        \proofitem{``$\implies$''} Let $f$ be a tolled dynamic equilibrium (in the sense of \zcref{eq:DefTolledDE}). Then, \zcref{cor:cvInverseOfrv} together with \zcref{cor:r AbsCont} implies that $\tilde c_v$ is a right-inverse of~$r_v$. We now show that every $\netTime \in \IR$ satisfying~\eqref{eq:DefTolledDE} also satisfies~\eqref{eq:TolledEqCharaActiveTimes}: Let $\netTime \in \IR$ such a time with $f^+_e(\netTime) > 0$. Then, we have $\tilde c_v(\netTime) = \eToll+\tilde c_w(T_e(\netTime))$ and, therefore,
            \[T_e(r_v(\tilde c_v(\netTime))) = T_e(\netTime) = r_w(\tilde c_w(T_e(\netTime))) = r_w(\tilde c_v(\netTime)-\eToll).\]
        Hence, $\sinkTime \coloneqq \tilde c_v(\netTime) \in \setActiveTimesR$ and, therefore, $\netTime = r_v(\tilde c_v(\netTime)) = r_v(\sinkTime) \in  r_v(\setActiveTimesR)$.

        \proofitem{``$\impliedby$''} Let $f$ be a feasible flow satisfying~\eqref{eq:TolledEqCharaActiveTimes} for all arcs $e=vw \in E$ and almost all $\netTime \in \IR$. We now show that any $\netTime \in \IR$ satisfying~\eqref{eq:TolledEqCharaActiveTimes} also satisfies~\eqref{eq:DefTolledDE}: Let $\netTime \in \IR$ such a time with $f^+_e(\netTime) > 0$. Then, there must be some $\sinkTime \in \setActiveTimesR$ with $r_v(\sinkTime) = \netTime$. Using the fact that $\tilde c_v$ is a left-inverse of~$r_v$ (see \zcref{lem:cvLeftinverseOfrv}) we compute
            \begin{align*}
                \tilde c_v(\netTime) 
                    &= \tilde c_v(r_v(\sinkTime)) 
                        = \sinkTime 
                        = \eToll + \sinkTime-\eToll 
                        = \eToll+ \tilde c_w(r_w(\sinkTime-\eToll)) \\
                    &= \eToll + \tilde c_w(T_e(r_v(\sinkTime))) 
                        = \eToll + \tilde c_w(T_e(\netTime)).
            \end{align*}
        Thus, $\netTime$ also satisfies~\eqref{eq:DefTolledDE}. \qedhere
    \end{structuredproof}
\end{proof}

The above \zcref[noref,nocap]{lem:TolledEqChara} characterizes tolled dynamic equilibria with a complementarity condition quantified over network times. There is also a natural analogue of this condition using sink based times. However, as in the case of untolled dynamic equilibria (see \cite[Remark in Section~2.5]{CCLDynEquil}), this does not yield an equivalent characterization. In our setting, though, in does still result in a necessary condition for equilibrium flows that will be useful later on:

\begin{corollary}\label{cor:TolledEqChara}
    For any tolled dynamic equilibrium flow~$f$ and every arc $e=vw \in E$ we have
        \begin{align}\label{eq:TolledEqCharaActiveEdges}
            f^+_e(r_v(\sinkTime)) > 0 \implies e \in \setActiveEdgesR(\sinkTime)
        \end{align}
    for almost all $\sinkTime \in \IR$. If all $r_v$ are absolutely continuous, then the reverse direction also holds.
\end{corollary}

\begin{proof}
    For a fixed arc~$e=vw$ we define 
        \[N \coloneqq \set{\sinkTime \in \IR | \sinkTime \text{ does not satisfy \eqref{eq:TolledEqCharaActiveEdges}}} = \set{\sinkTime \in \IR | f^+_e(r_v(\sinkTime)) > 0 \land e \notin \setActiveEdgesR(\sinkTime)}.\]
    Then, we have
        \begin{align*}
            r_v(N) &= \set{r_v(\sinkTime) | \sinkTime \in \IR: f^+_e(r_v(\sinkTime)) > 0 \land \sinkTime \notin \setActiveTimesR} \\
                &\symoverset{1}{\subseteq} \set{r_v(\sinkTime) | \sinkTime \in \IR: f^+_e(r_v(\sinkTime)) > 0 \land r_v(\sinkTime) \notin r_v(\setActiveTimesR)} \\ 
                &=\set{\netTime \in \IR | f^+_e(\netTime) > 0 \land \netTime \notin r_v(\setActiveTimesR)} \eqqcolon N',
        \end{align*}
    where \refsym{1} holds since $r_v$ is injective by \zcref{cor:r LowerBoundGrowth} and, hence,  $\sinkTime \notin \setActiveTimesR$ implies $r_v(\sinkTime) \notin r_v(\setActiveTimesR)$. 
    
    Now, $N'$ is exactly the set of times $\netTime \in \IR$, where \eqref{eq:TolledEqCharaActiveTimes} is violated. Hence, if $f$ is a tolled dynamic equilibrium, $N'$ must be a null set and so is $r_v(N)$. Moreover, $\tilde c_v$ is the left-inverse of $r_v$ (by \zcref{lem:cvLeftinverseOfrv}) and maps null sets to null sets (as absolutely continuous function -- see \zcref{lem:cv Lipschitz}). Thus, $N = \tilde c_v(r_v(N))$ is a null set as well.

    If $r_v$ is absolutely continuous, then we even have equality in~\refsym{1} (as $r_v$ is bijective then) and $r_v$ maps null sets to null sets. Hence, $N$ being a null set (i.e., \eqref{eq:TolledEqCharaActiveEdges} being true for almost all $\sinkTime \in \IR$) implies $N'$ being a null set (i.e., \eqref{eq:TolledEqCharaActiveTimes} being true for almost all $\sinkTime \in \IR$). This shows the reverse direction of the \zcref[noref,nocap]{cor:TolledEqChara}.
\end{proof}

Next, we show that for tolled dynamic equilibria, the node-labels $r_v$ already fully determine the set of active arcs as well as the arcs with queues on them and their respective size. This is, again, similar to untolled dynamic equilibria and source-based node labels (\cite[Proposition~2]{CCLDynEquil}).

\begin{lemma}\label{lem:CharaActiveAndResettingEdges}
	For any dynamic equilibrium we have for any arc $e=vw \in E$ and any time $\sinkTime \in \IR$ we have
	\begin{enumerate}[label=(\alph*)]
		\item $\wtST(\sinkTime)> 0 \implies e \in \setActiveEdgesR(\sinkTime)$, \thmpartlabel{lem:CharaActiveAndResettingEdges:QueueImpliesActive}
		\item $e \in \setActiveEdgesR(\sinkTime) \iff r_v(\sinkTime) + \eTTime \leq r_w(\sinkTime-\eToll)$,\thmpartlabel{lem:CharaActiveAndResettingEdges:CharaActive}
		\item $\wtST(\sinkTime) = [r_w(\sinkTime-\eToll) - r_v(\sinkTime) - \eTTime]^+$ and\thmpartlabel{lem:CharaActiveAndResettingEdges:QueueViaR}
        \item $\wtST(\sinkTime)> 0 \iff r_v(\sinkTime) + \eTTime < r_w(\sinkTime-\eToll)$.\thmpartlabel{lem:CharaActiveAndResettingEdges:CharaResetting}
	\end{enumerate}
\end{lemma}

This proof is similar to the one of \cite[Proposition~2]{CCLDynEquil}.

\begin{proof}
	\begin{structuredproof}
		\proofitem{\ref{lem:CharaActiveAndResettingEdges:QueueImpliesActive}} Since $\wtST$ is continuous, any time~$\sinkTime$ with non-zero queue also has a neighborhood with non-zero queue. By \zcref{lem:QueueingEdgesHaveMinimumInflow} there is inflow into arc~$e$ for some set of positive measure in every subinterval of the neighborhood, so by \zcref{lem:TolledEqChara} they all intersect $r_v(\setActiveTimesR)$. This set is closed (as preimage of the closed set $\setActiveTimesR$ under the continuous function~$\tilde c_v$ -- see \zcref{lem:cv Lipschitz,cor:cvInverseOfrv}) and, therefore $r_v(\varphi) \in r_v(A_e)$, i.e., $e \in E'_r(\varphi)$.
        
		\proofitem{\ref{lem:CharaActiveAndResettingEdges:CharaActive}} For any arc $e \in \setActiveEdgesR(\sinkTime)$ we have
		\[r_v(\sinkTime)+\eTTime \leq T_e(r_v(\sinkTime)) = r_w(\sinkTime-\eToll).\]
		On the other hand, for any arc~$e$ with $r_v(\sinkTime) + \eTTime \leq r_w(\sinkTime-\eToll)$ we either have $\wtST(\sinkTime) > 0$ and, therefore, $e \in \setActiveEdgesR(\sinkTime)$ by part~\ref{lem:CharaActiveAndResettingEdges:QueueImpliesActive} or $\wtST(\sinkTime) = 0$. The latter, together with $r_v(\sinkTime) + \eTTime \geq r_w(\sinkTime-\eToll) - \wtST(\sinkTime)$ from \zcref{lem:rLabelsAlternativeEquation}, implies $T_e(r_v(\sinkTime)) = r_v(\sinkTime)+\eTTime = r_w(\sinkTime-\eToll)$ and, thus, $e \in \setActiveEdgesR(\sinkTime)$.
		
		\proofitem{\ref{lem:CharaActiveAndResettingEdges:QueueViaR}} We distinguish two cases: If $e \in \setActiveEdgesR(\sinkTime)$, then
			\[r_v(\sinkTime) + \wtST(\sinkTime) + \eTTime = T_e(r_v(\sinkTime)) = r_w(\sinkTime-\eToll)\]
		and, therefore, $0 \leq \wtST (\sinkTime) = r_w(\sinkTime-\eToll) - r_v(\sinkTime) - \eTTime$. Otherwise, we have $\wtST(\sinkTime)=0$ by part~\ref{lem:CharaActiveAndResettingEdges:QueueImpliesActive} and $r_w(\sinkTime-\eToll) - r_v(\sinkTime) - \eTTime \leq 0$ by part~\ref{lem:CharaActiveAndResettingEdges:CharaActive}. 

        \proofitem{\ref{lem:CharaActiveAndResettingEdges:CharaResetting}} Follows directly from part~\ref{lem:CharaActiveAndResettingEdges:QueueViaR}. \qedhere
	\end{structuredproof}
\end{proof}

\begin{corollary}\label{cor:WaitingTimeDerivative}
    For any dynamic equilibrium we have for any arc $e=vw \in E$ that $\wtST$ is absolutely continuous and satisfies
    \[\wtST'(\sinkTime) = \begin{cases}
        r_w'(\sinkTime-\eToll) - r_v'(\sinkTime), &\text{if } \wtST(\sinkTime) > 0 \\
        0, &\text{else}
    \end{cases} \quad\text{for almost all } \sinkTime \in \IR.\]
\end{corollary}

\begin{proof}
    The function $\wtST$ is absolutely continuous as concatenation of the absolutely continuous function~$\qmNT$ and the monotone, absolutely continuous function~$r_v$. Hence, it is differentiable for almost all $\sinkTime \in \IR$. Now consider for any point~$\sinkTime \in \IR$ of differentiability of $\wtST$ we distinguish two cases:
    \begin{proofbycases}
        \caseitem{$\wtST(\sinkTime) > 0$} Here, continuity of~$\wtST$ implies that $\wtST > 0$ also holds in some neighborhood of~$\sinkTime$. \zcreftp{lem:CharaActiveAndResettingEdges:QueueViaR} then directly implies
            \[\wtST'(\sinkTime) = r'_w(\sinkTime-\eToll)-r'_v(\sinkTime).\]
        \caseitem{$\wtST(\sinkTime) = 0$} Since $\wtST \geq 0$, this directly implies $\wtST'(\sinkTime)=0$ (see \cite[Lemma~9]{CCLDynEquil}). \qedhere
    \end{proofbycases}
\end{proof}

\subsection{Sink-Based Thin Flows}

Based on the characterizations of tolled dynamic equilibria in the last section, we can now derive another characterization based on derivatives of the cumulative arc inflow $x_e(\sinkTime) \coloneqq F^+_e(r_v(\sinkTime))$ and the node labels~$r_v$. 
We start by showing some basic properties of the former.

\restatexLipschitz*

\begin{proof}
    The node label $r_v$ is Lipschitz-continuous by \zcref{lem:r Lipschitz}. For $F^+_e$ we note that, due to flow conservation at node~$v$, $f^+_e$ is upper bounded by $u_0+\sum_{e' \in \edgesTo{v}}f^-_{e'}$ which itself is upper bounded by $u_0+\sum_{e' \in \edgesTo{v}}\eCap[e']$ since Vickrey flows respect capacity. Thus, $x_e$ is Lipschitz continuous as concatenation of two Lipschitz-continuous functions.
\end{proof}

\begin{lemma}\label{lem:TolledInFlowOutFlow}
     For any tolled dynamic equilibrium~$f$, every arc $e=vw \in E$ and every time $\sinkTime \in \IR$ we have
        \[F^+_e(r_v(\sinkTime)) = F^-_e(r_w(\sinkTime-\eToll)).\]
\end{lemma}

We show this using a similar approach to \cite[Theorem~1 (a)$\iff$(b)]{CCLDynEquil}.

\begin{proof}
	Consider any arc~$e=vw$ and generalized arrival time~$\sinkTime \in \IR$. We define
		\[\netTime \coloneqq \sup\set{\netTime' | T_e(\netTime') \leq r_w(\sinkTime-\eToll)}.\]
	Due to the continuity of~$T_e$, we then have $T_e(\netTime) = r_w(\sinkTime-\eToll)$. Moreover, by \zcref{lem:rLabelsBellmanEq}, we have
		\[r_v(\sinkTime) \geq \sup\set{\netTime' | T_e(\netTime') \leq r_w(\sinkTime-\eToll)}\]
	and, therefore $r_v(\sinkTime) \geq \netTime$. With this, we compute
		\begin{align*}
			F^+_e(r_v(\sinkTime)) 
				&= F^+_e(\netTime) + \int_{\netTime}^{r_v(\sinkTime)}f^+_e(\netTime')\diff\netTime' \\
				&\coverset[Prop.~3.19e)]{GrafThesis}{=} F^-_e(T_e(\netTime)) + \int_{\netTime}^{r_v(\sinkTime)}f^+_e(\netTime')\diff\netTime' \\
				&=F^-_e(r_w(\sinkTime-\eToll))  + \int_{\netTime}^{r_v(\sinkTime)}f^+_e(\netTime')\diff\netTime'.
		\end{align*}
    Now, it remains to show that $f^+_e(\netTime') = 0$ for almost all $\netTime' \in (\netTime,r_v(\sinkTime)]$. If $\netTime' \notin r_v(\IR)$ this follows directly from condition~\eqref{eq:TolledEqCharaActiveTimes} in \zcref{lem:TolledEqChara}. Otherwise, choose some $\sinkTime' \in \IR$ with $r_v(\sinkTime') = \netTime'$. Since $r_v$ is strictly increasing (see \zcref{cor:r LowerBoundGrowth}), this $\sinkTime'$ is unique and satisfies $\sinkTime' \leq \sinkTime$. Then, we have
        \[T_e(r_v(\sinkTime')) = T_e(\netTime') \symoverset{1}{>} r_w(\sinkTime-\eToll) \geq r_w(\sinkTime'-\eToll),\]
    where \refsym{1} holds due to our choice of~$\netTime$. This, then, implies $\sinkTime' \notin \setActiveTimesR$ and, therefore, $\netTime' \notin r_v(\setActiveTimesR)$. Hence, \eqref{eq:TolledEqCharaActiveTimes} implies $f^+_e(\netTime') = 0$ for almost all such $\netTime'$.
\end{proof}

\restatexeFlowConservation*

\begin{proof}
    For $v \in V\setminus\set{s,t}$ this follows from \zcref{lem:TolledInFlowOutFlow} together with $f$ satisfying flow conservation at~$v$:
        \begin{align*}
            \sum_{e \in \edgesTo{v}}x_e(\sinkTime+\eToll)
                &= \sum_{e=uv \in \edgesTo{v}}F^+_e(r_u(\sinkTime+\eToll)) 
                    \zoverset{lem:TolledInFlowOutFlow}{=} \sum_{e=uv \in \edgesTo{v}}F^-_e(r_v(\sinkTime)) \\
                &= \sum_{e=vw \in \edgesFrom{v}}F^+_e(r_v(\sinkTime)) 
                    = \sum_{e \in \edgesFrom{v}}x_e(\sinkTime). 
        \end{align*}
    For the source~$s$, flow conservation there gives us
        \begin{align*}
            &u_0\cdot r_s(\sinkTime) \CharF[\sinkTime \geq \crem[s][0]] + \sum_{e \in \edgesTo{s}}x_e(\sinkTime+\eToll) \\
                &\qquad\qquad=u_0\cdot r_s(\sinkTime) \CharF[\sinkTime \geq \crem[s][0]]+ \sum_{e=us \in \edgesTo{s}}F^+_e(r_u(\sinkTime+\eToll)) \\
                &\qquad\quad\!\zoverset{lem:TolledInFlowOutFlow}{=} u_0\cdot r_s(\sinkTime) \CharF[\sinkTime \geq \crem[s][0]]+ \sum_{e=us \in \edgesTo{s}}F^-_e(r_s(\sinkTime)) \\
                &\qquad\qquad= \CharF[\sinkTime \geq \crem[s][0]] \left(\int_0^{r_s(\sinkTime)}u_0\diff\netTime \right)+ \sum_{e=us \in \edgesTo{s}}\int_0^{r_s(\sinkTime)}f^-_e(\netTime)\diff\netTime \\
                &\qquad\qquad=  \int_0^{r_s(\sinkTime)} \left(  \CharF[\sinkTime \geq \crem[s][0]] u_0 + \sum_{e=us \in \edgesTo{s}}f^-_e(\netTime) \right) \diff\netTime \\
                &\qquad\qquad= \int_0^{r_s(\sinkTime)}\sum_{e=sw \in \edgesFrom{s}}f^+_e(\netTime)\diff\netTime \\
                &\qquad\qquad= \sum_{e \in \edgesFrom{s}}x_e(\sinkTime).
        \end{align*}
    The analogue statements about the derivatives then follow immediately.
\end{proof}

In the following, we present the thin flow equations based on latest departure labels. 

\restateThinFlowCharacterization*

\begin{proof}
    For the first part let $f$ be a tolled dynamic equilibrium. Then, the functions $x$ and $r$ are absolutely continuous (by \zcref{lem:xLipschitz} and \zcref{lem:r Lipschitz}, respectively) and, hence, their derivatives $x'$ and $r'$ exist almost everywhere.      
    From \zcref{cor:Fe+andFe-} we then directly get~\eqref{eq:TF_FlowCons}.
    Moreover, \eqref{eq:TF_InactiveEdges} is an immediate consequence of \zcref{cor:TolledEqChara}.

    For the conditions on the node-labels we use \zcref{lem:rLabelsAlternativeEquation}: \eqref{eq:TF_Sink} is a direct consequence of $r_t(\sinkTime) = \sinkTime$. For the other nodes (i.e., \eqref{eq:TF_NodeLabel}) we apply the differentiation rule for maxima (i.e., the exact analogue of the differentiation rule for minima in \cite[Lemma~2.3]{SeringThesis}) and the fact that, by definition of~$\setActiveEdgesR(\sinkTime)$, we have $r_v(\sinkTime)=r_w(\sinkTime-\eToll)-\eTTime-\wtST(\sinkTime)$ if and only if $e=vw \in \setActiveEdgesR(\sinkTime) = E'$:
    \begin{align*}
        r'_v(\sinkTime) &= \Bigl(\max_{e=vw}r_w(\sinkTime-\eToll) -\tau_e-\wtST (\sinkTime)\Bigr)' \\
            &= \max_{e=vw\in E'}r'_w(\sinkTime-\eToll) -\frac{\qmNT'(r_v(\sinkTime))}{\eCap}r_v'(\sinkTime)\\
            &\symoverset{1}{=} \max_{e=vw\in E'}\begin{cases}
                r'_w(\sinkTime-\eToll) - \left(\frac{f_e^+(r_v(\sinkTime))}{\nu_e} - 1\right)r_v'(\sinkTime) & e\in E^*\\
                r'_w(\sinkTime-\eToll) - \left(\max\{\frac{f_e^+(r_v(\sinkTime))}{\nu_e} - 1,0\}\right)r_v'(\sinkTime) & e\in E'\setminus E^*\\
            \end{cases}\\
            &\symoverset{2}{=} \max_{e=vw\in E'}\begin{cases}
                r'_w(\sinkTime-\eToll) - \frac{x_e'(\sinkTime)}{\nu_e} +r_v'(\sinkTime) & e\in E^*\\
                r'_w(\sinkTime-\eToll) - \max\{\frac{x'_e(\sinkTime)}{\nu_e} - r_v'(\sinkTime),0\} & e\in E'\setminus E^*\\
            \end{cases},
    \end{align*}
    where \refsym{1} holds due to \cite[Lemma~3.1~(vii)]{SeringThesis}: 
        \[\frac{\qmNT'(\netTime)}{\eCap} = \begin{cases}
            \frac{f_e^+(\netTime)}{\nu_e} - 1, & \text{if } \qmNT(\netTime)>0 \\
            \max\Set{\frac{f_e^+(\netTime)}{\nu_e} - 1,0},&\text{else}
        \end{cases}\]
    and \refsym{2} since, by definition of $x_e$ and chain rule, we have $x'_e(\sinkTime) = f^+_e(r_v(\sinkTime))r'_v(\sinkTime)$.
    
    For \eqref{eq:TF_NodeLabelUsedEdge} it suffices to only consider times~$\sinkTime$ where $e=vw$ is active (as this is true for almost all times with $x'_e(\sinkTime)>0$ by \zcref{cor:TolledEqChara}), i.e., 
        \begin{align}\label{eq:EdgeActive}
            r_v(\sinkTime) + \wtST(\sinkTime) + \eTTime = r_w(\sinkTime-\eToll).
        \end{align}
    As all functions involved in this equality are absolutely continuous, we can further restrict ourselves to times where all their derivatives exist. We now distinguish to sets of times:
    \begin{proofbycases}
        \caseitem{$(r_v(\sinkTime) + \wtST(\sinkTime) + \eTTime)' \neq (r_w(\sinkTime-\eToll))'$} Then, due to continuity of both sides, there is a proper interval $(\sinkTime,\sinkTime+\eps)$ such that \eqref{eq:EdgeActive} does not hold on any point in this interval. Since \eqref{eq:EdgeActive} holds at~$\sinkTime$, these intervals (for different $\sinkTime$) cannot overlap. Thus, there can be only countably many times in this case and, consequently, they form a null set.
        \caseitem{$(r_v(\sinkTime) + \wtST(\sinkTime) + \eTTime)' = (r_w(\sinkTime-\eToll))'$} Here, we can use the same steps as for \eqref{eq:TF_NodeLabel} to get
            \begin{align*}
               r'_v(\sinkTime) 
                    &= r'_w(\sinkTime-\eToll) - \wtST'(\sinkTime) 
                        = r'_w(\sinkTime-\eToll) - \frac{\qmNT'(r_v(\sinkTime))}{\eCap}r'_v(\sinkTime) \\
                    &= \begin{cases}
                            r'_w(\sinkTime-\eToll) - \frac{x_e'(\sinkTime)}{\nu_e} +r_v'(\sinkTime) & e\in E^*\\
                            r'_w(\sinkTime-\eToll) - \max\{\frac{x'_e(\sinkTime)}{\nu_e} - r_v'(\sinkTime),0\} & e\in E'\setminus E^*\\
                        \end{cases}.
            \end{align*}
    \end{proofbycases}

    The second part of the \zcref[noref,nocap]{lem:ThinFlowCharacterization} is a direct consequence of \zcref{cor:TolledEqChara}: Under the given assumption of absolute continuity for~$r_v$, \eqref{eq:TF_InactiveEdges} directly implies that $f$ is a tolled dynamic equilibrium.  
\end{proof} 

\restateBoundsOnX*

\begin{proof}
    By \zcref{lem:ThinFlowCharacterization} it suffices to only consider times $\sinkTime \in \IR$ where $x'$ and $r'$ satisfy the sink-based thin flow equations. If $x'_e(\sinkTime) = 0$, there is nothing to prove (as $r'_v(\sinkTime) \geq 0$ by \zcref{cor:r LowerBoundGrowth}). Otherwise, \eqref{eq:TF_NodeLabelUsedEdge} gives us
        \begin{align*}
            r'_v(\sinkTime) 
                &= \begin{cases}
                    r'_w(\sinkTime-\eToll)-\frac{x'_e(\sinkTime)}{\eCap} + r'_v(\sinkTime), &e \in E^*_r(\sinkTime) \\
                    r'_w(\sinkTime-\eToll)-\max\Set{\frac{x'_e(\sinkTime)}{\eCap} - r'_v(\sinkTime),0}, &e \in E'_r(\sinkTime) \setminus E^*_r(\sinkTime) 
                \end{cases}\\
                &\leq r'_w(\sinkTime-\eToll)-\frac{x'_e(\sinkTime)}{\eCap} + r'_v(\sinkTime).
        \end{align*}
    For $e \in E^*_r(\sinkTime)$ this implies $0 = r'_w(\sinkTime-\eToll)-\frac{x'_e(\sinkTime)}{\eCap}$, while for all other arcs $e \in E'_r(\sinkTime) \setminus E^*_r(\sinkTime) $ it still gives us $0 \leq r'_w(\sinkTime-\eToll)-\frac{x'_e(\sinkTime)}{\eCap}$. Finally, \eqref{eq:TF_InactiveEdges} directly implies $x'_e(\sinkTime) = 0$ whenever $e \in E \setminus E'_r(\sinkTime)$.
\end{proof}

\section{Non-finite Convergence}\label{sec:nonfinite_conv}

\begin{theorem}\label{thm:non_finite_convergence}
    There exists an instance with $u_0 < \mincut$ for which a tolled dynamic equilibrium does not reach the steady state in finite time.
\end{theorem}

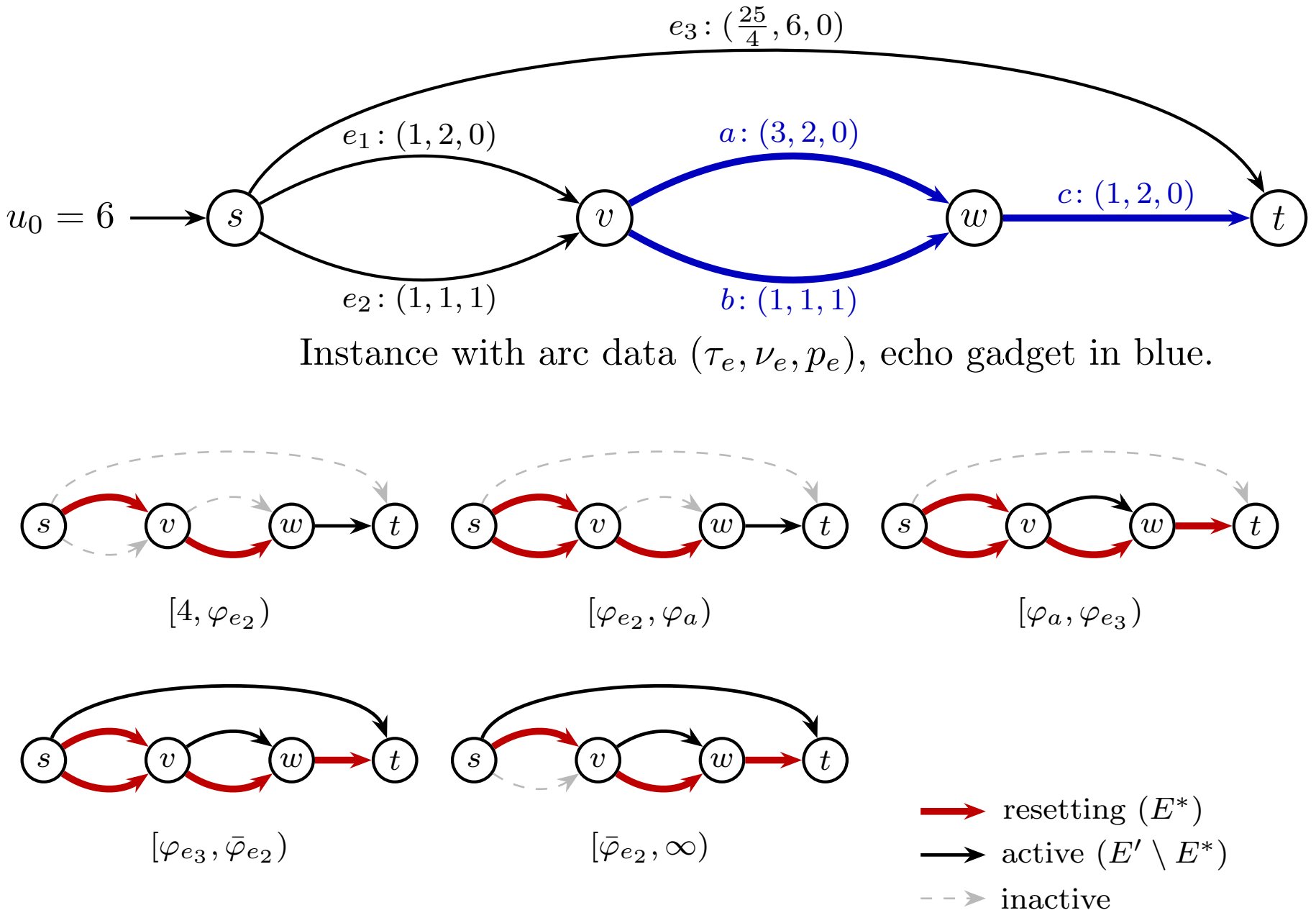
\begin{figure}\centering
    \begin{adjustbox}{max width=\textwidth}
  \begin{minipage}{122mm}\centering
\instancepic\\[4mm]
\begin{tabular}{@{}c@{\hspace{3mm}}c@{\hspace{3mm}}c@{}}
\phasepic{$[4,\sinkTime_{e_2})$}{R}{I}{I}{I}{R}{A} &
\phasepic{$[\sinkTime_{e_2},\sinkTime_a)$}{R}{R}{I}{I}{R}{A} &
\phasepic{$[\sinkTime_a,\sinkTime_{e_3})$}{R}{R}{I}{A}{R}{R}\\[1mm]
\phasepic{$[\sinkTime_{e_3},\bar\sinkTime_{e_2})$}{R}{R}{A}{A}{R}{R} &
\phasepic{$[\bar\sinkTime_{e_2},\infty)$}{R}{I}{A}{A}{R}{R} &
\begin{tikzpicture}[baseline=-6mm]
\draw[R](0,0)--(.6,0)node[right,font=\scriptsize,black]{resetting ($E^*$)};
\draw[A](0,-.4)--(.6,-.4)node[right,font=\scriptsize]{active ($E'\setminus E^*$)};
\draw[I](0,-.8)--(.6,-.8)node[right,font=\scriptsize,black]{inactive};
\end{tikzpicture}
\end{tabular}
\end{minipage}
    \end{adjustbox}
    \caption{Instance for which the equilibrium does not reach steady state in finite time. Shown below are the configurations of active and resetting arcs in five phases. The last phase lasts forever, nevertheless, queues do not stabilize.}\label{fig:NonFiniteConvergence}
\end{figure}

Essentially, the network shown in \zcref{fig:NonFiniteConvergence} consists of a loading gadget (arcs $e_1,e_2,e_3$) and an echo gadget (arcs $a,b,c$).
The loading gadget's purpose is to place queues on $b$ and $c$, as well as to provide capacity exceeding $u_0$ (via arc~$e_3$). Once the queues have been loaded, convergence taking infinite time is then due to flow shifting between the \stpath[v]s $(a,c)$ and $(b,c)$ in the echo gadget which, in turn, shifts an ever decreasing but never vanishing amount of queue mass between $b$ and on $c$.

\begin{proof}
    Consider the network in \zcref{fig:NonFiniteConvergence}. We claim that the flow described in \zcref{tab:equilibrium_vals} is a tolled dynamic equilibrium that never reaches a steady state. In this table we describe the flow via the (piecewise constant) derivatives of its corresponding functions $r_v$, $x_e$ and $\wtST$. Together with the starting values $r_s(4)=0$, $r_v(4)=1$, $r_w(4)=3$, $x_c(4)=1$, $x_e(4)=0$ for all $e \neq c$ and $\wtST(4)=0$ for all $e$, these uniquely define the continuous functions $r_v$, $x_e$ and $\wtST$.

    \begin{table}[!ht]\centering\small\renewcommand{\arraystretch}{1.45}\setlength{\tabcolsep}{4pt}
        \begin{tabular}{@{}l@{}c@{}w{c}{\colw}@{}c@{}@{}c@{}w{c}{\colw}@{}c@{}@{}c@{}w{c}{\colw}@{}c@{}|@{}c@{}w{c}{\colw}@{}c@{}@{}c@{}w{c}{\colw}@{}c@{}@{}c@{}w{c}{\colw}@{}c@{}|@{}c@{}w{c}{\colw}@{}c@{}@{}}
            $\sinkTime\in$ & & \rotatebox{70}{$[4,\tfrac{13}3)$} &  & & \rotatebox{70}{$[\tfrac{13}3,\sinkTime_{e_2})$} &  & & \rotatebox{70}{$[\sinkTime_{e_2},\sinkTime_a)$} &  & & \rotatebox{70}{$I_1$} &  & & \rotatebox{70}{$I_2\cap[0,\sinkTime_{e_3})$} &  & & \rotatebox{70}{$I_2\cap[\sinkTime_{e_3},\infty)$} &  & & \rotatebox{70}{$I_k,\ k\ge3$} & \\\midrule
            $r_s'$ & \hspace{\hsp} & $\tfrac16$ & \hspace{\hsp} & \hspace{\hsp} & $\tfrac19$ & \hspace{\hsp} & \hspace{\hsp} & $\tfrac16$ & \hspace{\hsp} & \hspace{\hsp} & $\tfrac{11}{36}$ & \hspace{\hsp} & \hspace{\hsp} & $\tfrac{31}{72}$ & \hspace{\hsp} & \hspace{\hsp} & \hfill\enspace$1$\enspace\dfill\null & \rule[0.55ex]{\hsp}{0.4pt} & \rule[0.55ex]{\hsp}{0.4pt} & \dfill\null & \hspace{\hsp}\\
            $r_v'$ & \hspace{\hsp} & $\tfrac12$ & \hspace{\hsp} & \hspace{\hsp} & \hfill\enspace$\tfrac13$\enspace\dfill\null & \rule[0.55ex]{\hsp}{0.4pt} & \rule[0.55ex]{\hsp}{0.4pt} & \dfill\null & \hspace{\hsp} & \hspace{\hsp} & $\tfrac34$ & \hspace{\hsp} & \hspace{\hsp} & \hfill\enspace$\tfrac{11}{12}$\enspace\dfill\null & \rule[0.55ex]{\hsp}{0.4pt} & \rule[0.55ex]{\hsp}{0.4pt} & \dfill\null & \hspace{\hsp} & \hspace{\hsp} & $1-q_c'$ & \hspace{\hsp}\\
            $r_w'$ & \hspace{\hsp} & \dfill\null & \rule[0.55ex]{\hsp}{0.4pt} & \rule[0.55ex]{\hsp}{0.4pt} & \dfill\enspace$1$\enspace\dfill\null & \rule[0.55ex]{\hsp}{0.4pt} & \rule[0.55ex]{\hsp}{0.4pt} & \dfill\null & \hspace{\hsp} & \hspace{\hsp} & $\tfrac34$ & \hspace{\hsp} & \hspace{\hsp} & \hfill\enspace$\tfrac{11}{12}$\enspace\dfill\null & \rule[0.55ex]{\hsp}{0.4pt} & \rule[0.55ex]{\hsp}{0.4pt} & \dfill\null & \hspace{\hsp} & \hspace{\hsp} & $1-q_c'$ & \hspace{\hsp}\\\midrule
            $x_{e_1}'$ & \hspace{\hsp} & $1$ & \hspace{\hsp} & \hspace{\hsp} & \hfill\enspace$\tfrac23$\enspace\dfill\null & \rule[0.55ex]{\hsp}{0.4pt} & \rule[0.55ex]{\hsp}{0.4pt} & \dfill\null & \hspace{\hsp} & \hspace{\hsp} & $\tfrac32$ & \hspace{\hsp} & \hspace{\hsp} & \hfill\enspace$\tfrac{11}6$\enspace\dfill\null & \rule[0.55ex]{\hsp}{0.4pt} & \rule[0.55ex]{\hsp}{0.4pt} & \dfill\null & \hspace{\hsp} & \hspace{\hsp} & $2-2q_c'$ & \hspace{\hsp}\\
            $x_{e_2}'$ & \hspace{\hsp} & \hfill\enspace$0$\enspace\dfill\null & \rule[0.55ex]{\hsp}{0.4pt} & \rule[0.55ex]{\hsp}{0.4pt} & \dfill\null & \hspace{\hsp} & \hspace{\hsp} & \hfill\enspace$\tfrac13$\enspace\dfill\null & \rule[0.55ex]{\hsp}{0.4pt} & \rule[0.55ex]{\hsp}{0.4pt} & \dfill\null & \hspace{\hsp} & \hspace{\hsp} & \hfill\enspace$\tfrac34$\enspace\dfill\null & \rule[0.55ex]{\hsp}{0.4pt} & \rule[0.55ex]{\hsp}{0.4pt} & \dfill\null & \hspace{\hsp} & \hspace{\hsp} & $0$ & \hspace{\hsp}\\
            $x_{e_3}'$ & \hspace{\hsp} & \dfill\null & \rule[0.55ex]{\hsp}{0.4pt} & \rule[0.55ex]{\hsp}{0.4pt} & \dfill\null & \rule[0.55ex]{\hsp}{0.4pt} & \rule[0.55ex]{\hsp}{0.4pt} & \dfill\enspace$0$\enspace\dfill\null & \rule[0.55ex]{\hsp}{0.4pt} & \rule[0.55ex]{\hsp}{0.4pt} & \dfill\null & \rule[0.55ex]{\hsp}{0.4pt} & \rule[0.55ex]{\hsp}{0.4pt} & \dfill\null & \hspace{\hsp} & \hspace{\hsp} & $\tfrac{41}{12}$ & \hspace{\hsp} & \hspace{\hsp} & $4+2q_c'$ & \hspace{\hsp}\\
            $x_a'$ & \hspace{\hsp} & \dfill\null & \rule[0.55ex]{\hsp}{0.4pt} & \rule[0.55ex]{\hsp}{0.4pt} & \dfill\enspace$0$\enspace\dfill\null & \rule[0.55ex]{\hsp}{0.4pt} & \rule[0.55ex]{\hsp}{0.4pt} & \dfill\null & \hspace{\hsp} & \hspace{\hsp} & $\tfrac54$ & \hspace{\hsp} & \hspace{\hsp} & \hfill\enspace$\tfrac{13}{12}$\enspace\dfill\null & \rule[0.55ex]{\hsp}{0.4pt} & \rule[0.55ex]{\hsp}{0.4pt} & \dfill\null & \hspace{\hsp} & \hspace{\hsp} & $1+q_c'$ & \hspace{\hsp}\\
            $x_b'$ & \hspace{\hsp} & \dfill\null & \rule[0.55ex]{\hsp}{0.4pt} & \rule[0.55ex]{\hsp}{0.4pt} & \dfill\enspace$1$\enspace\dfill\null & \rule[0.55ex]{\hsp}{0.4pt} & \rule[0.55ex]{\hsp}{0.4pt} & \dfill\null & \rule[0.55ex]{\hsp}{0.4pt} & \rule[0.55ex]{\hsp}{0.4pt} & \dfill\null & \hspace{\hsp} & \hspace{\hsp} & \hfill\enspace$\tfrac34$\enspace\dfill\null & \rule[0.55ex]{\hsp}{0.4pt} & \rule[0.55ex]{\hsp}{0.4pt} & \dfill\null & \hspace{\hsp} & \hspace{\hsp} & $1-3q_c'$ & \hspace{\hsp}\\
            $x_c'$ & \hspace{\hsp} & \dfill\null & \rule[0.55ex]{\hsp}{0.4pt} & \rule[0.55ex]{\hsp}{0.4pt} & \dfill\enspace$1$\enspace\dfill\null & \rule[0.55ex]{\hsp}{0.4pt} & \rule[0.55ex]{\hsp}{0.4pt} & \dfill\null & \hspace{\hsp} & \hspace{\hsp} & \dfill\null & \rule[0.55ex]{\hsp}{0.4pt} & \rule[0.55ex]{\hsp}{0.4pt} & \dfill\enspace$2$\enspace\dfill\null & \rule[0.55ex]{\hsp}{0.4pt} & \rule[0.55ex]{\hsp}{0.4pt} & \dfill\null & \rule[0.55ex]{\hsp}{0.4pt} & \rule[0.55ex]{\hsp}{0.4pt} & \dfill\null & \hspace{\hsp}\\\midrule
            $q_{e_1}'$ & \hspace{\hsp} & $\tfrac13$ & \hspace{\hsp} & \hspace{\hsp} & $\tfrac29$ & \hspace{\hsp} & \hspace{\hsp} & $\tfrac16$ & \hspace{\hsp} & \hspace{\hsp} & $\tfrac49$ & \hspace{\hsp} & \hspace{\hsp} & $\tfrac{35}{72}$ & \hspace{\hsp} & \hspace{\hsp} & $-\tfrac1{12}$ & \hspace{\hsp} & \hspace{\hsp} & $-q_c'$ & \hspace{\hsp}\\
            $q_{e_2}'$ & \hspace{\hsp} & \hfill\enspace$0$\enspace\dfill\null & \rule[0.55ex]{\hsp}{0.4pt} & \rule[0.55ex]{\hsp}{0.4pt} & \dfill\null & \hspace{\hsp} & \hspace{\hsp} & $\tfrac16$ & \hspace{\hsp} & \hspace{\hsp} & $\tfrac1{36}$ & \hspace{\hsp} & \hspace{\hsp} & $\tfrac{23}{72}$ & \hspace{\hsp} & \hspace{\hsp} & $-\tfrac14$ & \hspace{\hsp} & \hspace{\hsp} & $0$ & \hspace{\hsp}\\
            $q_b'$ & \hspace{\hsp} & $\tfrac12$ & \hspace{\hsp} & \hspace{\hsp} & \hfill\enspace$\tfrac23$\enspace\dfill\null & \rule[0.55ex]{\hsp}{0.4pt} & \rule[0.55ex]{\hsp}{0.4pt} & \dfill\null & \hspace{\hsp} & \hspace{\hsp} & $\tfrac14$ & \hspace{\hsp} & \hspace{\hsp} & \hfill\enspace$-\tfrac16$\enspace\dfill\null & \rule[0.55ex]{\hsp}{0.4pt} & \rule[0.55ex]{\hsp}{0.4pt} & \dfill\null & \hspace{\hsp} & \hspace{\hsp} & $-2q_c'$ & \hspace{\hsp}\\
            $q_c'$ & \hspace{\hsp} & \dfill\null & \rule[0.55ex]{\hsp}{0.4pt} & \rule[0.55ex]{\hsp}{0.4pt} & \dfill\enspace$0$\enspace\dfill\null & \rule[0.55ex]{\hsp}{0.4pt} & \rule[0.55ex]{\hsp}{0.4pt} & \dfill\null & \hspace{\hsp} & \hspace{\hsp} & $\tfrac14$ & \hspace{\hsp} & \hspace{\hsp} & \hfill\enspace$\tfrac1{12}$\enspace\dfill\null & \rule[0.55ex]{\hsp}{0.4pt} & \rule[0.55ex]{\hsp}{0.4pt} & \dfill\null & \hspace{\hsp} & \hspace{\hsp} & $\tfrac14\,3^{1-k}$ & \hspace{\hsp}\\
            \bottomrule\\ 
        \end{tabular}
        
        \caption{The derivatives of the equilibrium flow. Horizontal lines indicate constant values. In the last column
        $q_c'=\tfrac14\,3^{1-k}$ on $I_k$.}\label{tab:equilibrium_vals}
    \end{table}

\begin{table}[ht]\centering\small\renewcommand{\arraystretch}{1.9}
    \begin{tabular}{@{}l@{}c@{}w{c}{\colw}@{}c@{}@{}c@{}w{c}{\colw}@{}c@{}@{}c@{}w{c}{\colw}@{}c@{}|@{}c@{}w{c}{\colw}@{}c@{}@{}c@{}w{c}{\colw}@{}c@{}@{}c@{}w{c}{\colw}@{}c@{}|@{}c@{}w{c}{\colw}@{}c@{}@{}}
    $\varphi\in$ & & \rotatebox{70}{$[4,\tfrac{13}3)$} &  & & \rotatebox{70}{$[\tfrac{13}3,\varphi_{e_2})$} &  & & \rotatebox{70}{$[\varphi_{e_2},\varphi_a)$} &  & & \rotatebox{70}{$I_1$} &  & & \rotatebox{70}{$I_2\cap[0,\varphi_{e_3})$} &  & & \rotatebox{70}{$I_2\cap[\varphi_{e_3},\infty)$} &  & & \rotatebox{70}{$I_k,\ k\ge3$} & \\\midrule
    $f_{e_1}^+$ & \hspace{\hsp} & \hfill\enspace$6$\enspace\dfill\null & \rule[0.55ex]{\hsp}{0.4pt} & \rule[0.55ex]{\hsp}{0.4pt} & \dfill\null & \hspace{\hsp} & \hspace{\hsp} & $4$ & \hspace{\hsp} & \hspace{\hsp} & $\tfrac{54}{11}$ & \hspace{\hsp} & \hspace{\hsp} & $\tfrac{132}{31}$ & \hspace{\hsp} & \hspace{\hsp} & $\tfrac{11}6$ & \hspace{\hsp} & \hspace{\hsp} & $2-2q_c'$ & \hspace{\hsp}\\
    $f_{e_2}^+$ & \hspace{\hsp} & \hfill\enspace$0$\enspace\dfill\null & \rule[0.55ex]{\hsp}{0.4pt} & \rule[0.55ex]{\hsp}{0.4pt} & \dfill\null & \hspace{\hsp} & \hspace{\hsp} & $2$ & \hspace{\hsp} & \hspace{\hsp} & $\tfrac{12}{11}$ & \hspace{\hsp} & \hspace{\hsp} & $\tfrac{54}{31}$ & \hspace{\hsp} & \hspace{\hsp} & $\tfrac34$ & \hspace{\hsp} & \hspace{\hsp} & $0$ & \hspace{\hsp}\\
    $f_{e_3}^+$ & \hspace{\hsp} & \dfill\null & \rule[0.55ex]{\hsp}{0.4pt} & \rule[0.55ex]{\hsp}{0.4pt} & \dfill\null & \rule[0.55ex]{\hsp}{0.4pt} & \rule[0.55ex]{\hsp}{0.4pt} & \dfill\enspace$0$\enspace\dfill\null & \rule[0.55ex]{\hsp}{0.4pt} & \rule[0.55ex]{\hsp}{0.4pt} & \dfill\null & \rule[0.55ex]{\hsp}{0.4pt} & \rule[0.55ex]{\hsp}{0.4pt} & \dfill\null & \hspace{\hsp} & \hspace{\hsp} & $\tfrac{41}{12}$ & \hspace{\hsp} & \hspace{\hsp} & $4+2q_c'$ & \hspace{\hsp}\\\midrule
    $f_a^+$ & \hspace{\hsp} & \dfill\null & \rule[0.55ex]{\hsp}{0.4pt} & \rule[0.55ex]{\hsp}{0.4pt} & \dfill\enspace$0$\enspace\dfill\null & \rule[0.55ex]{\hsp}{0.4pt} & \rule[0.55ex]{\hsp}{0.4pt} & \dfill\null & \hspace{\hsp} & \hspace{\hsp} & $\tfrac53$ & \hspace{\hsp} & \hspace{\hsp} & \hfill\enspace$\tfrac{13}{11}$\enspace\dfill\null & \rule[0.55ex]{\hsp}{0.4pt} & \rule[0.55ex]{\hsp}{0.4pt} & \dfill\null & \hspace{\hsp} & \hspace{\hsp} & $\dfrac{1+q_c'}{1-q_c'}$ & \hspace{\hsp}\\
    $f_b^+$ & \hspace{\hsp} & $2$ & \hspace{\hsp} & \hspace{\hsp} & \hfill\enspace$3$\enspace\dfill\null & \rule[0.55ex]{\hsp}{0.4pt} & \rule[0.55ex]{\hsp}{0.4pt} & \dfill\null & \hspace{\hsp} & \hspace{\hsp} & $\tfrac43$ & \hspace{\hsp} & \hspace{\hsp} & \hfill\enspace$\tfrac9{11}$\enspace\dfill\null & \rule[0.55ex]{\hsp}{0.4pt} & \rule[0.55ex]{\hsp}{0.4pt} & \dfill\null & \hspace{\hsp} & \hspace{\hsp} & $\dfrac{1-3q_c'}{1-q_c'}$ & \hspace{\hsp}\\
    $f_c^+$ & \hspace{\hsp} & \dfill\null & \rule[0.55ex]{\hsp}{0.4pt} & \rule[0.55ex]{\hsp}{0.4pt} & \dfill\enspace$1$\enspace\dfill\null & \rule[0.55ex]{\hsp}{0.4pt} & \rule[0.55ex]{\hsp}{0.4pt} & \dfill\null & \hspace{\hsp} & \hspace{\hsp} & $\tfrac83$ & \hspace{\hsp} & \hspace{\hsp} & \hfill\enspace$\tfrac{24}{11}$\enspace\dfill\null & \rule[0.55ex]{\hsp}{0.4pt} & \rule[0.55ex]{\hsp}{0.4pt} & \dfill\null & \hspace{\hsp} & \hspace{\hsp} & $\dfrac{2}{1-q_c'}$ & \hspace{\hsp}\\
    \bottomrule\\
    \end{tabular}
    
    \caption{The inflow rates of the equilibrium, $f_e^+(r_u(\varphi))=x_e'(\varphi)/r_u'(\varphi)$ for $e=uw$. The columns refer to deadlines $\varphi$; the corresponding real time $r_u(\varphi)$ depends on the tail $u$. In the last column $q_c'=\tfrac14\,3^{1-k}$.}\label{tab:f}\vspace{4pt}
\end{table}

    The description of the flow can be partitioned in a sequence of phases, where a phase is defined by a constant set $\setActiveEdgesR(\sinkTime) \coloneqq \Set{e = vw \in E | T_e(r_v(\sinkTime)) = r_w(\sinkTime-\eToll)}$ of \emph{active arcs} and a constant set of \emph{resetting arcs} $E^* \coloneqq E^*_r(\sinkTime) \coloneqq \Set{e = vw \in E | \wtST(\sinkTime) > 0}$. The change from one phase to another is caused by so-called events: An arc becoming newly active/inactive or a queue starting to form/depleting.

    For the flow described in \zcref{tab:equilibrium_vals} these phases/events happen at the following sink times:
    \begin{itemize}
        \item $\sinkTime=4$: As $\crem[s][0]=4$, we only consider sink times starting from sink time~$4$. Here, the only active arcs are $e_1$, $b$ and $c$, where the former two immediately become queuing.
        \item $\sinkTime_{e_2}=\tfrac{64}{12}$: $e_2$ becomes active, causing $a$ to become active soon and thus eventually leading to a queue on arc $c$ as flow can then enter $v$ and thus also $w$ at rate $3$.
        \item $\sinkTime_a=\tfrac{67}{12}$: $a$ becomes active. From here on both $v$-$w$ arcs are used.
        \item $\sinkTime_{e_3}= \sinkTime_a+\tfrac{54}{41}$: $e_3$ becomes active. Eventually, this leads to the queue on $e_2$ depleting so that flow reaches $v$ at constant rate $\nu_{e_1}$.
        \item $\bar\sinkTime_{e_2} = \sinkTime_a + 2$: $e_2$ becomes inactive. 
    \end{itemize}

    From this time onwards, the set of active and resetting arcs does not change. The flow itself, however, still changes in time intervals of length~$1$. More precisely, we define the following partition of the time interval $[\sinkTime_a,\infty)$:
        \[I_k\coloneqq [\sinkTime_a+k-1,\sinkTime_a+k) \quad\text{for } k\geq 1.\]
    Then, for $k \geq 3$ all derivatives are constant whitin such an interval but change from one to the other. In particular, the queue on arc~$c$ behaves as follows:
    \begin{equation}\label{eq:nonfinite_example_queue}
        q_c'(\sinkTime)= \tfrac{1}{4} 3^{1-k} \quad \sinkTime\in I_k,\ k\geq 1.
    \end{equation}
    Thus, $q_c'(\sinkTime)=\tfrac13q_c'(\sinkTime-1)$ for $\sinkTime\in I_k$ with $k\geq2$, implying that the queue growth keeps slowing down, but never terminates. All other derivatives during these intervals can be expressed in terms of $q'_c$ as well (see \zcref{tab:equilibrium_vals}).

    \begin{claim}\label{claim:labels fit flow}
        The label and queue derivatives describe a feasible flow $f$.
    \end{claim} 
    \begin{proofClaim}
    Flow conservation can be checked directly from \zcref{tab:f}. Similarly, one can check that $f$ and $x'$ fit together by $f_e^+(r_{u_1}(\varphi))=x_e'(\varphi)/r_{u_1}'(\varphi)$ for $e=u_1u_2$.
        
        For the queues we show that the given values satisfy $$q_e'(\sinkTime) = \begin{cases}
            \frac{x_e'(\sinkTime)}{\nu_e}-r_{u_1}'(\sinkTime) & q_e(\sinkTime)>0\\
            [\frac{x_e'(\sinkTime)}{\nu_e}-r_{u_1}'(\sinkTime)]^+ & q_e(\sinkTime)=0
        \end{cases}$$ for $e=u_1u_2$. Inserting the values from \zcref{tab:equilibrium_vals}, we see that this equality holds for all arcs. All terms are evaluated at the same times, so comparing terms columnwise suffices. We only need to additionally check by integration that when $q_e' < 0$, the queues remain non-negative:
        \begin{structuredproof}
            \proofitem{Queue on $e_1$} The total decrease is $-\int_{\sinkTime_{e_3}}^\infty q_{e_1}'(z) )\diff z \leq \sum_{k\geq 2} \tfrac{1}{4}3^{1-k} = \tfrac18$, whereas at time~$\sinkTime_{e_3}$ we have $q_{e_1}(\sinkTime_{e_3}) \geq \int_{I_1} q_{e_1}'(z) \diff z= \tfrac{4}{9}$.
            \proofitem{Queue on $e_2$} Here, we only need to consider the time interval $[\sinkTime_{e_3},\bar\sinkTime_{e_2})$. Using $\sinkTime_{e_3}-\sinkTime_a=\tfrac{54}{41}$, $q_{e_2}(\sinkTime_{e_3})=\tfrac7{41}$. With rate $-\tfrac14$ it reaches $0$ exactly at $\sinkTime_a+2=\bar\sinkTime_{e_2}$, where $e_2$ stops carrying flow.
            \proofitem{Queue on $b$} This queue depletes from $\sinkTime_a+1$ on. At this time, we have $q_b(\sinkTime_a+1)>\tfrac14$, since the rates before $\sinkTime_a$ are positive and $I_1$ alone contributes $\tfrac{1}{4}$. The total decrease is $\tfrac16+2\sum_{k\ge3}\tfrac14 3^{1-k}=\tfrac14$.
        \end{structuredproof} 
        
        For the labels, we check that for every arc $e=u_1u_2$ the slack
        $\sigma_e(\varphi):=r_{u_1}(\varphi)+\tau_e+q_e(\varphi)-r_{u_2}(\varphi-p_e)
        $ is nonnegative, and some arc leaving each node has $\sigma_e=0$. We compute $\sigma$ at $\varphi=4$ from the empty network, where $r_s=\varphi-4$, $r_v=\varphi-3$, $r_w=\varphi-1$, as all queues are $0$, and its derivative $\sigma_e'=r_{u_1}'+q_e'-r_{u_2}'(\varphi-p_e)$ from the table and the queues above. For a queued arc, $\sigma_e'=x_e'/\nu_e-r_{u_2}'(\varphi-p_e)$.
        \begin{itemize}
        \item For $e\in \{e_1$, $b$, $c$\}, we have $\sigma_e(4)=0$. Moreover $\sigma_e'=0$ throughout: for $c$ before $\varphi_a$ since $r_w'=1$ and $q_c'=0$ and afterwards since $x_c'/\nu_c=r_t'=1$. For $e_1$ and $c$ this compares entries in the
        same column. For $b$, $x_b'(\varphi)=r_w'(\varphi-1)$ holds because $\varphi\mapsto\varphi-1$ maps $I_k$ onto $I_{k-1}$ for $k\ge2$, and $r_w'=1$ before $\varphi_a$. Hence $\sigma_{e_1}=\sigma_b=\sigma_c=0$.
        \item On arc $a$, we have $\sigma_a(4)=1$ which decreases to 0 by time $\varphi_a$: during that time window, $\sigma_a'=r_v'-r_w'<0$ and $\sigma_a(\varphi_a)=\tfrac{19}{12}+3-\tfrac{55}{12}=0$. Afterwards, $a$ stays active as $\sigma_a'=r_v'-r_w'=0$.
        \item On arc $e_3$, $\sigma_{e_3}(4)=\tfrac94$, decreasing to 0 until time $\varphi_{e_3}$: during that time window, $\sigma_{e_3}'=r_s'-1<0$, and $\sigma_{e_3}$ vanishes
        at $\varphi_{e_3}$ (integrate over $r_s'$). Afterwards $r_s'=1$, so $\sigma_{e_3}'=0$.
        \item On arc $e_2$, $\sigma_{e_2}(4)=1$ which decreases to 0 by time $\varphi_{e_2}$: $\sigma_{e_2}'=r_s'-r_v'(\varphi-1)<0$, and $\sigma_{e_2}(\varphi_{e_2})=\tfrac16+1-\tfrac76=0$. On $(\varphi_{e_2},\bar\varphi_{e_2})$ the arc is queued and
        $\sigma_{e_2}'=x_{e_2}'-r_v'(\varphi-1)=0$, using that $\varphi\mapsto\varphi-1$ maps $[\varphi_{e_2},\varphi_a+1)$ into
        $[\tfrac{13}3,\varphi_a)$, where $r_v'=\tfrac13$, and maps $I_2$ onto $I_1$. After $\bar\varphi_{e_2}$ the queue is empty
        and $\sigma_{e_2}'=1-r_v'(\varphi-1)>0$.
        \end{itemize}
        Thus, all slacks are nonnegative and for each node there exists a leaving arc with slack~$0$: $e_1$ at $s$, $b$ at $v$, and $c$ at $w$. Hence, the labels fit and the active and resetting arcs are as in \zcref{fig:NonFiniteConvergence}.
    \end{proofClaim}

    \begin{claim}
        The flow described above is a dynamic equilibrium.
    \end{claim}
    \begin{proofClaim}
        We check that every flow carrying arc is active. As the activity of arcs for the time intervals given in \zcref{fig:NonFiniteConvergence} was verified in \zcref{claim:labels fit flow}, one can check with \zcref{tab:equilibrium_vals} that flow only travels on active arcs. Applying \zcref{cor:TolledEqChara} yields the claim.
    \end{proofClaim}
    Thus, the example shows a dynamic equilibrium for which queues never stabilize.
\end{proof}

The example shown above also exhibits the following behavior which differs from that of untolled dynamic equilibria.
\begin{corollary}
    There exists an instance with $u_0 < \mincut$ for which the label derivatives and derivatives of the queuing times
    in a tolled dynamic equilibrium are non constant within a fixed configuration of active and resetting arcs.
\end{corollary}

\section{The Potential}
We introduce the following shorthand notation for any arc $e=vw$: $\gamma_e(\sinkTime)\coloneqq \frac{r_v'(\sinkTime)}{r_w'(\sinkTime-\eToll)}$.
\subsection{Proof of \zcref{lem:dual_feasible}}

\lemdualfeasible*

    \begin{proof}
        By definition $\tilde q(\sinkTime)$ is non-negative.
        We have $d_t(\sinkTime) = \sinkTime - r_t(\sinkTime) = 0$ and 
        
        \begin{align*}
            \eTTime + \eToll + \qt_e(\sinkTime) 
                &\geq \eTTime + \eToll + r_w(\sinkTime) - r_v(\sinkTime) - \eTTime - \eToll  \\
                &= (\sinkTime - r_v(\sinkTime)) - (\sinkTime - r_w(\sinkTime)) 
                    = d_v(\sinkTime) - d_w(\sinkTime).
        \end{align*}
        As in steady state $\bar q(\sinkTime) = \tilde q(\sinkTime)$, we have a feasible solution to~\eqref{D}.
    
       The primal problem~\eqref{P} is feasible as well since we can choose $y$ to be any (static) $s$,$t$-flow of value~$u_0$ (which exists by our assumption that $u_0$ is bounded by the minimum cut in the underlying static network). By weak duality, \eqref{D} is bounded.
    \end{proof}

\subsection{Proof of \zcref{thm:PotNondecreasing}}

\thmPotentialNondecreasing*

\begin{proof}
    Since the two components of $\fullpot = \potBasic - \potCorr$ are both absolutely continuous, it suffices to only consider times~$\sinkTime \geq \crem[s][0]$ where their derivatives exist and show that those are non-negative. 
    For~$\potBasic$, we start by computing the derivative of the contribution of an individual arc:

    \begin{claim}\label{claim:PotBasicLowerBoundperEdge}
        For any arc~$e =vw \in E$ we have $\bigl(\eCap\cdot \wtST(\sinkTime)\bigr)' = x'_e(\sinkTime) \cdot(1-\gamma_e(\sinkTime))$.
    \end{claim}

    \begin{proofClaim}
    Here, we distinguish two cases:
    \begin{proofbycases}
        \caseitem{$\wtST(\sinkTime)=0$} 
            In this case, \zcref{cor:WaitingTimeDerivative} implies $\wtST'(\sinkTime)=0$. Hence, it suffices to show that the set 
                \[N \coloneqq \Set{\sinkTime \in \IR | \wtST'(\sinkTime)=0, x'_e(\sinkTime) > 0, \gamma_e(\sinkTime) \neq 1}\]
            is a null set. As the given flow is a dynamic equilibrium, arc~$e$ is active for almost all $\sinkTime \in N$ (by \zcref{cor:TolledEqChara}). Hence, it suffices to only consider the set
                \[N' \coloneqq \Set{\sinkTime \in N | e \in \setActiveEdgesR(\sinkTime)}.\]
            Then, for any $\sinkTime \in N'$ we have $r_w(\sinkTime-\eToll) = T_e(r_v(\sinkTime)) = r_v(\sinkTime) + \wtST(\sinkTime) + \eTTime$ as well as $r'_w(\sinkTime-\eToll) \neq r'_v(\sinkTime) + \wtST'(\sinkTime)$. Thus, $e$ becomes inactive right after~$\sinkTime$ and, due to the continuity of all involved functions, stays inactive for some proper interval $(\sinkTime,\sinkTime')$. In particular, $(\sinkTime,\sinkTime') \cap N' = \emptyset$. Thus, $N'$ may only contain countably many points and, therefore, is a null set.
            \caseitem{$\wtST(\sinkTime)>0$} In this case we have $e \in E^*_r(\sinkTime)$ by \zcreftp{lem:CharaActiveAndResettingEdges:CharaResetting} and, therefore, 
                \begin{align*}
                    \bigl(\eCap\cdot \wtST(\sinkTime)\bigr)'
                        &\zoverset{cor:WaitingTimeDerivative}{=} \eCap\cdot (r'_w(\sinkTime-\eToll)-r'_v(\sinkTime))  \\
                        &= \eCap r'_w(\sinkTime-\eToll)\cdot\left(1-\frac{r'_v(\sinkTime)}{r'_w(\sinkTime-\eToll)}\right)  \\
                        &\zoverset{cor:bounds on x}{=} x'_e(\sinkTime)\cdot(1-\gamma_e(\sinkTime)).  \qedhere
                \end{align*}
        \end{proofbycases} 
    \end{proofClaim}
   
    With this claim at hand, we can now directly compute the derivative of~$\potBasic$ as follows:
        \begin{align*}
            \potBasic'(\sinkTime)  
                &= u_0\cdot\bigl(\sinkTime-r_s(\sinkTime)\bigr)' - \sum_{e=vw \in E}\bigl(\nu_e\cdot \wtST(\sinkTime)\bigr)' \\
                &\zoverset{claim:PotBasicLowerBoundperEdge}{=} u_0\cdot(1-r'_s(\sinkTime)) - \sum_{e=vw\in E} x'_e(\sinkTime) \cdot(1-\gamma_e(\sinkTime)) \\
                &= u_0\cdot(1-r'_s(\sinkTime)) + \sum_{e=vw\in E}x'_e(\sinkTime) (\gamma_e(\sinkTime)-1) 
        \end{align*}

    Next, we compute the derivative of $B$:
    
    \begin{claim}
        For almost every $\sinkTime \geq \crem[s][0]$ we have:\label{claim:B'bounded}
           \[    \potCorr'(\sinkTime) 
            = -u_0 r'_s(\sinkTime)\log r'_s(\sinkTime) + \sum_{e=vw\in E}x'_e(\sinkTime) \log \gamma_e(\sinkTime).\] 
    \end{claim}

    \begin{proofClaim}
    The proof is a direct computation:
        \begin{align*}
            \potCorr'(\sinkTime) 
                &=  \Bigl(\sum_{e=vw\in E}\int_\sinkTime^{\sinkTime+\eToll}x'_e(z)\log r_w'(z-\eToll)\diff z\Bigr)'\\
                &= \sum_{e=vw \in E} \bigl(x'_e(\sinkTime+\eToll)\log r'_w(\sinkTime) - x'_e(\sinkTime)\log r'_w(\sinkTime-\eToll)\bigr)\\
                &= \sum_{w \in V} \Bigl(\log r'_w(\sinkTime) \sum_{e\in \delta^-(w)} x'_e(\sinkTime+\eToll)\Bigr)-  \sum_{e=vw \in E}\bigl(x'_e(\sinkTime)\log r'_w(\sinkTime-\eToll)\bigr)\\
                &\symoverset{1}{=} \sum_{w\in V\setminus\set{s,t}} \Bigl(\log r'_w(\sinkTime) \sum_{e\in \delta^+(w)} x'_e(\sinkTime)\Bigr) + \log r'_s(\sinkTime) \Bigl(-u_0 r'_s(\sinkTime)+\sum_{e\in \delta^+(s)} x'_e(\sinkTime)\Bigr) \\
                    &\quad\quad\quad\quad + \log r'_t(\sinkTime)\sum_{e\in \delta^+(t)} x'_e(\sinkTime) -  \sum_{e=vw\in E}\bigl(x'_e(\sinkTime)\log r'_w(\sinkTime-\eToll)\bigr)\\
                &= -u_0 r'_s(\sinkTime)\log r'_s(\sinkTime) + \sum_{e=vw\in E}x'_e(\sinkTime) \log r'_v(\sinkTime)-  \sum_{e=vw}\bigl(x'_e(\sinkTime)\log r'_w(\sinkTime-\eToll)\bigr)\\
                &= -u_0 r'_s(\sinkTime)\log r'_s(\sinkTime) + \sum_{e=vw\in E}x'_e(\sinkTime) \bigl(\log r'_v(\sinkTime) - \log r'_w(\sinkTime-\eToll)\bigr) \\
                &= -u_0 r'_s(\sinkTime)\log r'_s(\sinkTime) + \sum_{e=vw\in E}x'_e(\sinkTime) \log \gamma_e(\sinkTime).
        \end{align*}
       The fourth equality~\refsym{1} follows from flow conservation of $x'$ (\zcref{cor:Fe+andFe-}) for all nodes except~$t$ and from the fact that $\log(r'_t(\sinkTime))=\log(1)=0$ for the sink~$t$.       
    \end{proofClaim}
   By combining these two derivatives we obtain the derivative of$~\fullpot$:
        \begin{align}\label{eq:FullPotentialLowerBound}
            \fullpot' (\sinkTime) = u_0\cdot(1-r'_s(\sinkTime)+r'_s(\sinkTime)\log r'_s(\sinkTime))+\sum_{\mathclap{e=vw\in E}}x'_e(\sinkTime) \left(\gamma_e(\sinkTime)-1- \log \gamma_e(\sinkTime)\right).
        \end{align}
   In order to lower bound this, we consider the function 
        \[g:\IR\to \IR, z \mapsto  z- 1 -\log(z).\] 
   Using the identity $1-z+z\log z = z(\tfrac 1z - 1 -\log \tfrac 1z) = zg(\frac{1}{z})$, we can rewrite \eqref{eq:FullPotentialLowerBound} as
        \[\fullpot' (\sinkTime)= u_0\cdot r'_s(\sinkTime)g(1/r'_s(\sinkTime))+\sum_{e=vw\in E}x'_e(\sinkTime) g(\gamma_e(\sinkTime)).\]
   As $r'_s \geq 0$ (\zcref{lem:r Lipschitz}) and $x_e'\geq 0$ for all $e\in E$, we only need to show that $g(z)\geq 0$ for $z >0$. This follows as $g$ is convex and attains its minimum value~$0$ at $z=1$. Thus, $\fullpot'\geq 0$, concluding the proof. 
\end{proof}

\subsection{The potential $\fullpot$ is upper bounded}

\thmpotbounded*

\begin{proof}

From primal feasibility and \zcref{lem:dual_feasible} we obtain boundedness of the function
    \[\potVar: \IR \to \IR, \sinkTime \mapsto u_0\cdot(\sinkTime-r_s(\sinkTime)) - \sum_{e=vw \in E}\nu_e[r_w(\sinkTime)-r_v(\sinkTime)-\eTTime-\eToll]^+\]
Together with the fact that $-B$ is bounded above as well (as we will show in \zcref{lm:B_bound}) this implies that $\fullpotVar \coloneqq \potVar - \potCorr$ is bounded above.

However, setting $\qt$ as in \zcref{lem:dual_feasible} and thus in $\potVar$ yields $\qt_e(\sinkTime) = [r_w(\sinkTime)- r_v(\sinkTime)-\eTTime-\eToll]^+$, whereas $\potBasic$ evaluates the queue as $q_e(\sinkTime) = [ r_w(\sinkTime-\eToll)- r_v(\sinkTime)-\eTTime]^+$. 
The difference between the two is that the contribution of the toll $\eToll$ shifts from the arc itself (in $\potVar$) to the deadline, i.e., into the argument of $r_w$ in $\potBasic$. This time shift only affects the terms associated with the queues. We can thus split the potential $\fullpot= \potBasic-\potCorr$ into three terms, the first two of which just amount to $\fullpotVar$:
    \begin{equation}\label{eq:3_part_potential}
        \fullpot (\sinkTime) \coloneqq \underset{\text{objective $\dualObj$ of \lpref{D}}}{\underbrace{u_0d_s- \sum_{e}{\nu_e}\cdot \qt_e }}-B+ \underset{\text{$\fullpot-\fullpotVar$}}{\underbrace{\sum_e \nu_e (\qt_e -q_e(\sinkTime))}}
    \end{equation}

We will now show that $\fullpot$ cannot be too much larger than $\fullpotVar$. Since $\fullpot$ is non-decreasing, this immediately implies that $\fullpot$ is bounded as well.

\begin{restatable}{lemma}{restateDifferenceBetweenPots}\label{lem:DifferenceBetweenPots}
    For any $\sinkTime\in \IR$ we have
    \MainbodyOrAppendix{$\potBasic(\sinkTime) - \potVar(\sinkTime) \leq C\sum_{e\in E} \nu_e \eToll$.}%
    {\[ \fullpot(\sinkTime) - \fullpotVar(\sinkTime) = \potBasic(\sinkTime) - \potVar(\sinkTime) \leq C\sum_{e\in E} \nu_e \eToll . \]}
\end{restatable}

\begin{proof}
    For any $\sinkTime \in \IR$ we have
        \begin{align*}
            &\fullpot(\sinkTime) - \fullpotVar(\sinkTime)
                = \potBasic(\sinkTime) - \potVar(\sinkTime) 
                    = -\sum_{e = vw \in E}\bigl(\eCap q_e(\sinkTime) - \nu_e [r_w(\sinkTime)-r_v(\sinkTime)-\eTTime-\eToll]^+\bigr) \\
                &\zoverset{lem:CharaActiveAndResettingEdges}{=} -\sum_{e = vw \in E}\nu_e\bigl([r_w(\sinkTime-\eToll)-r_v(\sinkTime)-\eTTime]^+ - [r_w(\sinkTime)-r_v(\sinkTime)-\eTTime-\eToll]^+\bigr) \\
                &\leq  -\sum_{e=vw \in E}\nu_e\bigl(r_w(\sinkTime-\eToll)-r_w(\sinkTime)\bigr) 
                    = \sum_{e=vw \in E}\eCap\bigl(r_w(\sinkTime)-r_w(\sinkTime-\eToll)\bigr)\\
                &\zoverset{cor:r Lipschitz}{\leq} C\sum_{e\in E}\nu_e \eToll. 
        \end{align*}
        The first inequality follows as $[b]^+-[a]^+\leq [b-a]^+$.
\end{proof}

\begin{lemma}\label{lm:B_bound}
    $B$ is lower bounded.
\end{lemma}

\begin{proof}
    Recall that \[\potCorr: \IR \to \IR, \sinkTime \mapsto \sum_{vw\in E}\int_\sinkTime^{\sinkTime+\eToll[vw]}x'_{vw}(z)\log r_w'(z-\eToll[vw])\diff z.\]
    We establish a lower bound on the integrand by distinguishing two cases:
    \begin{proofbycases}
        \caseitem{$r_w'(z-\eToll[vw]) \geq 1$} In this case, as $x'_{vw}\geq 0$, the integrand is nonnegative.
        \caseitem{$r_w'(z-\eToll[vw]) < 1$} Here, we use $x'_{vw}(z)\leq \eCap[vw] r_w'(z-\eToll[vw])$ from \zcref{cor:bounds on x}. Together with $\log r_w'(z-\eToll[vw]) < 0$, this implies 
            \[x'_{vw}(z)\log  r_w'(z-\eToll[vw]) \geq \eCap[vw] r_w'(z-\eToll[vw]) \log r_w'(z-\eToll[vw])\geq  -\frac{\eCap[vw]}{\e},\]
        where the last inequality follows as $\log (\e z)^{-1}\leq (\e z)^{-1}-1$ implies $-1-\log z \leq (\e z)^{-1}-1$ and so $z\log z \geq -\tfrac{1}{\e}$ for any $z>0$.
    \end{proofbycases}
    Thus, $B \geq -\sum_{vw \in E}\int_{\sinkTime}^{\sinkTime+\eToll[vw]}\frac{\eCap[vw]}{\e}=-\tfrac{1}{\e}\sum_{vw \in E}\eCap[vw]\eToll[vw].$
\end{proof}

    Boundedness of the dual objective function together with \zcref{lm:B_bound} imply that $\fullpotVar$ is bounded. 
\end{proof}

\subsection{Connections to the Potential Function in \cite{LongTermBehaviorOfDE}}\label{app:CCONonMonotone}

In this section we show how our potential~$\fullpot$ is related to the one used by Cominetti, Correa, and Olver in~\cite{LongTermBehaviorOfDE}. We first recall the definition of their potential function:

For any feasible flow $f$, any node $v \in V$ and any time $\sourceTime \in \IR$ we denote by
    \[\ell_v(\sourceTime) \coloneqq \min\set{T_P(\sourceTime) | P \in \PathSet[s,v]}\]
the earliest possible arrival time at node~$v$ when entering the network at time~$\sourceTime$. With this, Cominetti, Correa, and Olver \cite{LongTermBehaviorOfDE} define the following potential function for dynamic equilibria (without tolls)
    \[\potCCO: \IR \to \IR, \sourceTime \mapsto u_0\cdot(\ell_t(\sourceTime)-\sourceTime) - \sum_{e = vw \in E}\nu_e\cdot\eWTime(\ell_v(\sourceTime))\]
and show that it is monotone and bounded (\cite[Theorems~2 and~3]{LongTermBehaviorOfDE}).
The following example shows that this potential is not monotone anymore for tolled dynamic equilibria:

\begin{example}
    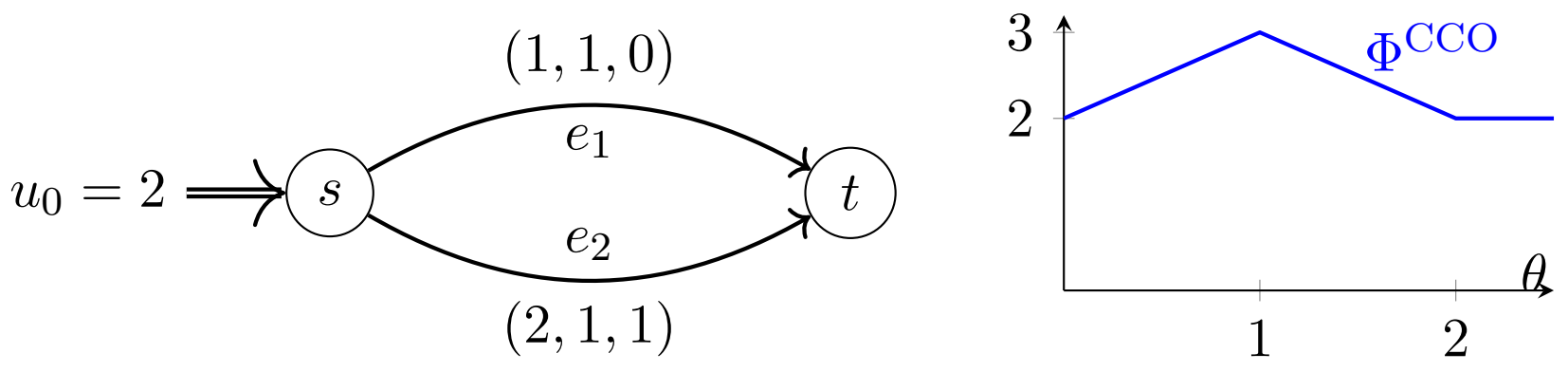
\begin{figure}\centering
        \begin{tikzpicture}[node distance=3cm]
            \node[vertex] (s) {$s$};
                \draw[edge,<-,double] (s) -- ++(-1,0) node[left]{$u_0=2$};
           
            \node[vertex,right=of s] (t) {$t$};
    
            \draw[edge] (s) to[bend left=30]  node[below]{$e_1$}node[above]{$(1,1,0)$} (t);
            \draw[edge] (s) to[bend right=30] node[above]{$e_2$}node[below]{$(2,1,1)$} (t);
          
        \end{tikzpicture}
        \hspace{1em}
        \begin{tikzpicture}[declare function = {
                potCCO(\x) = 2+and(\x>=0,\x<1)*\x + and(\x>=1,\x<2)*(2-\x);
    		}]
    		\begin{axis}[
    			axis lines=left,
    			ymin=0,ymax=3.2,
    			xmin=0,xmax=2.5,
    			ytick={2,3},
    			xtick={1,2},
                height=3.5cm,
                width=5cm]
    			\addplot[blue,domain=0:2.5,thick,samples=2500]{potCCO(\x)} node[pos=.8,above=20]{$\potCCO$};
                \node at (axis cs:2.4,.2){$\sourceTime$};
    		\end{axis}
        \end{tikzpicture}
        \caption{Example of non-monotonicity of the dual objective potential function from \cite{LongTermBehaviorOfDE}. Arc data are in the format $(\eTTime, \eCap, \eToll)$.}\label{fig:CCO_not_monotone}
    \end{figure}

    Consider the network given in \zcref{fig:CCO_not_monotone} (left) consisting of two parallel arcs with capacity~$1$. One of the arcs~($e_1$) has free-flow transit time of~$1$ and a toll of~$0$, while the other has free-flow transit time of~$2$ and a toll of~$1$. The network inflow rate is~$2$. The following is then a tolled dynamic equilibrium in this network: Until time $\theta=2$, all flow travels via arc~$e_1$, building up a queue with $\qmNT[e_1](2)=2$. From this point onwards, there is an equal flow split between the two arcs and the steady state is reached. 
    
    Since $e_2$ never grows a queue, it never contributes to the potential. The travel time to reach the sink is $1+\min\set{\sourceTime,2}$ for $e_1$ and $2$ for $e_2$, so the minimum travel time is $\min\set{1+\sourceTime,2}$. The queue volume on $e_1$ contributes $\min\set{\sourceTime,2}$. Thus, the potential evaluates as $\potCCO(\sourceTime) = u_0(\ell(\sourceTime)-\sourceTime)-\qmNT[e_1](\sourceTime)=2\min\set{1+\sourceTime,2}-\min\set{\sourceTime,2}$ which increases linearly with slope 1 on $[0,1]$ and then decreases with linear slope $-1$ on $[1,2]$ -- see \zcref{fig:CCO_not_monotone} (right).
\end{example}

Next, we show that for networks without tolls our potential coincides with the potential used in~\cite{LongTermBehaviorOfDE} after exchanging the reference point for all time measurements between source and sink. We start by establishing a natural connection between source- and sink-based labels:

\begin{lemma}\label{lem:rvandlv}
    Given an instance without tolls, i.e. $\eToll=0$ for all $e \in E$ and 
    a dynamic equilibrium. Then for any node~$v \in V$ we have
        \begin{align}
            (r_v\circ\ell_t)(\sourceTime) = \ell_v(\sourceTime)
        \end{align}
    for all times $\sourceTime$ where $v$ lies on an active \stpath{} wrt.\ source times and
        \begin{align*}
            (\ell_v\circ r_s)(\sinkTime) = r_v(\sinkTime)
        \end{align*}
    for all times $\sinkTime$ where $v$ lies on an active \stpath{} wrt.\ sink times.
\end{lemma}

\begin{proof}
    Let $\sourceTime$ be a time where $v$ lies on an active \stpath~$P$ with respect to source times. Then, both the \stsubpath[s][v]~$P_{\leq v}$ and the \stsubpath[v]~$P_{\geq v}$ are active (at times $\sourceTime$ and $\ell_v(\sourceTime)$, respectively). That is, we have
        \[\ell_t(\sourceTime) = T_P(\sourceTime), \quad \ell_v(\sourceTime) = T_{P_{\leq v}}(\sourceTime) \quad\text{and}\quad \tilde c_v(\ell_v(\sourceTime)) = T_{P_{\geq v}}(\ell_v(\sourceTime)).\]
    Hence, we have $\ell_t(\sourceTime) = \tilde c_v(\ell_v(\sourceTime))$. Together with \zcref{cor:cvInverseOfrv} this now implies
        \[r_v(\ell_t(\sourceTime)) = r_v(\tilde c_v(\ell_v(\sourceTime))) = \ell_v(\sourceTime).\]

    For the second part let $P$ be an \stpath{} containing~$v$ that is active at sink-time~$\sinkTime$, i.e., we have $T_e(r_v(\sinkTime)) = r_w(\sinkTime)$ for all arcs $e=vw \in P$. By iterative application this then immediately implies
        \[\ell_v(r_s(\sinkTime)) \leq T_{P_{\leq v}}(r_s(\sinkTime)) = r_v(\sinkTime).\]
    Now let $\sourceTime \in \IR$ be some time with $\ell_v(\sourceTime) = r_v(\sinkTime)$ (which exists by continuity of~$\ell_v$) and $P'$ an \stpath[s][v] with $T_{P'}(\sourceTime) = r_v(\sinkTime)$. Then, we have $T_{P_{\geq v}}(T_{P'}(\sourceTime)) = T_{P_{\geq v}}(r_v(\sinkTime)) = \sinkTime$. This implies $r_s(\sinkTime) \geq \sourceTime$ and, by monotonicity of~$\ell_v$, that $\ell_v(r_s(\sinkTime)) \geq \ell_v(\sourceTime) = r_v(\sinkTime)$.
\end{proof}

\begin{corollary}\label{lem:ConnectionToCCOPotential}
    For any dynamic equilibrium in a network without tolls we have
        \[\potCCO = \fullpot \circ \ell_t \quad\text{and}\quad \fullpot = \potCCO \circ r_s.\]
\end{corollary}

\begin{proof}
    We start by observing that in a network without tolls, the correction term~$\potCorr$ in our potential is always zero. Hence, we have $\fullpot = \potBasic$ and it suffices to show the claimed equalities for~$\potBasic$ instead of~$\fullpot$.
    
    For the first part, we have
        \begin{align*}
            (\potBasic\circ\ell_t)(\sourceTime)
                &= u_0\cdot(\ell_t(\sourceTime)-r_s(\ell_t(\sourceTime))) - \sum_{e=vw \in E} z_e(r_v(\ell_t(\sourceTime))) \\
                &\symoverset{1}{=} u_0\cdot(\ell_t(\sourceTime)-\ell_s(\sourceTime))) - \sum_{e=vw \in E} z_e(\ell_v(\sourceTime)) \\
                &= \potCCO(\sourceTime),
        \end{align*}
    where \refsym{1} holds due to \zcref{lem:rvandlv} and the fact that the source~$s$ as well as any non-zero queue always lies on an active path with respect to source-times (see \cite[Theorem~2]{CCLDynEquil}).

    For the second part, we have 
        \begin{align*}
            (\potCCO \circ r_s)(\sinkTime)
                &= u_0\cdot(\ell_t(r_s(\sinkTime))-r_s(\sinkTime)) - \sum_{e = vw \in E}z_e(\ell_v(r_s(\sinkTime))) \\
                &\symoverset{1}{=} u_0\cdot(r_t(\sinkTime)-r_s(\sinkTime)) - \sum_{e = vw \in E}z_e(r_v(\sinkTime)) \\
                &= \potBasic(\sinkTime),
        \end{align*}
    where, again, \refsym{1} holds due to \zcref{lem:rvandlv} and the fact that the sink~$t$ as well as any non-zero queue always lies on an active path with respect to sink-times. The latter, this time, follows since arcs with queue on them, always must have non-zero inflow (\cite[Lemma~2.4]{DEContUniqLT}) and this is only possible when this arc is active with respect sink times (\zcref{lem:TolledEqChara}).
\end{proof}

Due to the strict monotonicity of the node labels, this implies, in particular, that in a network without tolls, monotonicity of $\fullpot$ is equivalent to that of~$\potCCO$.

\section{Bounded Queues and Convergence to $Q^*$}
\subsection{Proof of \zcref{lm:freezing}}

\lemfreezing*

\begin{proof}
    Using \zcref{cor:WaitingTimeDerivative} we can express $\wtST'$ in terms of $r'_v$ and $r'_w$ and then use the upper bound~$C$ on~$r'_v$ from \zcref{cor:r Lipschitz} to get:
        \begin{align}
            \int_{\sinkTime}^{\sinkTime+h}\abs{\wtST'(z)}\diff z
                &\leq \int_{\sinkTime}^{\sinkTime+h}\mathbf{1}_{\set{\wtST>0}}(z)\abs{r'_w(z-\eToll)-r'_v(z)}\diff z \\
                    &= \int_{\sinkTime}^{\sinkTime+h}\mathbf{1}_{\set{\wtST>0}}(z)\abs{r'_w(z-\eToll)}\cdot\abs{1-\gamma_e(z)}\diff z \notag \\
                &\leq C\cdot\int_{\sinkTime}^{\sinkTime+h}\mathbf{1}_{\set{\wtST>0}}(z)\abs{1-\gamma_e(z)}\diff z. \label{eq:BoundingVariationOfQ}
        \end{align}
    
    We now show the desired statement via the following chain of implications of convergence for $\sinkTime\to \infty$:
    \begin{align}\label{eq:chain_implications_q_conv}
        \begin{split}
        \int_{\sinkTime}^{\sinkTime+h}\fullpot_e'(z)\diff z\to 0 &\implies\int_{\sinkTime}^{\sinkTime+h}\mathbf{1}_{\set{\wtST>0}}(z) g(\gamma_e(z))\diff z\to 0 \\
        &\implies \int_{\sinkTime}^{\sinkTime+h} \mathbf{1}_{\set{\wtST>0}}(z) \abs{1-\gamma_e(z)}\diff z \to 0.
        \end{split}
    \end{align}

    The first convergence statement holds as the tails of a finite integral (\zcref{cor:dissipation_finite}) over a nonnegative function (\zcref{not:potential_deriv_e}) tend to 0. 
    The first implication is obtained by exploiting the description
    $\Psi_e'(\sinkTime) = x_e'(\sinkTime) g(\gamma_e(\sinkTime)) $ by \zcref{not:potential_deriv_e}
    and lower bounding $x_e'$. For a point in time $z$ with $q_e(z)>0$ we have $x_e'(z)=\eCap r_w'(z -\eToll)$ by \zcref{cor:bounds on x}. With \zcref{lem:r Lipschitz}, we thus obtain $x_e'(z) \geq \eCap c$.
    
    We now assume for a contradiction that the second implication does not hold, i.e., for $\sinkTime\to \infty$
        \[\int_{\sinkTime}^{\sinkTime+h} \mathbf{1}_{\set{\wtST>0}}(z) \abs{1-\gamma_e(z)}\diff z \not\to 0.\]
    Then, there exists some $\eps > 0$ such that there is a sequence $\sinkTime_n \to \infty$ with 
        \[\int_{\sinkTime_n}^{\sinkTime_n+h} \mathbf{1}_{\set{\wtST>0}}(z) \abs{1-\gamma_e(z)}\diff z \geq \eps \text{ for all } n \in \INo.\]
    By the upper and lower bounds on $r'$ (\zcref{cor:r Lipschitz}), $\gamma_e(\sinkTime)=\frac{r_v'(\sinkTime)}{r_w'(\sinkTime-\eToll)}$ is bounded and, thus, we have $\abs{1-\gamma_e(z)} \leq M$ for some constant $M$. Now, define for any $\eps' > 0$ the set of times for which $|1-\gamma(\sinkTime)|$ is larger than~$\eps'$ and $q_e(\sinkTime)>0$ as $J_{\eps'} \coloneqq \set{\sinkTime' \in [\sinkTime_n,\sinkTime_n+h] | \abs{1-\gamma_e(\sinkTime')} \geq \eps' \text{ and }q_e(\sinkTime')>0}$. As $g$ is continuous, strictly convex and $|1-z|$ and $g(z)$ attain the value~$0$ only at $z=1$, there exists for every $\eps' > 0$ some $\delta(\eps') > 0$ such that  
        \[\abs{1-\gamma_e(\sinkTime)} \geq \eps' \implies g(\gamma_e(\sinkTime)) = \gamma_e(\sinkTime)-1-\log\gamma_e(\sinkTime) \geq \delta(\eps').\]
    Now, 
        \begin{align*}
            \eps    &\leq \int_{\sinkTime_n}^{\sinkTime_n+h}\mathbf{1}_{\set{\wtST>0}}(z) \abs{1-\gamma_e(z)}\diff z \\
                        &
                    = \int_{J_{\eps'}} \abs{1-\gamma_e(z)}\diff z + \int_{[\sinkTime_n,\sinkTime_n+h]\setminus J_{\eps'}} \mathbf{1}_{\set{\wtST>0}}(z) \abs{1-\gamma_e(z)}\diff z \\
                        &\leq \abs{J_{\eps'}}M + h\eps'
        \end{align*} 
    implies
        \[\abs{J_{\eps'}} \geq \frac{\eps -h\eps'}{M}.\]
    Therefore, 
        \begin{align*}
            \int_{\sinkTime_n}^{\sinkTime_n+h} \mathbf{1}_{\set{\wtST>0}}(z)g(\gamma_e(z))\diff z 
                &= \int_{J_{\eps'}} g(\gamma_e(z))\diff z + \int_{[\sinkTime,\sinkTime+h]\setminus J_{\eps'}} \mathbf{1}_{\set{\wtST>0}}(z) g(\gamma_e(z))\diff z  \\
                &\geq \abs{J_{\eps'}}\delta(\eps')\\
                    &\geq \frac{\eps -h\eps'}{M}\delta(\eps').
        \end{align*}
    Thus, choosing $\eps' \in (0,\eps/h)$ and fixing $m \coloneqq \frac{\eps -h\eps'}{M}\delta(\eps') >0$ gives us 
        \[\int_{\sinkTime_n}^{\sinkTime_n+h}\mathbf{1}_{\set{\wtST>0}}(z) g(\gamma_e(z))\diff z \geq m\]
    for all $n \in \INo$, contradicting 
        \[\int_{\sinkTime_n}^{\sinkTime_n+h}\mathbf{1}_{\set{\wtST>0}}(z) g(\gamma_e(z))\diff z\xrightarrow{\sinkTime \to \infty} 0.\]
    Thus, the second implication in \eqref{eq:chain_implications_q_conv} holds as well. Hence, \eqref{eq:BoundingVariationOfQ} together with \eqref{eq:chain_implications_q_conv} yields the \zcref[noref,nocap]{lm:freezing}.
\end{proof}

\subsection{Proof of \zcref{thm:Q*steadystates} and \zcref{thm:convergence_queues}}

In this section we first present the proof of \zcref{thm:Q*steadystates} and then prove \zcref{thm:convergence_queues}. For this sake some of the lemmas in the first proof are stated more general than needed, so that we can reuse them for the latter proof. 

\thmQsteadystates*

\begin{proof}
    First observe that for a dynamic tolled equilibrium in steady state for sufficiently late $\sinkTime$, we have $r_w(\sinkTime - p_e)=r_w(\sinkTime) -p_e$. This is where we require the equilibrium to be in steady state at time $\sinkTime-\max_{e\in E} p_e - \sum_{e\in E} p_e$. The shift by $\max_{e\in E}p_e$ is due to the time shift in the equation, the one by $\sum_{e\in E} p_e$ is needed as we pull back $r_t'=1$ via active simple paths with $r'_v(\sinkTime)=r_w'(\sinkTime-\eToll)$. Thus, 
    we have $q_e(\sinkTime)= [r_w(\sinkTime - p_e)-r_v(\sinkTime)-\eTTime]^+ = [r_w(\sinkTime)-r_v(\sinkTime)-\eTTime-\eToll]^+$. Thus, to prove \zcref{thm:Q*steadystates} we can use both descriptions interchangeably.
    
    By \zcref{lem:dual_feasible} and with the above consideration the defined dual solution is feasible.
    
    We aim to show optimality by complementary slackness. For this we need a feasible primal certificate. Unfortunately, we cannot simply use $x'(\sinkTime)$ as this does not satisfy flow conservation.
    
    Thus, we define the primal flow $\bar y$ as limit over averages over a time windows of length $h > 0$. Let $y^h_e(\sinkTime)\coloneqq  \frac{x_e(\sinkTime + h)-x_e(\sinkTime)}{h}$. We will show that each $y^h_e(\sinkTime)$ is approximately feasible and optimal. Approximate feasibility in terms of capacity constraints implies boundedness of $y^h_e$ and we thus obtain a convergent subsequence of any sequence $\sinkTime_n\to \infty$ whose limit is the optimality certificate for $(d,\bar q)$. 
    
     We first show approximate capacity constraints and flow conservation. For this we need a technical lemma that converts the bound on $d$ to a bound on the label difference.
    \begin{lemma}\label{claim:bound_r_diff}\label{lem:bound_r_diff}
        Let $h>0$ and $\UpD$ be the constant from \zcref{lm:queues_bounded}. Then $\max\{0, h - \UpD\} \leq r_v(\sinkTime+h)-r_v(\sinkTime)\leq h +\UpD$.
    \end{lemma}
    \begin{proof}
        By \zcref{lm:queues_bounded}, $0\leq d_v(\sinkTime')\leq \UpD$ for any $\sinkTime'$. Using monotonicity of~$r_v$ and the definition $d_v(\sinkTime)= \sinkTime-r_v(\sinkTime)$, thus, implies $0 \leq r_v(\sinkTime+ h)-r_v(\sinkTime) = \sinkTime + h -d_v(\sinkTime+h) - \sinkTime +d_v(\sinkTime) = h + ( d_v(\sinkTime)-d_v(\sinkTime+h)) $.
        The last term is at most $h+ \UpD$ and at least $h- \UpD$.
    \end{proof}
    
    We show the following lemma in a slightly more general form than needed here, as we want to reuse it later. We do not exploit that our tolled dynamic equilibrium is in steady state. Instead we just exploit that queues are bounded. 
     
     \begin{lemma}\label{lem:approx_capacity_and_flow_conversation}
        Consider any equilibrium flow $\bar f$ not necessarily in steady state and the corresponding $\bar x$, let $\sinkTime \geq \crem[s][0]$ and $h>0$, and define $\bar y^h_e(\sinkTime) \coloneqq  \frac{\bar x_e(\sinkTime + h)-\bar x_e(\sinkTime)}{h}$. Then, $\bar y^h_e(\sinkTime)$ is nonnegative, and approximately satisfies the capacity constraints and flow conservation: 
         \begin{itemize}
             \item we have $0 \leq \bar y^h_e(\sinkTime)\leq \nu_e(1+\tfrac{\UpD}{h})$ for all $e \in E$, and  
             \item there is some constant $\bar U > 0$ such that
                \[\abs{\sum_{e \in \edgesFrom{v}} \bar y^h_e(\sinkTime) -u_0 \CharF[v=s] - \sum_{e \in \edgesTo{v}} \bar y^h_e(\sinkTime)}\leq \frac{\bar U}{h}\]
            for all $v \in V \setminus\set{t}$ and
                \[\sum_{e \in \edgesFrom{t}} \bar y^h_e(\sinkTime) - \sum_{e \in \edgesTo{t}} \bar y^h_e(\sinkTime)\leq 0\]
            for the sink.
         \end{itemize}
     \end{lemma}
     \begin{proof}
        Non-negativity of $\bar y^h_e(\sinkTime)$ is an immediate consequence of $x_e$ being non-decreasing. 
        Next, we show the capacity bound. Using the definition of $\bar x_e$ together with \zcref{lem:TolledInFlowOutFlow} and the fact that the arc outflow rate $\bar f^-_e$ from arc~$e$ is bounded by~$\eCap$, we get
            \begin{align*}
                \bar x_e(\sinkTime + h)-\bar x_e(\sinkTime) 
                    &= \bar F^-_e(r_w(\sinkTime-\eToll+h))-\bar F^-_e(r_w(\sinkTime-\eToll)) \\
                    &\leq \eCap\cdot(r_w(\sinkTime-\eToll+h)-r_w(\sinkTime-\eToll). 
            \end{align*}
        Together with \zcref{lem:bound_r_diff}, this gives us
            \begin{align}\label{eq:BoundDiffXhX}
                \bar x_e(\sinkTime + h)-\bar x_e(\sinkTime) \leq \eCap\cdot(h+U).
            \end{align}
        Thus, we have $\bar y^h_e(\sinkTime)\leq \nu_e(1+\tfrac{\UpD}{h})$.    
        
        Next, we show approximate flow conservation. Since $\sinkTime+h> \sinkTime\geq \crem[s][0]$, both indicators equal one. Consider the following difference of flow conservation equations from \zcref{cor:Fe+andFe-} for $v \in V\setminus\set{t}$:
        
        \begin{align}
            &u_0 \cdot \left( r_s(\sinkTime+h) - r_s(\sinkTime) \right)\label{eq:FlowConsAtSUpperBound}\\
                &\quad\quad=  \sum_{e \in \edgesFrom{v}}(\bar x_e(\sinkTime+h) -\bar x_e(\sinkTime))- \sum_{e \in \edgesTo{v}}(\bar x_e(\sinkTime+h+\eToll) -\bar x_e(\sinkTime+\eToll) )\notag\\
                &\quad\quad=  \sum_{e \in \edgesFrom{v}}h \bar y^h_e(\sinkTime) - \sum_{e \in \edgesTo{v}}h \bar y^h_e(\sinkTime)\notag\\ 
                &\qquad\qquad - \sum_{e \in \edgesTo{v}}\Bigl(\bigl(\bar x_e(\sinkTime+h+\eToll) - \bar x_e(\sinkTime+h)\bigr) - \bigl(\bar x_e(\sinkTime+\eToll)-\bar x_e(\sinkTime)\bigr)\Bigr) \notag
        \end{align}
        Using \zcref{claim:bound_r_diff}, we can upper bound the term in~\eqref{eq:FlowConsAtSUpperBound} by $u_0\CharF[v=s](h + \UpD)$ and lower bound it by $u_0\CharF[v=s](h - \UpD)$.    
        Rearranging and dividing by~$h$ then gives us
        \begin{align*}
                &\abs{\sum_{e \in \edgesFrom{v}} \bar y^h_e(\sinkTime) -u_0 \CharF[v=s] - \sum_{e \in \edgesTo{v}} \bar y^h_e(\sinkTime)}\\
                &\quad\quad\leq u_0\frac{U}{h} \CharF[v=s] +  \frac{1}{h} \sum_{e \in \edgesTo{v}}\Bigl|\bigl(\bar x_e(\sinkTime+h+\eToll) - \bar x_e(\sinkTime+h)\bigr) - \bigl(\bar x_e(\sinkTime+\eToll)-\bar x_e(\sinkTime)\bigr)\Bigr | \\
                &\quad\quad \symoverset{1}{\leq} u_0 \frac{\UpD}{h}+ \sum_{e \in \edgesTo{v}}\Bigl(\nu_e \frac{\eToll  + \UpD}{h} - 0\Bigr) \eqqcolon \frac{\bar U}{h},
        \end{align*}
        where, for~\refsym{1}, we use~\eqref{eq:BoundDiffXhX} and non-decreasingness of $\bar x_e$. This yields that both differences lie in $[0,\nu_e( \eToll + \UpD)]$ and thus also their absolute difference.
        
       Finally, for $v=t$, we observe that in an equilibrium flow, we have $f^+_e = 0$ for all $e \in \edgesFrom{t}$. Thus, the claimed bound holds trivially.
    \end{proof}
    
    Clearly we can apply the lemma directly to $y^h$. Thus, these properties together give feasibility of a limit flow $\bar y$ for $h \to \infty$. We now show that $y^{h}$ approximately satisfies complementary slackness. 
    
    \begin{lemma}\label{lem:ComplementarySlack}
        The vector $y^{h}$ approximately satisfies complementary slackness for all $e=vw \in E$:
            \begin{align*}
                \bar q_e > 0 &\implies y_e^{h} \geq \nu_e(1-\tfrac{ \UpD}{h}) \\
                d_v- d_w < \eTTime+\eToll+\bar q_e &\implies y_e^{h}=0
            \end{align*}
    \end{lemma}
    \begin{proof}
        Note that, by definition, $y_e^{h}(\sinkTime) = \frac{x_e(\sinkTime+h)-x_e(\sinkTime)}{h} = \tfrac 1h \int_\sinkTime^{\sinkTime+h}x'_e(z)dz$. For the first part of the \zcref[noref,nocap]{lem:ComplementarySlack}, note that if $\bar q_e >0$, then $q_e(\sinkTime')>0$ for all $\sinkTime' \geq \sinkTime$. 
    
        Thus, by \zcref{cor:bounds on x} together with \zcreftp{lem:CharaActiveAndResettingEdges:CharaResetting}, $x_e'(\sinkTime') = \nu_e r_w'(\sinkTime'-\eToll)$ for almost all $\sinkTime'\geq \sinkTime$ and, therefore,
        \begin{align*}
            y_e^{h}(\sinkTime) 
                &= \frac {\nu_e}h \int_\sinkTime^{\sinkTime+h}r'_w(z-\eToll)\diff z\\
                &= \frac {\nu_e}h \int_\sinkTime^{\sinkTime+h}(1-d'_w(z-\eToll))\diff z\\
                &= \tfrac {\nu_e}h (h-[d_w(\sinkTime+h-\eToll)-d_w(\sinkTime-\eToll)])\\
                &\geq \tfrac {\nu_e}h (h- \UpD) \\
                &= \nu_e(1-\tfrac  \UpD h).
        \end{align*}
        Here, the second equality is by $d_v(\sinkTime)=\sinkTime-r_v(\sinkTime)$, and the inequality follows from the bound on $d$ in \zcref{lm:queues_bounded}.
    
        For the second part of the \zcref[noref,nocap]{lem:ComplementarySlack}, assume that the dual constraint is not tight, i.e., $d_v-d_w < \eTTime+\eToll+\bar q_e$. 
    
        In the language of $r$-labels this means
        \begin{align*}
            (\sinkTime- r_v(\sinkTime)) - (\sinkTime - r_w(\sinkTime)) &< \tau_e + p_e + q_e(\sinkTime)\\
            \Leftrightarrow r_w(\sinkTime) &< r_v(\sinkTime) + \tau_e + p_e + q_e(\sinkTime)\\
            \Leftrightarrow r_w(\sinkTime - p_e) + p_e&< r_v(\sinkTime) + \tau_e + p_e + q_e(\sinkTime)\\
            \Leftrightarrow r_w(\sinkTime' - p_e) &< r_v(\sinkTime') + \tau_e  + q_e(\sinkTime'),\\
        \end{align*}
    where $\sinkTime' \geq \sinkTime$ and the second last and last equivalence both exploited that in steady state labels change with a rate of 1 and queues remain constant.
    
        Thus, a non-tight dual constraint corresponds to an inactive arc by \zcreftp{lem:CharaActiveAndResettingEdges:CharaActive}. By \zcref{cor:TolledEqChara}, these arcs have no flow in equilibrium and thus $x_e'(\sinkTime) =0$ for all $\sinkTime' \geq \sinkTime$ which implies
        $y_e^{h}=0$ by definition.
    \end{proof}
    
    With this, we can now show existence, feasibility and optimality of a limit flow $\bar y$: Since the sequence~$y^h$ is bounded by \zcref{lem:approx_capacity_and_flow_conversation}, there must exist a convergent subsequence $y^{h_n}$ with some limit point~$\bar y$. As the capacity and flow conservation errors in \zcref{lem:approx_capacity_and_flow_conversation} vanish in the limit, $\bar y$ is a feasible flow. Hence, it is a feasible solution to the primal problem~\eqref{P}.

    Moreover, by \zcref{lem:ComplementarySlack}, it satisfies complementary slackness with $(d,\bar q)$ and, thus, certifies the latter's optimality for the dual problem~\eqref{D}: This, again, holds since, for $h \to \infty$, the error for the saturation bound vanishes while the second complementary slackness constraint is already satisfied for all points in the sequence. Thus, obtain optimality of $(d,\bar q)$ and, in particular, $\bar q \in Q^*$.
\end{proof}

Finally, we now turn to our convergence result. Here, we first show the technical version of our result stating, that for every sequence of times diverging to infinity, there exists a subsequence such that the corresponding waiting times converge to some steady state vector $\bar q \in Q^*$. This, immediately implies that for any tolled dynamic equilibrium the trajectory of waiting times approach the set~$Q^*$ (i.e., \zcref{thm:convergence_queues}).

\thmqueuesconverge*

\begin{proof}
    Again, we show this by giving a primal solution $\bar y$ of \lpref{P} together with a dual solution $(\bar d,\bar q)$ of \lpref{D} certifying optimality so that $\bar q\in Q^*$.
    First, as the queues remain bounded over any sequence of deadlines (by \zcref{thm:queues_bounded}), there exists a convergent subsequence by Bolzano-Weierstrass and, thus, a limit $\bar q$. To simplify the notation, we will assume that $\sinkTime_n$ is already that sequence. Moreover, we may assume that $\sinkTime_n \geq \crem[s][0]$ by discarding only finitely many terms.
    Based on this, we define 
        \[\bar d_v \coloneqq \min_{P\in \PathSet[v,t]}\sum_{e\in P}(\eTTime+\eToll+\bar q_e) \text{ for } v \in V\setminus\set{t} \quad\text{and}\quad \bar d_t=0.\]
    Then, the pair $(\bar d,\bar q)$ is dual feasible as $\bar q \geq 0$, $\bar d_t=0$ and $\bar d_v-\bar d_w \leq \eTTime+\eToll+\bar q_e$ for any $e=vw\in E$. The latter holds as $e$ together with the \stpath[w] attaining the minimum in $\bar d_w$ is one of the paths over which the minimum in $\bar d_v$ is taken.

    We now define the primal flow $\bar y$ via an average over a time window of length $h > 0$. Let $y_e^{n,h}\coloneqq  \frac{x_e(\sinkTime_n+h)-x_e(\sinkTime_n)}{h}$. By \zcref{lem:approx_capacity_and_flow_conversation}, each $y_e^{n,h}$ is approximately feasible. 
    
    Next, we show that $y^{n,h}$ approximately satisfies complementary slackness (i.e., the analogue to \zcref{lem:ComplementarySlack} in the proof of \zcref{thm:Q*steadystates}). For this, we require the following \zcref[noref,nocap]{claim:window_q_close}.
    
    \begin{claim}\label{claim:window_q_close}
        Recall that $\sumTolls \coloneqq \sum_{e \in E}\eToll$ and let $\mathcal I_n \coloneqq [\sinkTime_n-2\sumTolls,\sinkTime_n+h]$ for $n\in \mathbb N$. Then, we have
        \begin{align} 
            \max_e \sup_{\sinkTime \in \mathcal I_n}\abs{q_e(\sinkTime)-\bar q_e} &\xrightarrow{n \to \infty} 0 \label{eq:queues_limit_close} \\
            \sup_{\sinkTime \in [\sinkTime_n-\sumTolls,\sinkTime_n+h]}\abs{d_v(\sinkTime)-\bar d_v} &\xrightarrow{n \to \infty} 0 \label{eq:d_limit_close}
        \end{align}
    \end{claim}
    \begin{proofClaim}
        For \eqref{eq:queues_limit_close} we use triangle inequality to get
            \[\sup_{\sinkTime \in \mathcal I_n}\abs{q_e(\sinkTime)-\bar q_e} \leq \sup_{\sinkTime \in \mathcal I_n}\abs{q_e(\sinkTime)- q_e(\sinkTime_n)} + \abs{q_e(\sinkTime_n)-\bar q_e}.\]
        Here then, the first term converges to~$0$ due to \zcref{lm:freezing} while the second converges to~$0$ by definition of~$\bar q_e$.
        
        Before, we prove \eqref{eq:d_limit_close}, we first spell out the Bellman equation for $d$ which follows directly form \zcref{lem:rLabelsAlternativeEquation} and the definition  $d_v(\sinkTime)=\sinkTime -r_v(\sinkTime)$. 
        Plugging in this definition in \zcref{lem:rLabelsAlternativeEquation} yields
    
        \begin{equation}
            \label{eq:bellmann_d}
            d_v(\sinkTime) = \begin{cases} 0 & \text{if }v=t\\
            \min_{vw \in E} d_w(\sinkTime-\eToll) + \eTTime + \wtST(\sinkTime) +\eToll &\text{else.}
            \end{cases}
        \end{equation}
    
        Next, we introduce some notation. Given any \stpath[v] $P=e_1,.., e_k$, we define $\bar\sinkTime_0\coloneqq\sinkTime$ and $\bar\sinkTime_i\coloneqq\bar\sinkTime_{i-1} -p_{e_i}$ for $i \in [k]$.
        By \eqref{eq:queues_limit_close}, there is some $\eps_n$ with $\eps_n \to 0$ as $n \to \infty$ such that 
        $\abs{q_{e_i}(\bar\sinkTime_{i-1}) - \bar q_{e_i}} \leq \eps_n$.
        
        Let $P$ be a simple \stpath[v] attaining the minimum in the definition of $\bar d_v$, i.e. 
        \[
        \bar d_v = \sum_{e \in P} (\tau_e + \eToll + \bar q_e).
        \]
        For all $e_i=v_{i-1}v_i \in P$ we get from~\eqref{eq:bellmann_d} that 
        \[
        d_{v_{i-1}}(\bar\sinkTime_{i-1}) \leq d_{v_i}(\bar\sinkTime_i) + \tau_e +\eToll + q_e(\bar\sinkTime_{i-1}).
        \]
        Thus, with $P=e_1,...,e_k$ and iteratively applying the above, we obtain 
        \[d_v(\sinkTime) \leq \sum_{i=1}^k\bigl(\eTTime[e_i]+\eToll[e_i]+q_{e_i}(\bar\sinkTime_{i-1})\bigr) = \bar d_v + \sum_{i=1}^k\bigl(q_{e_i}(\bar\sinkTime_{i-1})-\bar q_{e_i}\bigr) \leq \bar d_v + \abs{V}\eps_n.\]
        
        Now, we choose a simple \stpath[v] $P'=e_1, ..., e_k$ such that at time $\bar\sinkTime_{i-1}$ the Bellman equation at arc $e_i=v_{i-1}v_i$ is attained with equality along~$P'$. We can choose the path to be simple as $r$ can only increase along tight paths, so nodes cannot be revisited. Then, applying the definition of $\bar d_v$, we get 
        \[d_v(\sinkTime) = \sum_{i=1}^k\bigl(\eTTime[e_i]+\eToll[e_i]+q_{e_i}(\bar\sinkTime_{i-1})\bigr) \geq \bar d_v + \sum_{i=1}^k\bigl(q_{e_i}(\bar\sinkTime_{i-1})-\bar q_{e_i}\bigr) \geq \bar d_v - \abs{V}\eps_n.\]
        
        Overall, $\abs{d_v(\sinkTime)-\bar d_v} \leq \abs{V}\eps_n$ for all $v \in V$. As $\eps_n$ tends to~$0$ as $n \to \infty$, the claimed convergence \eqref{eq:d_limit_close} follows.      
    \end{proofClaim}
    
    \begin{claim}\label{cl:ComplementarySlackness}
        The vector $y^{h,n}$ approximately satisfies complementary slackness, that is, for all $e=vw \in E$ and $h > 0$ we have:
            \begin{align*}
                \bar q_e > 0 &\implies y_e^{n,h} \geq \nu_e(1-\tfrac{U}{h}) \quad&\text{ for $n$ sufficiently large,}\\
                \bar d_v-\bar d_w < \eTTime+\eToll+\bar q_e &\implies y_e^{n,h}=0 \quad&\text{ for all but finitely many $n$}.
            \end{align*}
    \end{claim}

    The proof of this \zcref[noref,nocap]{cl:ComplementarySlackness} follows along very similar lines to that of \zcref{lem:ComplementarySlack}.
    
    \begin{proofClaim}
        Note that, by definition, $y_e^{n,h} = \frac{x_e(\sinkTime_n+h)-x_e(\sinkTime_n)}{h} = \tfrac 1h \int_{\sinkTime_n}^{\sinkTime_n+h}x'_e(z)\diff z$. For the first part of the \zcref[noref,nocap]{cl:ComplementarySlackness}, we use that by \zcref{claim:window_q_close}, $\bar q_e >0$ implies $q_e(\sinkTime)>0$ for $\sinkTime\in \mathcal I_n$ with $n$ sufficiently large. By \zcref{cor:bounds on x}, $x_e'(\sinkTime) = \nu_e r_w'(\sinkTime-\eToll)$ for almost every $\sinkTime\in \mathcal I_n$ for such~$n$. As in the proof of \zcref{lem:ComplementarySlack}, we can then bound $y_e^{n,h}$ as follows:
        \begin{align*}
            y_e^{n,h} 
                &= \frac {\nu_e}h \int_{\sinkTime_n}^{\sinkTime_n+h}r'_w(z-\eToll)\diff z\\
                &= \frac {\nu_e}h \int_{\sinkTime_n}^{\sinkTime_n+h}(1-d'_w(z-\eToll))\diff z\\
                &\geq \tfrac {\nu_e}{h} (h-\UpD) = \nu_e(1-\tfrac{\UpD}{h}).
        \end{align*}
        Here, the second equality is by $d_v(\sinkTime)=\sinkTime-r_v(\sinkTime)$, and the inequality follows from the bound on $d$ in \zcref{lm:queues_bounded}.
    
        For the second part of the \zcref[noref,nocap]{cl:ComplementarySlackness}, consider an arc~$e=vw$ with $\bar d_v-\bar d_w < \eTTime+\eToll+\bar q_e$ and assume, for contradiction, that we had $y_e^{n,h}>0$ for infinitely many~$n$. Then, by definition of $y_e^{n,h}=\tfrac 1h \int_{\sinkTime_n}^{\sinkTime_n+h}x'_e(z)dz > 0$ and $x_e'(\sinkTime)=0$ for almost all $\sinkTime$ such that $e\notin E'_r(\sinkTime)$, there exists a sequence $\sinkTime^{(n)} \in [\sinkTime_n,\sinkTime_n+h]$ such that $r_v(\sinkTime^{(n)})+\wtST(\sinkTime^{(n)})+\eTTime = r_w(\sinkTime^{(n)}-\eToll)$ for all such~$n$. Together with $d_v(\sinkTime)=\sinkTime-r_v(\sinkTime)$, this implies $d_v(\sinkTime^{(n)}) -d_w(\sinkTime^{(n)}-\eToll) = \eTTime + \eToll + q_e(\sinkTime^{(n)})$ and 
            \begin{align}\label{eq:ComplementarySlackness:Convergence}
                q_e(\sinkTime^{(n)})-d_v(\sinkTime^{(n)})+d_w(\sinkTime^{(n)}-\eToll)-(\bar q_e-\bar d_v+\bar d_w)\overset{n\to \infty}{\longrightarrow} 0.
            \end{align}
        This holds as $\abs{q_e(\sinkTime^{(n)})-\bar q_e}$, $\abs{d_v(\sinkTime^{(n)})-\bar d_v}$ and $\abs{d_w(\sinkTime^{(n)}-\eToll)-\bar d_w}$ all go to 0 as $n\to \infty$ by \zcref{claim:window_q_close} as all relevant evaluation times lie in $[\sinkTime_n-\sumTolls,\sinkTime_n+h]$.
        At the same time, we have $q_e(\sinkTime^{(n)})-d_v(\sinkTime^{(n)})+d_w(\sinkTime^{(n)}-\eToll) = -( \eTTime + \eToll)$ for all $n$ and, hence, the sequence in \eqref{eq:ComplementarySlackness:Convergence} is actually a constant sequence. In particular, we have $\bar d_v-\bar d_w = \eTTime+\eToll+\bar q_e$, which is a contradiction to our initial assumption. 
    \end{proofClaim}

        Now $(y_e^{n,h})_n$ is bounded for fixed $h$ and we obtain for the limit $y_e^h$ of a convergent subsequence that $y_e^h\leq \nu_e(1+\tfrac{U}{h})$ and that $|-u_0 \CharF[v=s] +\sum_{e \in \edgesFrom{v}} y_e^{h} - \sum_{e \in \edgesTo{v}} y_e^{h}|\leq \frac{\bar U} h .$ Again, in an equilibrium flow, we have $f^+_e = 0$ for all $e \in \edgesFrom{t}$ and the claimed bound holds.
         Now letting $h\to \infty$, we again consider the limit $\bar y$ of a convergent subsequence. Both the capacity and flow conservation errors vanish in the limit. Thus, $\bar y$ is a feasible flow.
         For optimality, observe that the error in complementary slackness constraints vanishes as well, so $\bar q \in Q^*$.
\end{proof}

\section{Paper Overview}\label{app:PaperOverview}

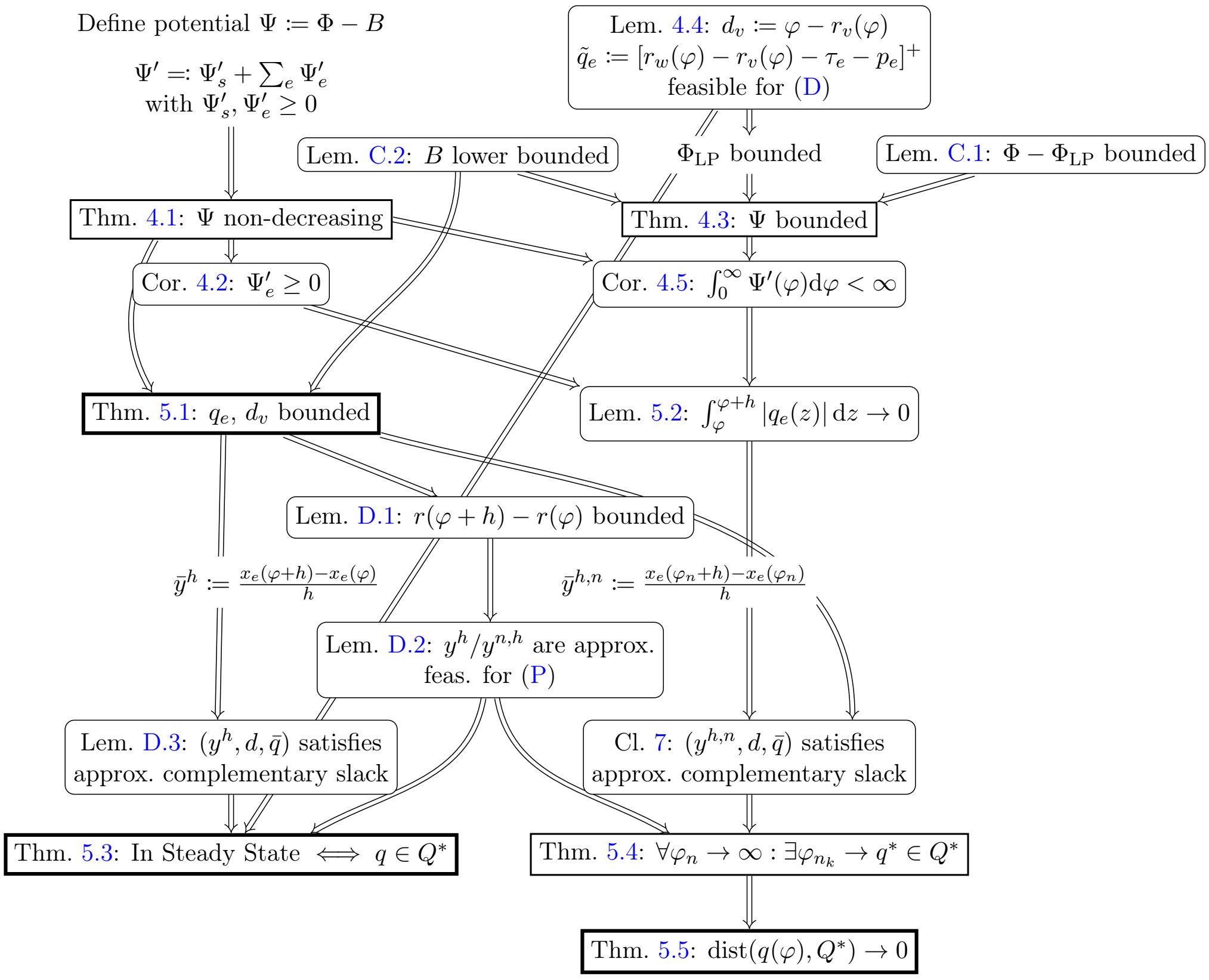
\begin{figure}[h]\centering
	\begin{adjustbox}{max width=\textwidth}
		{
\colorlet{backgroundGray}{black!5!white}

\tikzstyle{Thm} = [rectangle, draw, thick, text centered,fill=white]
\tikzstyle{MainThm} = [Thm, ultra thick,fill=white]
\tikzstyle{Lem} = [rectangle, draw, text centered, rounded corners,fill=white]
\tikzstyle{Text} = [fill=white]

\tikzstyle{Implies} = [draw, arrows=-implies, double distance=1.7pt,nfold]
\tikzstyle{ImpliesB} = [draw=white, line width=6pt]
\tikzstyle{UsedInB} = [draw=white, line width=4pt]
\tikzstyle{UsedIn} = [draw, ->]

\pgfdeclarelayer{background}
\pgfdeclarelayer{arrows}
\pgfsetlayers{background,arrows,main}
\begin{tikzpicture}[node distance = 4cm]
    \node[Thm](PsiMonotone){\zcref[abbrev]{thm:PotNondecreasing}: $\fullpot$ non-decreasing};
        \node[Lem,below of=PsiMonotone,node distance=1cm](PsieMonotone){\zcref[abbrev]{not:potential_deriv_e}: $\fullpot'_e \geq 0$};
        
    \node[Thm,right of=PsiMonotone,node distance=8cm](PsiBounded){\zcref[abbrev]{thm:pot bounded}: $\fullpot$ bounded};
        \node[Text,above of=PsiBounded,node distance=1cm](PotDBounded){$\Phi_{\mathrm{LP}}$ bounded};
        \node[Lem,left of=PotDBounded,node distance=4.5cm](BBounded){\zcref[abbrev]{lm:B_bound}: $\potCorr$ lower bounded};
        \node[Lem,right of=PotDBounded,node distance=4.5cm](PhiDiffBounded){\zcref[abbrev]{lem:DifferenceBetweenPots}: $\potBasic-\Phi_{\mathrm{LP}}$ bounded};
        \node[Lem,below of=PsiBounded,node distance=1cm](PotIncreaseBounded){\zcref[abbrev]{cor:dissipation_finite}: $\int_0^\infty\fullpot'(\sinkTime)\diff \sinkTime < \infty$};

     \node[MainThm,below of=PsiMonotone, node distance=3cm](QueuesBounded){\zcref[abbrev]{thm:queues_bounded}: $\wtST$, $d_v$ bounded};

     \node[Lem,below of=PsiBounded, node distance=3cm](QueueVariationBounded){\zcref[abbrev]{lm:freezing}: $\int_\sinkTime^{\sinkTime+h}\abs{\wtST(z)}\diff z \to 0$};

     \node[MainThm,below of=QueuesBounded, node distance=6.8cm,align=center](SteadyStateIffinQ){\zcref[abbrev]{thm:Q*steadystates}: In Steady State $\iff \wtST[] \in Q^*$};
        
     \node[Thm,below of=QueueVariationBounded, node distance=6.8cm](SteadyStateConvergenceTechn){\zcref[abbrev]{thm:queues_converge_subsequences}: $\forall \sinkTime_n \to \infty: \exists \sinkTime_{n_k} \to q^* \in Q^*$};

        \node[Lem,above of=SteadyStateIffinQ,node distance=1.5cm,align=center](ApproxComplSlack){\zcref[abbrev]{lem:ComplementarySlack}: $(y^h,d,\bar q)$ satisfies\\approx.\  complementary slack};
        \node[Lem,node distance=1.5cm,align=center](ApproxFeas)at($(SteadyStateIffinQ)!.5!(SteadyStateConvergenceTechn)+(0,3cm)$){\zcref[abbrev]{lem:approx_capacity_and_flow_conversation}: $y^h / y^{n,h}$ are approx.\\ feas. for~\eqref{P}};
        	\node[Text,above of=ApproxFeas, node distance=1.2cm,xshift=-3.3cm](DefYh){$\bar y^{h} \coloneqq  \frac{x_e(\sinkTime + h)-x_e(\sinkTime)}{h}$};
        	\node[Text,above of=ApproxFeas, node distance=1.2cm,xshift=3cm](DefYnh){$\bar y^{h,n} \coloneqq  \frac{x_e(\sinkTime_n + h)-x_e(\sinkTime_n)}{h}$};
        \node[Lem,above of=ApproxFeas,node distance=2.2cm,align=center](rDiffBounded){\zcref[abbrev]{lem:bound_r_diff}: $r(\sinkTime+h)-r(\sinkTime)$ bounded};
        
        \node[Lem,above of=SteadyStateConvergenceTechn,node distance=1.5cm,align=center](ApproxComplSlackB){\zcref[abbrev]{cl:ComplementarySlackness}: $(y^{h,n},d,\bar q)$ satisfies\\approx.\  complementary slack};

     \node[MainThm,below of=SteadyStateConvergenceTechn, node distance=1.5cm](SteadyStateConvergence){\zcref[abbrev]{thm:convergence_queues}: $\dist(\wtST[](\sinkTime),Q^*) \to 0$};

    \node[Lem,above of=PotDBounded,node distance=1.5cm,align=center](DFeas){\zcref[abbrev]{lem:dual_feasible}: $d_v \coloneqq \sinkTime - r_v(\sinkTime)$\\$\qt_e \coloneqq [r_w(\sinkTime)-r_v(\sinkTime)-\eTTime-\eToll]^+$\\ feasible for \eqref{D}};

    \node[Text,above of=DFeas,node distance=4cm]{\begin{minipage}[t]{0.25\textwidth}\centering
        Primal \eqref{P}:
        \begin{align*}
          \min_{f}\ & \sum_{e\in E}(\eTTime+\eToll)\, f_e \\
          \text{s.t.}\       & f\in\mathcal F_{u_0}\\
                             & 0\le f_e\le \eCap
        \end{align*}
        \end{minipage}%
        \hspace{.1cm}
        \begin{minipage}[t]{0.4\textwidth}\centering
        Dual \eqref{D}:
        \begin{align*}
          \max_{d\!,\ \qt}\ & u_0 d_s-\sum_{e\in E}\nu_e\,\qt_e \eqqcolon \Phi_{\mathrm{LP}} \\
          \text{s.t.}\ & d_t=0\\
                       & d_v-d_w\le \eTTime+\eToll+\qt_e \\
                       & \qt_e\ge 0 \quad 
        \end{align*}
        $Q^* \coloneqq \set{q^* | (q^*,d^*) \text{ opt.\ for } \eqref{D}}$
    \end{minipage}};

    \node[Text,above of=PsiMonotone,node distance=2cm,align=center](PotDeriv){$\fullpot' \eqqcolon \Psi'_s + \sum_e \Psi'_e$\\ with $\Psi'_s,\Psi'_e \geq 0$};

    \node[Text,above of=PotDeriv,node distance=1cm,align=center](PotDef){Define potential $\fullpot \coloneqq \potBasic - \potCorr$};

    \begin{pgfonlayer}{arrows}
        \draw[Implies](PotDeriv)--(PsiMonotone);
        \draw[Implies](PsiMonotone)--(PsieMonotone);

        \draw[Implies](PsiMonotone.east)--(PotIncreaseBounded.north west);
        \draw[Implies](PsiBounded)--(PotIncreaseBounded);

        \draw[Implies](BBounded)--(PsiBounded.north west);
        \draw[Implies](PotDBounded)--(PsiBounded);
            \draw[Implies](DFeas)--(PotDBounded);
        \draw[Implies](PhiDiffBounded)--(PsiBounded.north east);

        \draw[Implies](PsiMonotone.195)to[out=-120,in=120](QueuesBounded.165);
        \draw[Implies](BBounded)to[out=-90,in=60](QueuesBounded.15);

        \draw[Implies](PsieMonotone)--(QueueVariationBounded.north west);
        \draw[Implies](PotIncreaseBounded)--(QueueVariationBounded);

        \draw[Implies](QueuesBounded)--(rDiffBounded);
        \draw[Implies](rDiffBounded)--(ApproxFeas);
        
        \draw[Implies](QueuesBounded.250)--(ApproxComplSlack.110);

        \draw[Implies](ApproxComplSlack)--(SteadyStateIffinQ);
        \draw[Implies](ApproxFeas)to[out=-100,in=30](SteadyStateIffinQ.15);
        \draw[Implies](DFeas)--(SteadyStateIffinQ);

        \draw[Implies](QueuesBounded.south east)to[out=-15,in=90](ApproxComplSlackB.20);
        \draw[Implies](QueueVariationBounded)--(ApproxComplSlackB);

        \draw[Implies](ApproxComplSlackB)--(SteadyStateConvergenceTechn);
        \draw[Implies](ApproxFeas)to[out=-80,in=150](SteadyStateConvergenceTechn.165);

        \draw[Implies](SteadyStateConvergenceTechn)--(SteadyStateConvergence);

    \end{pgfonlayer}
\end{tikzpicture}
}
	\end{adjustbox}
	\caption{An overview over the main proof structure of this paper.}\label{fig:Overview}
\end{figure}

The diagram in \zcref{fig:Overview} provides an overview of the main proof structure of this paper.

\end{document}